\documentclass[preprint,floatfix] {revtex4} 
\newcommand{\rvec}{\mathrm {\mathbf {r}}} 
\newcommand{\xvec}{\mathrm {\mathbf {x}}} 
\newcommand{\Rvec}{\mathrm {\mathbf {R}}} 
\usepackage{graphicx}
\usepackage{subfigure}
\usepackage{xcolor}
\usepackage{enumerate}
\usepackage{braket}
\usepackage{amsmath}

\usepackage{color, soul}
\definecolor{darkblue}{rgb}{0,0,0.5}
\setulcolor{darkblue}

\begin{document}

\title{DFT calculations of atoms and molecules in Cartesian grids}

\author{Abhisek Ghosal}

\author{Amlan K.~Roy$^\dag$}
\altaffiliation{Email: akroy@iiserkol.ac.in, akroy6k@gmail.com. \\                                            
$^\dag$AKR dedicates this chapter to his kind-hearted uncle Dr.~Pranjalendu Roy, on the occasion of his 67th birthday. Dr. 
Roy is a noted zoologist and social activist, who has inspired the author greatly.} 
\affiliation{Department of Chemical Sciences\\
Indian Institute of Science Education and Research (IISER) Kolkata, 
Mohanpur-741246, Nadia, West Bengal, India}

\begin{abstract}
Density functional theory (DFT) has emerged as one of the most versatile and lucrative approaches in electronic structure 
calculations of many-electron systems in past four decades. Here we give an account of the development of a variational DFT method 
for atoms and molecules \emph{completely} in a Cartesian grid. The non-relativistic Kohn-Sham equation is solved by using an 
LCAO-MO ansatz. Atom-centered localized basis set, electron density, molecular orbitals, two-body potentials are directly 
constructed on the grid. We adopt a Fourier convolution method for classical Coulomb potentials by making 
an Ewald-type decomposition technique in terms of short- and long-range interactions. It produces quite accurate and competitive
results for various properties of interest, such as component energy, total energy, ionization energy, potential energy curve, 
atomization energy, etc. Both local and non-local functionals are employed for pseudopotential as well as full calculations. 
While most results are offered in a \emph{uniform} grid, initial exploratory attempts are made in a \emph{non-uniform} grid, which 
can significantly reduce the computational overhead. This offers a practical, viable alternative to atom-centered grid-based 
implementations, currently exploited by the majority of programs available world-wide.


\vspace{5mm}
{\bf Keywords:} Density functional theory, many-electron system, exchange-correlation functional, 
Cartesian grid. 

\end{abstract}
\maketitle

\section{Introduction}
The past several decades have witnessed a proliferation of \emph{ab initio} quantum mechanical methods for elucidation 
of electronic \emph{structure, properties and dynamics} of many-electron systems such as atoms, molecules, clusters, 
solids. With rapid advances in mathematical formalisms and smart numerical algorithms, as well as computer architecture 
and resources, 
lately, some of these have been extended to the case of materials science, nano-science and biological science, etc., 
with unprecedented success. The \emph{electronic} Schr{\"o}dinger equation (SE) of such an $N$-electron system is 
essentially a many-body problem involving space, spin and time coordinates. This usually causes enormous complexity
and hence, \emph{exact} analytical solutions are unavailable in vast majority cases, leaving aside a few countable
idealized model problems which offer such solutions. Thus almost invariably recourse must be taken to approximate 
methods. So a well-defined theoretical framework of approximation is indeed necessary to explore the electronic 
problem in a systematic way as accurate as possible. Nowadays, it is not a mere contemplation to obtain an optimal 
solution for a system containing 100 or more electrons, with physically and chemically meaningful accuracy. However, 
as the number of electrons increases, computational complexity grows 
up exponentially, which requires new paradigms of development regarding basic theory, methodology, mathematical 
algorithm, sophisticated computer code and practical computational implementation. The ultimate objective is to probe
bigger and bigger systems with as much accuracy and efficiency, as possible. Significant strides have been made in 
last few decades in these aforementioned areas, and in the coming years, a lot of progress will be continued in order 
to reach the eventual goal of theory predicting experimental results in an explanatory level relevant to practical 
applications.

Besides ground-state electronic structure, a lot of important aspects are involved in atomic physics to explain the 
experimental observations in a satisfactory level. Some of these, e.g., regard excited states of 
discrete and continuum spectra, ionization dynamics of Rydberg states, spectroscopic levels, oscillator strengths for 
electronic transitions, fine-structures due to spin-orbit coupling, external perturbations due to static probes, 
multi-photon ionizations, above-threshold ionizations, electron scattering, relativistic corrections to the 
Hamiltonian and so on. But in case of molecules, calculation of physical and chemical properties are somehow less
straightforward and clouded, simply because of the presence of more than one nucleus. At a first glance, it 
requires a rigorous understanding of the molecular potential energy surface through construction of the Hessian matrix, 
and all other molecular properties are inherently dependent on it. Such a surface consists of a large number of 
maxima and minima at unknown locations, and the global minimum corresponds to the ground-state structure, whereas the 
paths between the minima are the primary focus of chemical reactions. It is worthwhile to mention  
a few relevant molecular properties, such as dipole and multipole moments, vibrational frequencies, molecular 
dissociation energies, linear and non-linear responses to external static probes, molecular scattering through electron 
beams, molecular reaction dynamics, nonlinear reaction dynamics, photochemistry, etc. A significant amount of interest
concerns with modern spectroscopic techniques, namely, electronic spectroscopy, photo-electron spectroscopy, 
ro-vibrational spectroscopy, nuclear magnetic resonance, nuclear quadrupole resonance, electron spin resonance, 
M\"ossbauer spectroscopy etc.

An integral consideration in modern quantum chemistry is the obligatory computational work, which has profound impact on 
future theoretical development. In order to perform the complex mathematical operations for larger systems within a
suitable framework, it is imperative to have a systematic way of computation, and sufficient intricacy to capture the
characteristic subtle interactions. Sometimes the computational output may be very crude (overestimating or underestimating), 
and other times it could be within the bound accuracy, but that is mostly dictated by the nature of the practical problem at 
hand. Now, in order for a given computational method to be successful and viable, the two deciding factors are its 
\emph{accuracy and cost effectiveness}. In today's computational chemistry, it is desirable to achieve the energy of a 
chemical reaction within the 
bounds of chemical accuracy, which is less than 1 kcal/mole. And the main challenge is to lower computational cost without 
much compromise on accuracy for increasingly bigger and complicated systems. Starting from relatively less accurate 
and dependable empirical or semi-empirical schemes to modern \emph{ab initio} methods, all fall under the purview of current 
computational quantum chemistry repertoire. The latter covers a large array of methods including Hartree-Fock (HF), post-HF, 
multi-reference, quantum Monte-Carlo, density functional theory (DFT), etc., each providing a different computational cost 
depending on the particular problem. In order to broaden the domain of feasibility to larger systems, significant developments 
have been made towards the so-called linear-scaling approaches, which slash computational scaling very swiftly with 
molecular size. In this context, it is appropriate to mention some of the most popular and prominent electronic structure 
packages, which are routinely used by the scientific community, such as Gaussian, GAMESS, ACES, CADPAC, Dalton, Hyperchem, 
Jaguar, Molcas, Molpro, NWChem, PQS, Psi4, Q-Chem, Spartan, TURBOMOLE, UTChem, etc. On the other hand, one could take help of
molecular mechanics which uses classical mechanics to model a molecular, as well as large biological system or material 
cluster, having much lower computational cost, but is also less sophisticated and reliable. These are usually applied in the
field of molecular dynamics (in the context of thermodynamic and kinetic properties of a system) and potential energy 
minimization (e.g., to calculate binding constants, protein folding kinetics, protonation equilibria, active site 
coordinates). In recent years, computational chemistry has also been used vigorously in medicinal chemistry, leading to the 
discovery of new life-saving drugs. Essentially this helps to model the synthetic compounds in a systematic way, providing 
some critical checks on huge labor and chemical cost. A large number of beautiful and in-depth books, reviews and articles 
are available on the subject. Some of them are referred here \citep{yarkony95, szabo96, simon97, kohn99, helgaker00,
springborg00, young01, foulkes01, lewars03, cramer04, martin04, hoffman07, jensen07, kaisas07, borden11, springborg12a,
springborg14}.

In order to solve the electronic SE for an interacting system, two distinct well-established theoretical formalisms have 
gained popularity and credence in the last few decades, with contributions from physicists and chemists alike. We begin our 
discussion with a comparatively simple HF theory, which in a sense, is founded on a mean-field approximation. In essence, 
the instantaneous inter-electronic interaction is accounted for, in an average way, where each electron sees an average field 
because of the presence of the other electrons, leading to the optimization of a single Slater determinantal wave function consisting 
of individual spin orbitals. Though it offers a crucial first step towards the problem, its application is limited due to 
a complete neglect of electron correlation effects. Post HF formalisms are mainly concerned on how this subtle effects could 
be incorporated in an efficient manner. This family includes a variety of versatile formalisms, such as M\o ller-Plesset 
perturbation theory (MPn) \citep{saebo93}, configuration interaction (CI) \citep{szabo96}, coupled-cluster ansatz (they come in 
many flavours like CCSD, CCSD(T), CCSDT, CCSDTQ, etc.) \citep{bartlett89}, multi-reference perturbation theory (such as CASPT2) 
\citep{roos96}, etc. These methods offer potentially authentic and reliable results, but are quite difficult to be implemented 
computationally for large $N$, 
primarily because of their unfavorable scaling. They differ in scaling, cost, efficiency and accuracy; naturally 
a major research effort for such methods is directed towards betterment of these issues. They have their own space of 
applicability depending upon a given problem and often provide a benchmark of accuracy for the system under investigation.

The second approach relies on utilizing the limited information coming from electron density, density matrix or 
Green's function in an optimized way with the help of the variational principle. Amongst them, however, DFT
has appeared as the most versatile and outstanding method in electronic structure calculation. 
Many beautiful books and reviews are available on the subject; some recent ones are \citep{parr89, chong95, seminario96,
joubert98, dobson98, koch01, young01, fiolhais03, gidopoulos03, hu08, cramer09, cohen11, burke12, roy12, becke14, jones15}. 
The basic variable, the single-particle electron density, $\rho(\rvec)$,
is a real-valued and experimentally realizable quantity, in contrast to the traditional complex-valued $N$-electron wave 
function. Furthermore,
the most attractive part is that it incorporates the complicated electron correlation effects in a transparent and elegant way. 
Being a single determinant approach, it is computationally more tractable; the burden remains at 
same level as HF. However, this apparent simplicity and novelty is somehow tainted and compensated by the hitherto unavailable,
all-important, universal, exchange-correlation (XC) density functional. Although uncovering its exact form constitutes one of 
the most active and fertile area of research, unfortunately it still remains elusive; hence it must be approximated. Success 
of DFT lies heavily on the choice of these approximate XC functional. Historically, the journey started with the so-called 
local density approximation (LDA) in 1930 by Dirac \citep{dirac30} using the form of exchange assuming a uniform electron gas 
model. Later on, a working LDA correlation part was developed \citep{vosko80, perdew92}. However, the LDA XC functional suffers 
from serious problems in atoms, molecules and solids, as the density of such systems are far apart from the uniform electron gas. 
Since it is essential to have a good knowledge of the gradient of the electron density, a whole family of so-called generalized 
gradient approximation (GGA)-based functionals were proposed over the years. A few prominent ones are: B88 \citep{becke88a}, 
LYP \citep{lee88}, PBE \citep{perdew96} etc. In the next step, higher-order derivatives of the density including the kinetic energy 
density lead to the development of meta-GGA kind of functionals, such as B88C \citep{becke88}, Becke-Rousse \citep{becke89}, 
TPSS \citep{tao03} etc. Next crucial step lead to hybrid functionals, which bring exact HF exchange into the 
picture. Depending on the co-factor of different kinds of XC functionals like LDA, GGA with latter, a variety of hybrid 
functionals have been published, which include one of the most popular and versatile candidates in quantum chemistry, namely, 
B3LYP \citep{becke88a, lee88}. The LDA functional typically does not have an impressive energetic performance, but interestingly it 
executes better performance 
than GGA functional regarding geometry optimization. Generally it is true that hybrid functionals work better than LDA 
due to incorporation of HF exchange. But in case of transition-metal chemistry and response properties, hybrid functionals 
show rather quite poor performance. Also, a majority of these approximate functionals 
has some sort of deficiency regarding activation energies of chemical reactions, which is somehow circumvented in the 
recently developed range-separated, density fitting, adiabatic-connection, local hybrid functionals or those 
involving unoccupied orbitals and eigenvalues, etc. These are found to be largely good for dispersion and van der Waals 
interactions. A vast amount of literature exists on the topic; the interested reader may look 
at the lucid reviews \cite{cohen11, becke14, jones15} and references therein.  

Triumphs of DFT is now well-established in quantum chemistry and condensed matter physics; so much so that nowadays 
it is the most dominant and 
visible workforce. For large scale computation of materials, two distinct numerical approaches are engaged to solve the 
relevant many-electron equation. The condensed matter community chooses plane-wave basis sets, where periodic boundary conditions 
are appropriate. On the other hand, chemists prefer a localized atom-centred Gaussian basis set, because of being fondly 
attached with molecules (non-periodic systems), where the problematic multi-center integrals can be evaluated analytically. With
advances in computational facilities, modern DFT can provide results with benchmark accuracy accompanied with experimental 
results at least for small molecules in gaseous phase. Throughout the past several decades, a wide variety of noteworthy 
applications were made regarding molecular properties (including structures, thermo-chemistry, various spectroscopic 
quantities, responses to external perturbations), bulk and surface properties of solids, band-gaps and optical properties of 
solids, interactions of small molecules with surfaces with focus on structures, binding energies and catalytic chemistry, 
modelling of photochemical reactions, etc. Of late, extensive applications were reported in some other areas, such as
nano-technology, semiconductor quantum dots, quantum confinement imposed by various kinds of 
nano structures. Practical applications from biological fields include modelling of enzymatic catalysis and active-site 
chemistry, cooperative activity in backbone of hydrogen bonding, modelling of beta-sheet formation, enzyme functioning etc. 

The present chapter gives an account of the work done in our laboratory for static, non-relativistic ground states in 
many-electron atoms/molecules. Within the Born-Oppenheimer approximation and the Hohenberg-Kohn-Sham framework, an 
implementation of DFT in a Cartesian Coordinate grid (CCG) is offered. By using a linear combination of Gaussian functions,
molecular orbitals (MO) and quantities like basis functions, electron density as well as 
classical Hartree and non-classical XC potentials are constructed on a 3D real CCG directly \cite{roy08,
roy08a,roy09,roy10,roy11}. No additional auxiliary basis is used for the charge density. A Fourier 
convolution method, involving a combination of Fast Fourier transform (FFT) and inverse FFT \cite{martyna99,minary02} is 
used to obtain the
Coulomb potential quite accurately. Analytical one-electron Hay-Wadt-type effective core potentials 
\cite{wadt85,hay85}, made of sums of Gaussian type functions, are used to represent the inner core electrons 
whereas energy-optimized truncated Gaussian bases are used for valence electrons. Viability and suitability of this 
simpler grid is compared and contrasted with a routinely used atom-centered grid (ACG) for a decent number 
of atoms, molecules. This is demonstrated by presenting the total energy, energy components, orbital energy, 
potential energy curve, atomization energy for a bunch of local and non-local XC functionals.

\section{The methodology}
\subsection{Electron density as basic variable}
In the usual wave function-based approach, the difficulty to solve the many-electron SE increases rapidly with the number 
of electrons present in the system. In order to reduce the dimensionality of the problem, one can introduce the concept
of reduced density matrices \citep{lowdin95}. One- and two-particle reduced density matrices, which 
are adequate for the evaluation of expectation values of one- and two-particle operators, are defined as, 
\begin{equation}
\rho_1(\xvec_1^{'}|\xvec_1)=N\int \cdots \int \psi(\xvec_{1}^{'},\xvec_2, \cdots ,\xvec_{N}) \psi^{\star}
(\xvec_{1},\xvec_2, \cdots ,\xvec_{N}) d\xvec_{2} \cdots d\xvec_{N}
\end{equation}
\begin{equation}
\tau_{2}(\xvec_{1}^{'},\xvec_{2}^{'}|\xvec_{1},\xvec_{2})=\frac{N(N-1)}{2} \int \cdots \int\psi 
(\xvec_{1}^{'},\xvec_{2}^{'},\xvec_3, \cdots ,\xvec_{N}) 
\psi^{\star}(\xvec_{1},\xvec_{2},\xvec_3, \cdots ,\xvec_{N}) d\xvec_{3} \cdots d\xvec_{N}
\end{equation}
where each of ($\{\xvec_{i}\}; \ i=1,2,3, \cdots ,N$) consists of both spatial coordinates $\{\rvec_{i}\}$ as well as 
spin coordinates ($\sigma_{i}$). Now, the spin-less, single-particle density $\rho(\rvec)$, which is the diagonal element 
of one-particle density matrix, is given as,
\begin{equation}
\rho(\rvec)=N \int \cdots \int \psi(\rvec \sigma,\xvec_{2}, \cdots ,\xvec_{N}) 
\psi^{\star}(\rvec \sigma, \xvec_{2}, \cdots ,\xvec_{N}) \ d \sigma d\xvec_{2} \cdots d\xvec_{N}.
\end{equation}
In case of fermions, it actually describes a three-dimensional distribution of electrons in a system. In sharp 
contrast to the complex-valued wave function, $\rho(\rvec)$ is a function of only three coordinates, irrespective of 
the number of electrons present in it. As it is a fundamental physical observable and can be determined experimentally, 
this enables one to test directly the veracity of chemical calculations and approximations. The beauty of DFT lies in 
formulating a many-particle problem within a single-particle framework with the aid of density being considered as a 
central quantity. 

The first seminal work in this direction was provided independently by 
Thomas \citep{thomas27} and Fermi \citep{fermi27, fermi28}. They proposed a model for calculating atomic properties based 
on $\rho(\rvec)$ alone. For the kinetic energy, a local density approximation of a uniform, homogeneous electron gas 
model was suggested. On the other hand, the electron-electron repulsion energy was obtained from the classical Coulomb potential. 
This leads to the familiar Thomas-Fermi (TF) Euler-Lagrange equation for the density as, 
\begin{equation}
 \mu_{\mathrm{TF}}=\frac{\delta E_{\mathrm{TF}}[\rho (\rvec)]}{\delta\rho(\rvec)}=\frac{5}{3} 
 C_{\mathrm{F}}\rho(\rvec)^{\frac{2}{3}}+\int 
d\rvec^{\prime}\frac{\rho(\rvec^{\prime})}{|{\rvec-\rvec^{\prime}}|}+v_{\mathrm{ext}}(\rvec),
\end{equation}
where $C_{\mathrm{F}} = \frac{3}{10}(3\pi^{2})^{\frac{2}{3}}$, $v_{\mathrm{ext}}(\rvec)$ signifies the external potential at 
point $\rvec$ 
consisting of nuclear-electron attraction and $\mu_{\mathrm{TF}}$ is, as usual, the Lagrange multiplier, to be identified as chemical 
potential. Thus we have the first genuine density functional for the energy of an interacting system. Though TF theory 
approximately describes charge density, electrostatic potential and total energy (at least 
for atoms and periodic solids), there are serious deficiencies due to a lack of proper description of the charge density at 
outer regions of atoms. Further, it is unable to produce the periodic variation of many properties with changing atomic number 
Z, due to the lack of a shell structure in the atoms. Besides, it does not bind atoms (neutral or ionic) to form molecules 
and solids. In a way, its utility is very limited to atomic systems.

Within this approximate framework, Dirac \citep{dirac30} first made a correction to the TF energy, incorporating exchange 
effects into it. He derived an exchange term from exchange energy of a homogeneous electron gas of density 
$\rho (\rvec)$ by reforming the HF theory solely in terms of \emph{density function}. This gives the \emph{Thomas-Fermi-Dirac}
(TFD) equation as,   
\begin{equation}
E_{\mathrm{TFD}}[\rho(\rvec)] = C_{\mathrm{F}}\int\rho(\rvec)^{\frac{5}{3}} d\rvec+\int\rho(\rvec)v_{\mathrm{ext}} (\rvec) d\rvec-
C_{\mathrm{x}}\int\rho(\rvec)^{\frac{4}{3}} d\rvec+\frac{1}{2}\int\int\frac{\rho(\rvec)\rho(\rvec^\prime)}{|\rvec-\rvec^{\prime}|} \ 
d\rvec d\rvec^{\prime},
\end{equation}
where $C_{\mathrm{x}} = \frac{3}{4}(\frac{3}{\pi})^{\frac{1}{3}}$. The second term corresponds to the external potential energy 
containing the nuclear-electron attraction, and the last expression refers to the classical electrostatic Hartree repulsion 
energy. While
the sheer simplicity of replacing the troublesome many-electron SE by a \emph{single} equation in terms of $\rho(\rvec)$ 
\emph{alone} is manifestly appealing, underlying approximations are crude and grossly inadequate to be of any practical
use in quantum chemistry. 

As overwhelming complications are met to advance beyond the primitive levels of the TFD model, this line of thought went
into oblivion until 1964, when the future hopes were rekindled in a pioneering work by Hohenberg and Kohn (HK) 
\cite{hohenberg64}. It laid a rigorous theoretical foundation of modern-day DFT by asking a plain obvious question as, 
whether the information contained in $\rho(\rvec)$ is sufficient for elucidating a many-electron system completely. 
Following an amazingly simple proof, they answered the question in an affirmative way. In a landmark paper
\citep{hohenberg64}, they first proved that a non-degenerate ground state of an $N$-particle system, moving under the 
influence of their mutual Coulomb repulsion and an external interacting potential $v_{\mathrm{ext}}(\rvec)$, is solely 
characterized 
by single-particle density $\rho(\rvec)$, i.e., $v_{\mathrm{ext}}(\rvec)$, $\psi (\rvec)$ and hence all other ground-state 
properties are uniquely determined by $\rho(\rvec)$ alone (or as unique functionals of $\rho(\rvec)$). They also showed that, 
for a given $v_{\mathrm{ext}}(\rvec)$, the energy functional is minimum corresponding to the true density. Further, the total energy 
and density of a given system can be determined variationally by minimizing the functional $E[\rho (\rvec)]$ subject to 
the normalization condition, 
$\int \rho(\rvec) d\rvec = N$, 
as a constant, through the following equation, 
\begin{equation}
\delta \bigg\{E[\rho (\rvec) ]-\mu \left[ \int \rho(\rvec) d\rvec -N \right] \bigg\}=0, 
\end{equation}
where $N$ is a measure of the total number of electrons present. This is the central equation of DFT providing a deterministic 
route for $\rho(\rvec)$. Later, it has been established that the HK theorem is equally valid for degenerate ground states 
and has been extended to excited states.

Thus the one-to-one mapping between density and energy, the basic preamble of the HK theorem, can be succinctly 
expressed by conveniently writing the energy functional as follows,  
\begin{eqnarray}
E[\rho] & = & T[\rho (\rvec)]+  V_{\mathrm{ee}} [\rho (\rvec)] + V_{\mathrm{ne}}[\rho (\rvec)]    = 
F[\rho] + \int v_{\mathrm{ext}} (\rvec) \rho(\rvec)  d\rvec,   \\
F[\rho] & = &  T [\rho] +V_{\mathrm{ee}} [\rho].  \nonumber
\end{eqnarray} 
Here $F[\rho]$ signifies the so-called \emph{universal} energy density functional, independent of $N$ and $v_{\mathrm{ext}} (\rvec)$, 
while $T, V_{\mathrm{ne}}$, $V_{\mathrm{ee}}$ refer to the usual kinetic, electron-nuclear attraction and electron-electron repulsion energy 
operators in the Hamiltonian respectively. The original HK theorem implies that ground-state properties are functionals 
of $v$-representable densities only and not of any other trial densities. Later this restriction was lifted; 
thus widening its domain of applicability from $v$-representable to $N$-representable densities (one that can be obtained 
from an antisymmetric wave function). 

In order to make sure that a given density is indeed the true ground-state density, now second HK theorem gives the 
crucial variational theorem, 
\begin{equation}
E[\rho(\rvec)] \geq E[\rho_0 (\rvec)], 
\end{equation}
where $E [\rho_0]$ corresponds to the ground-state energy of a Hamiltonian with $v_{\mathrm{ext}}(\rvec)$ as external 
potential, and 
$\rho_0$ is the true ground-state density. In other words, the functional attains its minimum value with respect to all 
allowed densities if and only if our input density corresponds to the true ground-state density. These two theorems, although,
validate previous efforts of TFD to utilize $\rho (\rvec)$ as a basic fundamental variable to describe many-electron systems, 
emphasis must be given to the fact that it is merely a \emph{proof of existence} only. It confirms the much sought-after
unique mapping between density and energy, in principle; however it remains absolutely quiet on any guidelines for 
construction of such a functional. Even though $\rho(\rvec)$ is sufficient, the relation is delicate, intricate, and 
engenders tremendous complexity. Another major discomfiture lies in the fact that minimization of $E[\rho]$ is, in general, 
an unusually challenging numerical task. Nature of actual calculations before and after HK theorem, does not change  
in any noticeable way; for they are as hard as before. No clear-cut clue is available to approximate the unknown 
functional. Since this does not provide any simplification over MO theory, the ultimate step still remains to be the 
solution of an excessively clumsy SE equation. It is of very little comfort for chemistry and physics, leading to 
virtually no actual progress for realistic calculation. Another unpleasant predicament is that the density 
variation principle holds true only for \emph{exact} functionals. This implies that, in contrast to standard 
wave-function-based approaches (such as HF or CI), where wave functions are rigorously variational, within the realm of 
HK DFT, energy offered by a trial functional has no physical meaning whatsoever.

\subsection{The single-equation approach}
At this stage, we will briefly mention about the so-called \emph{single-equation} approach in DFT, which functions 
exclusively in terms of $\rho(\rvec)$ alone precluding the consideration of orbitals.  
Thus this preserves the true spirit of DFT without sacrificing its associated physics. A real crux of problem for
development of such methods lies exclusively on the unavailability of accurate (not to speak of exact) kinetic and 
XC functionals. Nevertheless two major attempts are noteworthy. The first one invokes a link between quantum fluid 
dynamics and DFT through $\rho(\rvec)$. The former was proposed long times ago by Madelung, de Broglie, and Bohm based 
on a hydrodynamic formulation of quantum mechanics. It is more complicated than the Schr\"odinger 
picture, in a sense that the former requires solving a set of nonlinear partial differential equations, in contrast to 
linear equations in the latter. However the charm is that this route has a conceptual 
appeal, for the electron cloud is treated as a ``classical fluid'' moving under 
the influence of classical Coulomb forces and an additional quantum potential. Moreover, amplitude and phase of the
wave function (treated explicitly in the fluid formulation as independent variables) are more slowly varying in time 
than the wave function itself, which leads to computational advantages and offers additional physical insights of 
quantum dynamics or the quantum trajectory. Thus all the electrons, assumed to be distributed over a 3D space-like 
continuous fluid, being governed by two basic equations in terms of local variables, $\rho (\rvec,t)$ 
and the current density $\mathbf{j} (\rvec,t)$, can be eventually described by a set of equations, \emph{viz.}, (atomic 
units used unless otherwise mentioned):\\
i) an equation of continuity: 
\begin{equation} 
\frac{\partial \rho(\rvec,t)} {\partial t} + \nabla \cdot
\mathbf{j} (\rvec, t) = 0,
\end{equation} 
and ii) an Euler-type equation of motion:
\begin{equation} 
\frac{\partial \chi (\rvec,t)}{\partial t} + \frac {1}{2}(\nabla \chi)^2
+ \frac{\delta \mathrm{G} [\rho]}{\delta \rho}
+\frac{\delta {\mathrm{E_{el-el}}}[\rho]}{\delta \rho}+\mathit{v_{\mathrm{ext}}}(\rvec,t) = 0,
\end{equation} 
where $\mathbf{j} (\rvec ,t)= \rho \nabla \chi (\rvec,t)$, with $\chi(\rvec,t)$ being the velocity potential. 
$\mathrm{E_{el-el}}$ represents the inter-electronic Coulomb repulsion energy, $v_{\mathrm{ext}}(\rvec,t)$ is 
the external potential containing electron-nuclear attraction and any time-dependent (TD) interaction present in the 
system, while $\mathrm{G}[\rho]$ is a universal functional comprising of kinetic and XC energy contributions.
Equations~(9) and (10) can be combined into a single equation by defining a complex-valued TD hydrodynamical wave
function for the entire time-evolving system as (in polar form):
\begin{equation}
\Psi (\rvec,t)=\rho(\rvec,t)^{1/2} e^{i\chi(\rvec,t)},
\end{equation}
and eliminating $\chi(\rvec,t)$ from them. The result is a generalized nonlinear SE: 
\begin{equation}
\left[-\frac{1}{2}\nabla^2+\mathit{v_{\mathrm{eff}}}([\rho];\rvec,t)\right]
\Psi(\rvec,t)= \mathit{i}\frac{\partial\Psi(\rvec,t)}{\partial t},
\end{equation}
where $\rho(\rvec,t) = |\Psi(\rvec,t)|^2$, and the effective potential $\mathit{v_{\mathrm{eff}}} ([\rho];\rvec,t)$ contains 
both classical and quantum potentials. This procedure has been successfully applied to a variety of interesting static 
and dynamic many-electron situations. Some of these are electronic structure calculations \citep{roy02b}, 
ion-atom collisions \citep{deb89, dey98a}, atoms, molecules in presence of strong TD
electric and/or magnetic fields \citep{dey98, roy02a, sadhukhan11, vikas11, vikas13,
sadhukhan16} including high-harmonic generation and multi-photon ionization, etc.  Many more details could be found in 
the above references and therein. 

The other one is the so-called orbital-free DFT \citep{wang00}, whose realization most decisively depends on 
the availability of correct, good-quality kinetic energy density functionals. Reportedly, it is possible to 
compute all components of energy functionals effectively in momentum space within a linear scaling process, which is 
certainly advantageous in terms of simplicity and computational cost. During the past several decades, an immense amount of 
work has been attempted towards designing suitable kinetic energy functionals \citep{wang98} including a density dependent 
kernel \citep{wang99} and non-local functionals \citep{ho08}, to mention a few; mostly on periodic system. Efforts
were also made to supplement the density-dependent kernel to non-periodic systems, with the aid of finite element modelling 
and coarse graining \citep{gavini07}. Later, its scope was enlarged to covalent systems and semiconductors \citep{huang10}
as well. Up to now, applications are predominantly limited to metallic system. 

\subsection{The Kohn-Sham method}
While the single-equation route of the previous subsection provides an elegant and attractive direction to deal with an 
interacting system and has witnessed numerous successes, insurmountable difficulties (as delineated in II.A) need to 
be overcome for its routine practical application. Unsurprisingly, a massive amount of today's work actually follows 
an alternative formalism, originally suggested by Kohn-Sham (KS). The prevailing unyielding state of affairs after
publication of the HK theorem dramatically changed a year later, in a path-breaking contribution \cite{kohn65}, 
that suggested a manageable way to approach the hitherto unknown universal functional. In order to mitigate the problem, 
they introduced the clever concept of a \emph{fictitious}, non-interacting system built from a set of orbitals (KS orbitals) 
such that the major part of the kinetic energy can be computed exactly. The remaining fairly small portion is absorbed in our 
non-classical contribution of the electron-electron repulsion, which is also unknown. However, the advantage is that electrons 
now move in an effective KS single-particle potential, while the mapped auxiliary system
yields the same ground-state density as our real interacting system, but this simplifies the actual calculation tremendously. 

This is accomplished by setting up a non-interacting reference system (designated with suffix `s') with the Hamiltonian 
containing an effective local potential $v_{\mathrm{s}}(\rvec)$ expressed as a sum of one-electron operators,  
\begin{equation}
H_{\mathrm{s}} = -\frac{1}{2} \sum_{i}^{N}\nabla_{i}^{2} + \sum_{i}^{N}v_{\mathrm{s}}(\rvec_{i}).
\end{equation} 
Note that there is no electron-electron interaction in the Hamiltonian. It is well known that the exact wave function of a 
system of non-interacting fermions can be represented as Slater determinants of individual one-electron eigenfunctions, and 
energies are sums of one-electron eigenvalues. In complete analogy to the HF equation, thus, one can write an eigenvalue equation 
for KS spin-orbitals as, 
\begin{equation}
\hat{f}^{\mathrm{KS}} \psi_{i} = \epsilon_{i}\psi_{i}, 
\end{equation}
where the one-electron KS operator $\hat{f}^{KS}$ is defined by, 
\begin{equation}
\hat{f}^{\mathrm{KS}} = -\frac{1}{2} \nabla^{2} + v_{\mathrm{s}}(\rvec).
\end{equation}
Hence the crucial step is to map our artificial non-interacting system to the interacting, real one through an effective potential 
$v_{\mathrm{s}}(\rvec)$ which plays a pivotal role. This is ensured by choosing $v_{\mathrm{s}}(\rvec)$ such that the density resulting 
from a summation of moduli of square of orbitals ($\big\{\psi_{i}\big\}$), is exactly the same as the ground-state density of our 
real system ($\sigma$ signifies spin), i.e., 
\begin{equation}
\rho_{\mathrm{s}}(\rvec) = \sum_i^N \sum_{\sigma} | \psi_i (\rvec, \sigma)|^2 = \rho_0(\rvec). 
\end{equation}

So the leading part of the kinetic energy of the real system can be restored adequately as a sum of individual kinetic 
energies of the reference system with same density as real system, 
\begin{equation}
T_{\mathrm{s}} = -\frac{1}{2} \sum_{i}^{N}\nabla_{i}^{2}.
\end{equation}
However, since the non-interacting kinetic energy is not equal to the true kinetic energy, they suggested that the residual, 
kinetic energy ($T_\mathrm{c}$), often a small contribution, is submerged to the unknown, non-classical component of 
electron-electron repulsion, namely, 
\begin{eqnarray}
F[\rho] & = & T_{\mathrm{s}}[\rho]+J[\rho]+E_{\mathrm{xc}}[\rho],  \nonumber \\
E_{\mathrm{xc}}[\rho] & = & (T[\rho]-T_{\mathrm{s}}[\rho]) + (E_{\mathrm{ee}}[\rho]-J[\rho]) = T_{\mathrm{c}}
[\rho]+E_{\mathrm{nc}}[\rho].
\end{eqnarray}
Associated terms have following meanings:  $J[\rho]$ is the known classical part of $E_{\mathrm{ee}} [\rho]$, whereas 
$E_{\mathrm{xc}}[\rho]$ 
contains everything that is unknown, i.e., non-classical electrostatic effects of $E_{\mathrm{ee}} [\rho]$ as well as the 
difference
between the true kinetic energy $T[\rho]$ and $T_{\mathrm{s}}[\rho]$. Then the expression of total energy $E[\rho]$ of 
our real system can be cast in the following way, 
\begin{equation}
E[\rho]=\int v_{\mathrm{ext}}(\rvec) \rho(\rvec) d\rvec+ J[\rho]+ T_{\mathrm{s}}[\rho]+E_{\mathrm{xc}}[\rho]
\end{equation}
This allows us to write the celebrated KS orbital equation in canonical form (henceforth atomic units employed unless 
otherwise stated), 
\begin{equation}
\bigg[-\frac{1}{2} \nabla^{2} + v_{\mathrm{eff}}(\rvec)\bigg]\psi_{i}(\rvec)=\epsilon_{i}\psi_{i}(\rvec)
\end{equation}
where the ``effective" potential contains following terms,
\begin{equation}
v_{\mathrm{eff}}(\rvec)=v_{\mathrm{ext}}(\rvec)+\int\frac{\rho(\rvec^{\prime})}{|\rvec-\rvec^{\prime}|}d\rvec^{\prime}+
v_{\mathrm{xc}}(\rvec).
\end{equation}
In the above equation, $v_{\mathrm{xc}} [\rho(\rvec)]$ signifies the functional derivative,
$\frac{\delta E_{\mathrm{xc}}[\rho (\rvec)]}{\delta\rho (\rvec)},$ with respect to charge density. So the KS equation, 
although \emph{exact} in principle, is structurally similar to the HF equation, but less complicated than the 
former. Both equations must be solved iteratively by self-consistent field. It has a profound effect on the quantum community 
because of its ability to account for XC effects in a rigorous, quantitative and transparent manner with a cost level of 
HF theory. However, it may be noted that, although being local in nature, the KS potential may be formally innocuous and less 
complicated than 
its counterpart in HF approximation, it may have a rather not-so-straightforward and non-local dependence on $\rho(\rvec)$.  

Thus, as explained above, for practical purposes, minimization of our explicit energy functional, although possible, 
\emph{in principle}, is hardly ever recommended. A far more viable and feasible route invokes solution of the KS 
Eq.~(20), which, however, interestingly brings back an orbital picture, and strictly speaking, does not function 
solely in terms of density. But, the good thing is that we have a mechanism to incorporate the intractable many-body 
effects formally \emph{exactly}, within a single-particle theory.  

\subsection{Grid consideration}
For realistic solution of the KS equation, it is necessary to deal with mathematically non-trivial integrals that cannot be 
evaluated analytically and pose certain amount of challenge. Even in finite basis-set expansion methods, there does not exist 
any explicit analytical formulas for requisite XC integrals. In absence of such procedures, one is left with no option but for 
numerical calculation. It is well acknowledged that such a discrete procedure for multi-center integrals in 3D space is not 
straightforward. The task becomes all the more formidable from a consideration of the fact that, in order to achieve a 
satisfactory level of chemical accuracy, a generous grid size and a longer computation time is often mandatory, as a prohibitively 
huge number of operations is needed. A vital problem for constructing such integration route arises due to cusps in the 
density and the singular nature of the Coulombic potential. Naturally, in order to pursue high-accuracy calculations within a 
reasonable number of quadrature grids, one needs efficient and sophisticated numerical integrators which can capture the
forms of density at a satisfactory level. This paves the way for a considerable number of integrators having varied degrees of  
performance. Amongst them, two distinct, well recognized partitioning schemes have shown significant promise. The Voronoi 
cellular approach was originally proposed by te Velde and Baerends \citep{te92}, in which the space is divided into 
non-overlapping regions of simple geometry. On the other hand, fuzzy cells avenue, or commonly known as ACG, 
was initially recommended by Becke \citep{becke88}; later championed by several others groups \citep{gill93, 
treutler95, delley90, lindh01}. Though the basic idea was implemented for completely numerical, non-basis set calculations
of polyatomic molecules, now it has reached the status of standard for linear combination of atomic orbitals 
(LCAO)-MO DFT approach, which is convenient while using spherical grids. 

The basic strategy is to divide the space into fuzzy (atom-centered overlapping analytically continuous regions) cells. In order 
to integrate over each cell, a spherical grid (corresponding to radial and angular quadrature grid) centered on each atom is 
the natural choice. The integral ($I$) of an arbitrary integrand $F(\rvec)$ over whole region 
is then converted to discrete numerical summations over individual atomic regions $j$:
\begin{equation}
I=\int F(\rvec) d\rvec=\sum_{j} I_{j}=\sum_{j} \int F_{j}(\rvec) d\rvec, 
\end{equation} 
where 
\begin{equation}
F_{j}(\rvec) = w_{j}(\rvec)F(\rvec)
\end{equation}
i.e., a multi-center integration therefore translates to a sum of single-center integrations over each atomic nucleus, whereas 
the $w_{j}$'s correspond to relative weight functions over each atomic center respectively. 
In order to make the discrete cell fuzzy, certain restrictions are imposed on $w_{j}$, so that,
\begin{equation}
\sum_{j} w_{j}(\rvec) = 1 \quad \forall \rvec,  
\end{equation}
and each $w_j (\rvec)$ takes value unity in the neighborhood of its own nucleus, but vanishes near any other 
nucleus in a well-behaved manner. 

Traditional cellular separation of molecular space is best realized by using confocal elliptical coordinates
($\lambda, \mu, \phi$). Taking center $i$ as reference, one considers other centers $j \neq i$ and establishes $\lambda_{ij}$, 
$\mu_{ij}$, $\phi_{ij}$ on the foci $i,j$. Of particular interest is,  
\begin{equation}
\mu_{ij}=\frac{\rvec_{i}-\rvec_{j}}{R_{ij}}
\end{equation}
where $r_{i}, r_{j}$ identify distances to nuclei $i,j$ respectively, and $R_{ij}$ the internuclear distance. In terms 
of an auxiliary function, $s(\mu_{ij}$), normalized weight functions, $w_{i} (\rvec)$, are defined as:
\begin{equation}
w_{i}(\rvec)=\frac{W_{i}(\rvec)} {\sum_{j} W_{j}(\rvec)}; \quad W_{i}(\rvec)=\prod_{j\neq i} s(\mu_{ij}).
\end{equation}
The most recommended form of $s(\mu_{ij}$) is a cut-off function as given below,
\begin{equation}
s_{k}(\mu) = \frac{1}{2} \left[ 1-f_{k}(\mu) \right],
\end{equation} 
where $f_{k}(\mu)$ indicates a $k$-th order iterative polynomial. The forms of polynomial, $f_{k}(\mu)$ and $k$ are chosen 
such that
$w_{i}(\rvec)$ satisfies the restriction imposed by Eq.~(24). There is no hard and fast rule to evaluate the value of $k$; it is 
chiefly decided through experience ($k=3$ is used in some earlier works \citep{becke88}). Instead of polynomial, if one 
chooses $s(\mu_{ij})$ as a step function, the fuzzy scheme transforms into Voronoi scheme. Moreover, a size adjustment procedure 
has been advocated for hetero-nuclear molecules. The above prescription is most suited for molecules; an alternative approach towards 
periodic system is nicely presented in \citep{franchini13}.

The single-center integral, $I_{j}$ in Eq.~(22) is expressed in a spherical coordinate system as:
\begin{equation}
I_{j} = \int_{o}^{\infty} \int_{0}^{\pi}\int_{0}^{2\pi} F_{j}(r_{j},\theta_{j},\phi_{j})  \ r_{j}^{2} sin\theta_{j} \  
dr_j d\theta_{j} d\phi_{j}.
\end{equation} 
Now we can distribute the 3D integration into 2D integration of variables $\Omega(\theta, \phi)$ and 1D integration of radial 
variable $r$ in the following fashion, by applying some appropriate $n$-point quadrature formula having weight factor $\omega$ 
for both angular and radial parts,
\begin{equation}
F_{j}(r_{j}) = \int_{0}^{\pi} \int_{0}^{2\pi} F_{j}(r_{j},\theta_{j},\phi_{j})  \ sin\theta_{j}d\theta_{j} d\phi_{j} \ 
\approx \ \sum_{i=1}^{n_{j}^{\Omega}} \omega_{i}^{\Omega} F_{j}(r_{j},\Omega_{i}), 
\end{equation}
and
\begin{equation}
I_{j} = \int_{0}^{\infty}F_{j}(r_{j}) r_{j}^{2} dr_{j}  \ \approx \ \sum_{i=1}^{n_{j}^{r}} \omega_{i}^{r} F_{j}(r_{i}).
\end{equation}
Angular integration is eventually carried out on numerous sizes of a Lebedev grid \citep{lebedev92, lebedev99} 
(also called octahedral grid), which is distributed on the surface of a unit sphere in such a way
that the resulting point distribution remains invariant with respect to inversion. This grid is a consequence of various 
quadrature formulas \citep{mclaren63} of remarkably high order, where all the spherical harmonics ($Y_{l,m}$) and its squares are 
integrated accurately. For optimal performance, it is best that the total number of grid points satisfies a certain condition as:
\begin{equation}
\mathrm{No.~ of~points } \     \approx \  \frac{(L+1)^2}{3}, 
\end{equation} 
where L is the maximum degree of the spherical harmonics ($0 \leq l \leq L$). Open-ended quadrature schemes are generally 
advisable; this is achieved via product 
type of grids quite easily in polar coordinates ($\Omega$) \citep{murray93}. Efficiency can be further improved using a Lobatto 
grid, which uses a different criterion for the total number of points $\left( \approx \frac{L(L+1)^{2}}{2} \right)$, for optimal 
realization \cite{treutler95}. However, the majority of modern KS-DFT programs (within basis-set approach) like 
Gaussian, NWChem, Q-Chem, GAMESS etc., exploit the Lebedev grid. 

For radial integration, a fairly decent number of implementations are conceivable. These are typically carried out with the aid 
of some suitable quadrature rules. Typically, a transformation is applied to map radial grid onto the quadrature grid
varying the number of sampling points on the grid. Generally, a grid-pruning procedure is also administered in conjunction, to 
curtail the number of sampling points \citep{el04, chien06}. Some commonly used quadrature grids are: Gauss-Chebyshev 
integration of second kind, Euler-Maclaurin summation (extended trapezoidal) formula, log-squared quadrature, 
Gauss-Legendre, Gauss-Laguerre quadrature etc., \citep{becke88, murray93, treutler95, mura96, gill03, kakhiani09}. 
The Gauss-Chebyshev integration formula is given as, 
\begin{equation}
\int_{-1}^{1} (1-x^{2})^{\frac{1}{2}} f(x) dx  \approx  \frac{\pi}{n+1}\sum_{i=1}^{n} 
\sin^{2}\bigg(\frac{i\pi}{n+1}\bigg)f(x_{i});  \ \ \ \ \ \
x_{i}  =  \cos\bigg (\frac{i\pi}{n+1}\bigg ),   
\end{equation}
whereas the Euler-Maclaurin scheme is executed through,  
\begin{equation}
\int_{0}^{1} f(x) dx \approx \frac{1}{n+1} \sum_{i=1}^{n} f(x_{i}); \ \ \ \ \ \
x_{i} = \frac{i}{n+1}.
\end{equation}
Subsequently, a double exponential formula has been proposed, and later refined by several other groups \citep{mori85, mori01, 
muhammad03} to generate a radial grid. For large $N$, it becomes beneficial in the way of providing fast convergence 
rate.  This is achieved by adopting an equally meshed trapezoidal rule for numerical integration over an arbitrary interval that 
transforms into an infinite integral by variable transformation. It has several variants depending on the mapping transformation 
and was invoked for finite, semi-finite, and infinite integrals. Another promising approach involves construction 
of an adaptive grid, pioneered by \citep{krack98, perez94, termath96} which is automatically generated for a given accuracy 
corresponding to a chosen basis function. Some of the prominent DFT programs like NRLMOL, deMon2k routinely use such 
grids.

It is well-known that the current enviable status of DFT that it enjoys largely depends on basis-set calculations. While such 
studies are heavily dominated by ACG, real-space grid has been invoked for fully numerical, basis-set free DFT methods. 
Apart from ACG, some scattered works exist for other grids in literature, e.g., an adaptive Cartesian grid with a hierarchical 
cubature method \citep{challacombe00}, a transformed sparse-tensor product grid \cite{rodriguez08}, a Fourier Transform Coulomb 
\citep{fustimolnar02, fustimolnar03} method interpolating density 
from ACG to a more regular grid that helps to enhance its efficiency. In the latter, a difficulty arises due to 
computation of XC matrix elements, which is circumvented successfully with help of a multi-resolution method 
\citep{brown06}. Here, in this work, we report a simple fruitful DFT implementation \cite{roy08, roy08a, roy09, roy10, roy11} 
within the LCAO-MO framework which solely uses CCG. So far, our results have been reported in a \emph{uniform} grid. All relevant 
quantities such as basis functions, electron densities, MOs as well as various two-electron potentials (Hartree and 
XC) are directly set up on a real 3D Cartesian grid simulating a cubic box as, 
\begin{equation}
r_{i}=r_{0}+(i-1)h_{r}, \quad i=1,2,3,....,N_{r};\quad \mathrm{for} \quad r\in \{x,y,z\},
\end{equation} 
where $h_{r}$,$N_{r}$ signify grid spacing and total number of grid points respectively ($r_{0}=-\frac{N_{r}h_{r}}{2}$). 

A major concern in the grid-based approach constitutes an accurate estimation of the classical electrostatic repulsion potential. 
For finite systems, the simplest and crudest way to calculate $v_{\mathrm{h}}(\rvec)$ is through direct numerical integration on the 
grid. For smaller systems, this is a 
feasible option; in all other cases, it is generally tedious and cumbersome. However, the most rewarding and widespread 
approach is through a solution of the corresponding Poisson equation, 
\begin{equation}
\nabla^2 v_{\mathrm{h}}(\rvec) = -4 \pi \rho(\rvec).
\end{equation}
The usual way to solve this is by conjugate gradients \citep{saad03} or through multi-grid solvers \citep{brandt77}.
As an alternative, the current work exploits a conventional Fourier convolution method originally suggested by \citep{martyna99, 
minary02, skylaris02} and adapted in the context of molecular modelling \citep{hine11, chang12}. The basic principle can be 
formulated as:
\begin{equation}
v_{\mathrm{h}}(\rvec)= \mathrm{FFT}^{-1}\{v_{\mathrm{h}}^{\mathrm{c}}(\mathbf{k})\rho(\mathbf{k})\} \quad \mathrm{and} \quad 
\rho(\mathbf{k})=\mathrm{FFT}\{\rho(\rvec)\},
\end{equation}
where $v_{\mathrm{h}}^{\mathrm{c}}(\mathbf{k})$ and $\rho(\mathbf{k})$ stand for Fourier integrals of the Coulomb interaction 
kernel and density respectively, in the cubic box. The quantity $\rho(\mathbf{k})$ can be easily computed by using a discrete 
Fourier 
transformation of its real space value. Thus our primary concern here lies in calculation of the Coulomb interaction kernel 
which has a singularity in real space. For this, we utilize an Ewald summation-type approach \citep{chang12}, expanding the 
Hartree kernel into long-range and short-range components,  
\begin{equation}
v_{\mathrm{h}}^{\mathrm{c}}(\rvec)=\frac{ \mathrm{erf}(\alpha r)}{r}+\frac{ \mathrm{erfc}(\alpha r)}{r} 
\equiv v^{\mathrm{c}}_{\mathrm{h}_{\mathrm{long}}}(\rvec)+v^{\mathrm{c}}_{\mathrm{h}_{\mathrm{short}}}(\rvec),
\end{equation}
where erf(x) and erfc(x) signify the error function and its complementary function respectively. The Fourier transform of 
the short-range part can be treated analytically, whereas the long-range portion needs to be computed directly from FFT of 
real-space values. A convergence parameter $\alpha$ is used to adjust the range of 
$v^{\mathrm{c}}_{\mathrm{h}_{\mathrm{short}}}(\rvec)$, 
such that the error is minimized. Following the conjecture of \cite{martyna99}, here we employ $\alpha \times L = 7$ 
($L$ denotes length
of cubic box), which produces quite accurate results. For a non-uniform grid, $L$ is chosen as the smallest side of the 
simulating box. Some other routes which calculate the long-range part efficiently are fast multipole 
\citep{beatson97}, multi-level summation \citep{skeel02}, fast Fourier-Poisson \citep{york94} method, etc.

\subsection{Basis-set free methods}
This sub-section gives an overview of fully numerical, real-space methods, used for solving the pertinent KS 
equation without needing any explicit basis set. Typically, the wave function 
is directly sampled on a discrete grid through a variety of representations; each having its own pros and cons. 
Most of these attempts also invoke pseudopotentials to freeze the inner core. One central problem is the formation of an 
appropriate finer mesh, which normally calls for an enormous number of grid points, making calculations of the electronic 
structure (especially for large systems) very expensive in terms of both time and computer memory. Apparently, there are 
ways to minimize discretization 
error systematically by controlling size, shape and spacing of the 3D mesh. It is worth mentioning two notable works, 
\emph{viz.,} a higher-order real-space pseudopotential method in uniform CCG \cite{chelikowsky94a, chelikowsky94b}, 
as well as real-time propagation of KS orbitals \cite{mundt07,marques12}. Often, they are easily amenable to linear 
scaling schemes. Recently, extensions have been made to Graphical Processing Unit interface \citep{andrade13} 
and domain-decomposition techniques \cite{chelikowsky09} for parallel computation. Some excellent reviews are available on the 
subject (see, for example, \cite{beck00, natan08, bowler12, bernholc08}). 

Broadly speaking, real-space techniques can be categorized into three major classes, namely, finite difference (FD), finite
element (FE) and wavelets \citep{mohr14,genovese08}. In all cases, the target discretized differential equation produces 
structured and highly banded matrices, which can be solved readily using efficient multi-scale techniques. The potential 
operator is diagonal in coordinate space and the Laplacian operator is nearly non-local; that makes them particularly suited for 
parallel computing. In a uniform orthogonal 3D CCG of spacing $h$, e.g., the Laplacian operator, within an FD approximation, can be 
expressed approximately in 1D ($x$ direction) as a part of 3D: 
\begin{equation}
\frac{d^{2}\phi(x_{i})}{dx^{2}} \approx \frac{1}{h^{2}}[\phi(x_{i-1})-2\phi(x_{i})+\phi(x_{i+1})]-
\frac{1}{12}\frac{d^{4}}{dx^{4}}\phi(x_{i})h^{2}+\mathcal{O}(h^{4}).
\end{equation}
The last two terms account for truncation errors; the first contribution being second order in $h$ with a pre-factor involving
the fourth derivative of $\phi(x)$. Here grid spacing, $h$, is one of the key parameters that controls convergence. This
approximation is valid in the limit of $h \rightarrow 0$, but for practical purposes, it is unfeasible to go below a certain
limit, which leads to the evolution of higher-order FD schemes \cite{chelikowsky94a}. Most commonly, regular CCG 
is engaged and FD methods result from a Taylor series expansion of our desired function around the grid points. 
However, it is non-variational, because the error can be of either sign depending on the derivatives and value of $\phi (x)$. 
It is often quite difficult to 
achieve convergence due to this lack of variationality. While its chief advantage lies in a simplicity of representation
and easy implementation, sometimes it is rather troublesome to construct flexible meshes which can reproduce physical 
geometry satisfactorily. There are reports of a Mehrstellen discretization technique as an 
alternative route to high-order FD methods \citep{briggs96, fattebert00}. Several workers \citep{alemany07, Iwata10, 
fujimoto10, heiskanen01}, have suggested use of FD techniques with in amalgamation of projector augmented wave 
methods \citep{mortensen05}. Applications of the FD method is quite impressive; mention may be made about some of the notable  
ones for clusters and other finite systems \cite{modine95, lee00, chelikowsky94a, chelikowsky94b}. 

On the other hand, FE-based methods are well known from the engineering field for a long time, and have been implemented on 
applications of electronic structure theory \citep{pask05}. Customarily, one uses some suitable non-orthogonal basis functions,
such as piecewise polynomials which are non-zero over a local region of space, and also divide the unit cell into elements. 
For solution of Poisson equation, one can expand the potential, $v(x)$, in a basis:
\begin{equation}
v(x) = \sum_{i} u_{i}\chi_{i}(x), 
\end{equation}
where $u_{i}$'s are expansion coefficients corresponding to actual function values on mesh points and $\chi_{i}$ are basis 
functions. A similar kind of treatment can be performed on the charge density, and the resulting matrix equation will be similar 
to a one-dimensional Poisson equation. At least for uniform meshes, there is a close correspondence between FD and 
FE real-space representation. Unlike the FD method however, it has a variational foundation, and provides greater flexibility 
regarding the unit cell construction. A multitude of efficient basis sets have been suggested for 3D electronic structure 
calculations for diverse problems \citep{tsuchida95}, like cubic-polynomial basis \cite{white89}, tetrahedral 
discretization with orders $p=1-5$ \citep{ackermann94}, cubic functions \citep{pask99}, Lobatto-Gauss basis set with orders 
$p=5-7$ \citep{yu94}, B-spline basis \citep{goringe97} etc. 

Standard iterative processes suffer from so-called \emph{critical slow down} \cite{beck00}; a common complication that 
occurs when more grid points are used to obtain an enhanced accuracy on a fixed domain and thus makes it less efficient on finer 
meshes. In order to facilitate better convergence of self-consistent iteration procedure, a few advancements have been 
made. Most prominent of them is the family of multi-grid iteration techniques \cite{briggs95, briggs96, beck00}, which ideally
utilize information from multiple lengths scales to alleviate this problem. Another promising route involves engaging an adaptive 
grid in conjunction with multi-grid \citep{gygi95, modine95}. This has found applications within 
FD-type representations, but maximum effort was devoted to develop efficient solvers for FE representations 
\citep{braess90, brenner94}. 

Another useful source of real-space techniques is based on the wavelet basis method having a more complicated matrix structure 
than either the FD or FE representation \citep{arias99}. Such basis sets are semi-cardinal, local, and not only that, they 
are conceived with multi-resolution properties. One crucial advantage is that calculations can be performed with different kinds of 
boundary conditions. Daubechies wavelet \citep{genovese08} is one of the most favored and frequently used, which is local 
in both real and Fourier space. These are easily adaptable to parallel programming and are employed in connection with linear 
scaling methods \cite{mohr14} for electronic structure calculations. Several DFT codes like ABINIT, ONETEP, CONQUEST, CP2K, 
and SIESTA etc., successfully implement wavelet schemes. 

\subsection{Basis set and LCAO-MO ansatz} 	
As emphasized earlier, the basis-set approach remains by far the most convenient and pragmatic line of approach towards the 
solution of KS DFT. It dominates in every field of science whether being chemical, physical or biological 
applications employed to atoms, molecules or solid etc. The inspiration mainly comes from the success of 
basis-set related methodologies in traditional wave function theory, such as HF and post-HF. This is a coupled 
integro-differential equation where kinetic and potential energies are defined by a differential and integral operator. It
must be solved numerically iteratively leading to a self-consistent set of orbitals $\{\psi_i(\rvec)\}$. Essentially, the 
unknown KS MOs are expanded in terms of $K$ suitably chosen, known basis functions $\{\chi_{\mu} (\rvec); \mu = 1,2,3,....,K\}$, 
conventionally called atomic orbitals, in a manner analogous to that in the Roothaan-HF method, such as, 
\begin{equation}
\psi_{i} (\rvec) =\sum_{\mu =i}^{K}C_{\mu i} \chi_{\mu} (\rvec), \quad i=1,2,3,....K. 
\end{equation} 
The electron density then takes the following expression in this basis, 
\begin{equation}
\rho(\rvec)=\sum_{i=1}^{N}\sum_{\mu=1}^{K}\sum_{\nu=1}^{K}C_{\mu i}C_{\nu i}\chi_{\mu}(\rvec)\chi_{\nu}(\rvec).
\end{equation}
In principle, one requires a \emph{complete} basis set ($K=\infty$) in order to get an exact expansion of MOs, but in reality 
it is not feasible. Therefore, suitable truncation is required for realistic computational purposes; it suffices to work with a 
finite basis set usually.

It is of paramount importance to choose appropriate functions which reproduce KS MOs as precisely as possible. The numerical 
accuracy of KS DFT is quite sensitive to the choice and design of the basis set for a particular problem, as an
incomplete basis set inducts certain constraints on the relaxation of the density through KS orbitals. Considerable developments
have taken place towards their construction as well as effects on assorted physico-chemical properties. There is no 
restriction, as such, for the effective implementation of a specific basis set, but it is required to fulfill certain 
conditions which may serve as some guidelines: (a) it should have proper behavior according to the physics of the problem 
(b) it should be friendly in terms of ease and efficiency of computation of requisite integrals (c) number of functions in 
basis-set expansion should be scaled down as much as possible without sacrificing any accuracy. A sizeable number of elegant, 
flexible, versatile basis sets has been proposed over the past several decades, from different perspectives such as: 
exponential \citep{slater30}, Gaussian \citep{boys50}, polynomial, cube functions, wavelets \citep{cho93}, numerical 
functions \citep{delley90}, plane waves \citep{ihm79} and many others. However, the problem still now remains open, as 
there is no universal basis set applicable for all methods or chemical systems. The interested reader may consult many 
nice reviews and books available on the topic \citep{szabo96, helgaker00, cramer04, jensen07}.

For a periodic system (e.g., solid), the preferred choice is a plane wave basis whereas for non-periodic systems 
(e.g., molecules, clusters) atom-centred localized basis functions, such as Slater Type Orbitals (STO) or 
Gaussian Type Orbitals (GTO) stand out to be two most attractive options. STOs assume the following functional form:
\begin{equation}
\chi_{\zeta,n,l,m}(r,\theta,\varphi)=N_c Y_{l,m}(\theta,\varphi)r^{n-1}e^{-\zeta r}, 
\end{equation} 
where $N_c$ is the normalization constant and $Y_{l,m}$ defines spherical harmonics. It is a well acknowledged fact that multi-center 
(3c or 4c) two-electron 
integrals can not be evaluated analytically with STOs, making them somehow less favorable for molecular system. Thus their 
application principally remains confined to atomic or diatomic systems or in semi-empirical approximations, where two- or 
three-center integrals are generally neglected.

On the other hand, GTOs are expressed in Cartesian coordinates as:
\begin{equation}
\chi_{\zeta, n_{x}, n_{y}, n_{z}}(x,y,z)=N_c x^{n_{x}} y^{n_{y}} z^{n_{z}} e^{-\zeta r^{2}}, 
\end{equation}
where $n_{x}$, $n_{y}$ and $n_{z}$ define the angular momentum such that $\lambda=n_{x}+n_{y}+n_{z}=0,1,2,\cdots,$ 
($s,p,d, \cdots$ functions), while the orbital exponent $\zeta$ governs its compactness (large $\zeta$) or diffuseness (small $\zeta$).
In spite of the fact that GTOs face problems to describe the proper behavior near and far from the nucleus than STOs, 
they are almost universally accepted for quantum chemistry calculations. To a great measure, this is due to their 
ability to provide closed, analytic formulas for multi-center integrals through Gaussian product theorem. In principle, both 
STOs and GTOs qualify to form a complete basis, but to reach a certain accuracy, ordinarily more GTOs are needed (roughly at least 
three times) than STOs. In practical purpose, instead of using individual GTOs as basis function, it is customary to use 
fixed linear combination of GTOs, called contracted GTOs, defined as:
\begin{equation}
\chi^{\mathrm{contracted}}_{\mu}(\rvec-\mathbf{R_{A}})=\sum_{p=1}^{L}d_{p\mu}\chi_{p}(\zeta_{p\mu},\rvec-\mathbf{R_{A}}).
\end{equation}
Here $d_{p\mu}$ are called contraction coefficients, $L$ signifies length of contraction, whereas ``p" stands for individual 
primitive Gaussians from which contracted Gaussian functions are formed. The normal procedure to find out primitives is to 
optimize exponents and contraction coefficients so as to obtain variational lowest-energy states through an atomic calculation 
(within HF, CI or so). However there are instances in literature where they are explicitly optimized through KS procedure 
using LDA XC functionals. Historically, basis sets were designed for applications in wave-function based methodology, and 
fortunately it turns out that, for most common molecular property calculations (like equilibrium geometry, energy), results are 
pleasantly insusceptible towards the manner through which these exponents and coefficients have been determined. That offers 
a welcome possibility to take advantage of such basis sets, initially constructed for traditional methods, to be directly usable 
in DFT calculations with confidence.  
            
Now inserting Eq.~(40) in Eq.~(20), multiplying the left side of the resulting equation with $\chi_{\mu}^{\star}(\rvec)$, 
then integrating over the whole space, followed by some algebraic manipulation generates the following KS matrix equation, in 
parallel to the HF case,
\begin{equation}
\mathbf{F}^{\mathrm{KS}}\mathbf{C}=\mathbf{S}\mathbf{C} \mathbf{\epsilon},
\end{equation}  
where $\mathbf{F}$ and $\mathbf{S}$ imply the $K \times K$ real, symmetric total KS and overlap matrices respectively. 
The eigenvector matrix $\mathbf{C}$ contains the basis-set expansion coefficients $\mathbf{C}_{\mu i}$ and the diagonal matrix 
$\mathbf{\epsilon}$ holds the orbital energies $\epsilon_{i}$. It could be readily solved by standard numerical techniques of 
linear algebra. The individual elements of KS matrix can be written as:
\begin{eqnarray}
F_{\mu \nu}^{\mathrm{KS}} & = & \int \chi_{\mu} (\rvec) \left[ h^{\mathrm{core}}+v_{\mathrm{hxc}}(\rvec) \right] \chi_{\nu} 
(\rvec)
\mathrm{d} \rvec = H_{\mu \nu}^{\mathrm{core}} + \langle \chi_{\mu}(\rvec) | v_{\mathrm{hxc}} | \chi_{\nu} 
(\rvec)
\rangle \\
 & = & H_{\mu \nu}^{\mathrm{core}} + J_{\mu \nu} +V_{\mu \nu}^{\mathrm{XC}},  \nonumber 
\end{eqnarray}
where $H_{\mu\nu}^{\mathrm{core}}$ represents the core bare-nucleus Hamiltonian matrix element consisting of the kinetic 
energy and nuclear-electron attraction, thus accounting for one-electron energies. The one-electron matrix elements can 
be evaluated analytically with the help of well-established recursion relations \citep{obara86} for Gaussian bases.
Second term $v_{\mathrm{hxc}} (\rvec)$ contains all two-electron interactions involving classical Coulomb repulsion and 
the non-classical XC potential. $J_{\mu\nu}$ denotes the matrix element of the classical Hartree potential defined in 
Eq.~(21), while the remaining term, $V_{\mu\nu}^{\mathrm{XC}}$ supplies the XC contribution into two-body matrix element, 
whose construction remains one of the most vital steps in the whole KS DFT process. In 
absence of any analytical method, two-body matrix elements can be either calculated numerically or fitted by an 
auxiliary set of Gaussian functions, as suggested by \citep{sambe75, dunlap79a, dunlap79b}. In our work, we employed direct 
numerical integration on the CCG to obtain these matrix elements:
\begin{equation}
\braket{\chi_{\mu}(\rvec)|v_{\mathrm{hxc}}(\rvec)|\chi_{\nu}(\rvec)} = h_x h_y h_z \sum_{\mathrm{grid}} 
\chi_{\mu}(\rvec) v_{\mathrm{hxc}}(\rvec) \chi_{\nu}(\rvec).
\end{equation} 
For gradient-corrected functionals, the non-local XC contribution of the KS matrix is 
implemented by a finite-orbital basis expansion, without requiring to evaluate the density Hessians. Thus, in such cases,
the XC contribution is written in a convenient working form, as suggested in \cite{pople92}, 
\begin{equation}
F_{\mu \nu}^{\mathrm{XC \alpha}}= \int \left[ \frac{\partial f}{\partial \rho_{\alpha}} \chi_{\mu} \chi_{\nu} +
 \left( 2 \frac{\partial f}{\partial \gamma_{\alpha \alpha}} \nabla \rho_{\alpha} + 
    \frac{\partial f} {\partial \gamma_{\alpha \beta}} \nabla \rho_{\beta} \right) 
    \cdot \nabla (\chi_{\mu} \chi_{\nu}) \right] d\rvec, 
\end{equation}
where $\gamma_{\alpha \alpha} = |\nabla \rho_{\alpha}|^2$, $\gamma_{\alpha \beta} = \nabla \rho_{\alpha} \cdot \nabla 
\rho_{\beta}$, $\gamma_{\beta \beta} = |\nabla \rho_{\beta}|^2$ and $f$ is a function only of local quantities $\rho_{\alpha}$, 
$\rho_{\beta}$ and their gradients. 

\subsection{Pseudopotential}
In today's repertoire of computational chemistry/physics electronic structure theory, a pseudopotential approximation is a very 
useful and powerful concept which is extensively used for atoms (especially with heavy nuclei), molecules (containing one/more
transition metal/lanthanide/actinide), solid state, etc. \citep{huang57, cohen75, leggett01}. The idea was originally introduced 
a long time ago by Hellmann, Fermi and Gombas independently; thereafter applied and popularized by several 
other groups \citep{cohen84, schwerdtfeger11}. It simplifies by dividing the space of many electrons into core and valence 
categories. To a very good approximation, because of their strong binding with the nucleus, inner-shell electrons form 
a passive ``inert" core (retaining an atomic-like configuration) which supposedly plays a less significant role in the 
understanding of chemical binding. Consequently, a surprising majority of chemical properties can be followed very satisfactorily 
by considering only valence electrons. In effective terms, this implies to replace the strong Coulomb potential between nucleus 
and tightly bound core electrons (as a part of the exact Hamiltonian) by a non-local, smoother potential acting on the valence 
electrons. That means, the spectrum of the resulting pseudo-Hamiltonian is as close to the exact one and this is also true for 
the pseudo eigenfunction. It diminishes 
computational cost in larger molecular systems dramatically, simply by cutting down the basis set, and thus avoiding any integral 
arising from core electrons. A lot of developments have been recorded regarding the suitability and effectiveness  
of this approach, compared to the more complete and rigorous, all-electron calculations \citep{maron98, schwerdtfeger95, 
schwerdtfeger00}. Moreover, relativistic effects can be brought into a many-electron picture efficiently through 
pseudopotential schemes \citep{batyrev01, schwerdtfeger11}. 

Thus a molecular pseudo-Hamiltonian $H_v$ can be factored as:
\begin{equation}
H_v = -\frac{1}{2} \sum_{i}^{n_v} \nabla_{i}^{2} + \sum_{i < j}^{n_v} \frac{1}{\rvec_{ij}} + \sum_{i}^{n_v} \sum_{a}^{N_c} 
\bigg[v_{a}^{\mathrm{pp}}(\rvec_{ai})-\frac{Q_a}{\rvec_{ai}} \bigg] + \sum_{a<b}^{N_c} \frac{Q_a Q_b}{\rvec_{ab}}, 
\end{equation} 
where $n_v$ denotes number of valence electrons, $N_c$ the number of core (nuclei) ones and $Q_a$ signifies residual charge 
of core $a$. Indices $(i,j)$ and $(a,b)$ run over all valence electrons and all cores respectively; while $v^{\mathrm{pp}}$, 
the pseudopotential operator takes account of the fact that valence space is repulsive in short range and attractive in long 
range. Therefore, the basic challenge is to find out a decent and computer-proficient approximation to $v^{\mathrm{pp}}$ comparable 
with all-electron calculations. It is to be noted that the distinction between core and valence shells is essential for the 
success of the pseudopotential approximation and hence, it is beneficial to work within an orbital-based theory. Under this 
situation, our desired KS equation could be rewritten as:
\begin{equation}
\bigg[ -\frac{1}{2} \nabla^{2} + v_{\mathrm{ion}}^{\mathrm{pp}}(\rvec) + v_{\mathrm{h}}[\rho(\rvec)] + 
v_{\mathrm{xc}}[\rho(\rvec)] \bigg ] \psi_i(\rvec) = 
\epsilon_i \psi_i(\rvec),
\end{equation}
where $v_{\mathrm{ion}}^{\mathrm{pp}}$ designates ionic pseudopotential for the system,    
\begin{equation}
v_{\mathrm{ion}}^{\mathrm{pp}} = \sum_{R_a} v_{\mathrm{ion},a}^{\mathrm{pp}}(\rvec-\Rvec_a).
\end{equation}
In above equation, $v_{\mathrm{ion}, a}^{\mathrm{pp}}$ represents the ion-core pseudopotential associated with atom $a$, situated at 
$\Rvec_a$; whereas $v_\mathrm{h}[\rho(\rvec)]$ denotes the classical electrostatic potential among the valence electrons and 
$v_{\mathrm{xc}}[\rho(\rvec)]$, 
as usual defines the non-classical part of the many-electron Hamiltonian. 

At present, two possible approaches are predominant for pseudopotential approximations in molecular applications, namely (i) model 
core potential \citep{huzinaga71} and (ii) semi-local approximation. In the former, $v^{\mathrm{pp}}$ is formulated as,
\begin{equation}
v^{\mathrm{pp}}(\rvec) = \sum_k A_k \rvec^{n_k} e^{-\alpha_k \rvec^{2}} + \sum_c B_c \ket{\phi(\rvec)} \bra{\phi(\rvec)}
\end{equation}
where $A_k$, $n_k$ and $\alpha_k$ are all adjustable parameters and $B_c$ is a chosen parameter such that $B_c$ = $-2\epsilon_c$, 
the core energies ($k$ runs over Gaussian expansion). It is possible to get HF orbital energies and corresponding radial 
functions \citep{zeng09} by adjusting above equation. An advantage of this representation is that the inner-shell structure of the
pseudo-valence orbitals is conserved and closely resembles the all-electron valence orbitals; as a result, relativistic and 
spin-orbit effects can be efficiently included into it \citep{fedorov02}. Incorporation of non-local core-valence exchange 
of the all-electron Fock operator into this approximation leads to \emph{ab initio} model potential approximation 
\citep{andzelm85}, which has been profitably implemented into molecular program packages such as Molcas, GAMESS-US etc. 

On the other hand, within a semi-local approximation \citep{schwarz71, kahn72, redondo77, preuss81, schwerdtfeger82, kahn84}, 
$v^{\mathrm{pp}}$ is expressed as, 
\begin{align}
v^{\mathrm{pp}}(\rvec) = v_{\mathrm{local}}(\rvec) + \sum_{l=0}^{l_{\mathrm{max}}} v_l(\rvec)P_{l} \nonumber \\
v^{\mathrm{pp}}(\rvec) =  \sum_k A_k \rvec^{n_k} e^{-\alpha_k \rvec^{2}} + \sum_{l=0}^{l_{\mathrm{max}}} \sum_k B_{lk}  
\rvec^{n_{lk}} e^{-\beta_{lk} 
\rvec^{2}} \sum_{m=-l}^{l} \ket{lm} \bra{lm}, 
\end{align}
where $A_k, B_{lk}, n_k, n_{lk}, \alpha_{lk}$ and $\beta_{lk}$ are all adjustable parameters. The last term in above equation 
contains a projection operator $P_l$ which projects onto the Hilbert sub-space with angular momentum $l$. Unlike Eq.~(52), it 
does not contain any core function and hence, uses orbitals that are smooth and node-less in short-range of the radial function. 
This approximation has found profound applications, especially in relativistic and spin-orbit effects \citep{mcmurchie81,  
fuentealba83, lee09} and, has been implemented in several molecular 
program packages like Gaussian90, Molpro, Turbomole and also in solid-state program CRYSTAL. Depending on the manner in which 
adjustable parameters in Eq.~(53) are ascertained, there are two leading categories in this case, namely, (i) energy-consistent 
\citep{thierfelder10} and (ii) shape-consistent \citep{hay85a}. In the former 
scenario, these parameters are settled down through a fitting procedure (often least square) \cite{figgen09} to a large 
number of pre-calculated transitions in valence space. In latter plan, this is done in such way that the valence orbitals of 
different symmetries resemble the shape of the corresponding all-electron orbitals very closely after a certain cut-off radius and 
also match the orbital energies with high accuracy. This scheme was formulated and implemented by a number of research groups 
\citep{kahn72, redondo77, kahn84, hay85a, hay85b, wadt85}. Additionally it is usually faster and more efficient than the
former, though it requires a large number of Gaussian fitting functions for inversion of the Fock equation involved. Moreover,
the first procedure requires a vast number of valence spectra to fulfill the shape-consistent requirement. On the other hand, 
there is a family of non-conserving pseudopotentials \citep{hamann79} (slightly different from shape-consistent), 
which has been found to be most suitable for plane-wave codes within DFT. In our current development, we adopt the latter form
as proposed by \citep{hay85a, wadt85}.

\section{Results and discussion}
Before we proceed for a discussion of results, mention may be made about some of the technical aspects. Our desired KS equation 
is solved following the usual self-consistent procedure by imposing three convergence criteria namely, (i) electronic energy 
differences between two successive iterations lies below $10^{-6} $ a.u. (ii) maximum absolute deviation in the potential is 
less than $10^{-5}$ a.u. and (iii) standard deviation in the density matrix remains below $10^{-5}$ a.u. The generalized matrix-eigen 
value problem is solved using standard LAPACK routine \citep{anderson99} efficiently and accurately. Computation of FFT is done
by means of the standard FFTW3 package \cite{fftw05}. 

\begingroup
\squeezetable
\begin{table}     
\caption{\label{tab:table1} Energy components and total number of electrons, $N,$ in various grids, for Cl$_2$, 
at $R=4.2$ a.u., along with reference results. All quantities are in a.u. It is adopted from \cite{roy08}.}
\begin{ruledtabular}
\begin{tabular} {lccccccccc}
Set  & A & B & C & D &   E        &  F         &    G       &  H   &  Ref. \cite{schmidt93} \\
\cline{1-10}
$h_r$                      &  0.3       &   0.4      &   0.2      & 0.3        & 0.4        & 0.1        
                           & 0.2        & 0.1        &            \\
$N_r$                      & 32 & 32 & 64 & 64 & 64 & 128 & 128 & 256 &  \\
$\langle T \rangle$        &  11.00750  & 11.17919   &  11.18733  & 11.07195   & 11.06448   & 11.18701   
                           & 11.07244   & 11.07244   & 11.07320    \\
$\langle V^{\mathrm{ne}}_{\mathrm{t}} \rangle$ & $-$83.43381& $-$83.68501& $-$83.70054& $-$83.45722& $-$83.44290& $-$83.69988
                           & $-$83.45810& $-$83.45810& $-$83.45964 \\
$\langle V^{\mathrm{ee}}_{\mathrm{t}} \rangle$ & 32.34338   & 31.22265   & 31.22885   & 31.00981   & 31.00306   & 31.22832   
                           & 31.01000   & 31.01000   &  31.01078   \\
$\langle E_{\mathrm{nu}} \rangle$   &  11.66667  &  11.66667  & 11.66667   & 11.66667   & 11.66667   & 11.66667   
                           & 11.66667   & 11.66667   &  11.66667 \\
$\langle V \rangle$        & $-$39.42376& $-$40.79570& $-$40.80503& $-$40.78074& $-$40.77317& $-$40.80489
                           & $-$40.78144& $-$40.78144&  $-$40.78219\\
$\langle E_{\mathrm{el}} \rangle$   & $-$40.08293& $-$41.28318& $-$41.28437& $-$41.37545& $-$41.37535& $-$41.28455
                           & $-$41.37566& $-$41.37566&  $-$41.37566\\
$\langle E \rangle$        & $-$28.41626& $-$29.61651& $-$29.61770& $-$29.70878& $-$29.70868& $-$29.61789
                           & $-$29.70900& $-$29.70900&  $-$29.70899\footnotemark[1] \\
 $N$                       & 13.89834   & 13.99939   &  13.99865  & 14.00002   &  14.00003  & 13.99864   
                           &  14.00000     &  13.99999         & 13.99998  \\
\end{tabular}
\end{ruledtabular}
\footnotetext[1]{Refers to the grid-DFT calculation; corresponding grid-free DFT value is $-$29.71530 a.u.} 
\end{table}
\endgroup

We begin by scrutinizing the stability and convergence for a representative homo-nuclear diatomic molecule Cl$_2$, at an 
internuclear distance of 4.2 a.u. For this, the non-relativistic ground-state total energy as well as various energy components
are reported in Table~I for eight chosen grid Sets A-H, varying in $N_r$ and $h_r$ ($r \in \{x,y,z\}$). 
The LDA XC potential, corresponding to homogeneous electron-gas correlation of Vosko-Wilk-Nusair \cite{vosko80}, is used in 
this table, which is quoted from \cite{roy08}. All results in this and the following tables compare these
with corresponding reference values obtained from the standard GAMESS suite of quantum chemistry program \citep{schmidt93} 
incorporating the same XC functional, basis set and effective core potential. All pseudopotential calculations in this work, utilize
the Hay-Wadt-type \cite{hay85} valence basis set in which the orbital is split into inner and outer components (derived by two and 
one primitive Gaussian functions respectively). Reference values are quoted for both ``grid'' and ``grid-free'' calculations; 
former has a default ``army'' grade grid with Euler-MacLaurin quadratures for radial integration and Gauss-Legendre quadrature 
for angular integrations. The latter \citep{glaesemann98} follows an auxiliary basis-set expansion procedure for molecular 
integrals instead of a quadrature grid; it has the ability to bypass any error associated with a finite grid, though it 
suffers from an apparently inherent weakness of incompleteness. Following expectation values of energy components are 
reported for comparison, \emph{viz.,} kinetic energy $\langle T \rangle$, total nucleus-electron potential energy 
$\langle V_{\mathrm{t}}^{\mathrm{ne}} \rangle$, total two-electron potential energy $\langle V_{\mathrm{t}}^{\mathrm{ee}} \rangle $, 
total potential energy 
$\langle V \rangle $($= \langle V_{\mathrm{t}}^{\mathrm{ne}} \rangle + \langle V_{\mathrm{t}}^{\mathrm{ee}} \rangle + 
\langle E_{\mathrm{nu}} \rangle $) where 
$ \langle E_{\mathrm{nu}} \rangle $ is nuclear repulsion energy, plus total integrated electron density $N$. Individual components 
of $ \langle E_{\mathrm{hxc}} \rangle $ are not produced as they can not be directly compared (unavailable from output). Clearly,  
Set A energies deviate maximum from reference values, presumably because our box is too small to account for all 
important interactions; this is also reflected in a poor $N$ value. As grid spacing is increased
from 0.3 of Set A to 0.4 in Set B keeping $N_r$ fixed at 32, with a corresponding enlargement of box size, all quantities
improve considerably. Sets C, F also furnish similar kind of accuracy as the box length is same; however both are
inadequate to reach the correct value. It is interesting to note that Sets B, C, F all correspond to the same box size; but the
quality of results is slightly better in case of Set F, as it offers a finer grid structure ($h_{r}=0.1$). A similar conclusion 
can be drawn for Sets E, G too. Results from Sets D, E are quite comparable to reference values. In order to 
prove stability, some further calculations are performed by elongating the box in Sets F-H in columns 7-9. This discussion 
evidently establishes that either of four Sets D, E, G, H offer the desired accuracy; while D, E are capable of providing 
sufficient accuracy for all practical purposes. A similar analysis was done for HCl \cite{roy08} (at $R=2.4$ a.u.) in various 
grids, giving analogous pattern. Set B seems to be quite reasonable for reproducing literature results up to third decimal
place, although it was inadequate for Cl$_2$. All the sets produce results of similar accuracy; Set B amongst them does 
it slightly poorly in terms of component energies and $N$. As one passes from B-C-D, a gradual improvement in results is 
recorded, as expected. Sets C and F indicate agreement with each other as in Cl$_2$. For Sets D and E, $N$ remains practically 
unchanged and 
energy increases by $\approx$0.0002 a.u. Additional calculations with $N_r=128, h_r=0.3$ suggests complete agreement with Set G. 
As in Cl$_2$, the best three sets for HCl are again D, E and G; with the first two being fairly accurate for most calculations. 

\begingroup
\squeezetable
\begin{table}      
\caption{\label{tab:table2} Variation of the energy components and $N$ with respect to the grid parameters for Cl$_2$ and
HCl with reference values. BLYP results in a.u.}
\begin{ruledtabular}
\begin{tabular} {lrrrrrr}
    & \multicolumn{3}{c}{Cl$_2$ ($R=4.2$ a.u.)}  &  \multicolumn{3}{c}{HCl ($R=2.4$ a.u.)}      \\
\cline{2-4} \cline{5-7} 
Set                        &     A      &    B       & Ref.~\cite{schmidt93}     &   A  &   B   &  Ref.~\cite{schmidt93} \\
$N_r$                      &    64      &   128      &             &   64        &    128      &             \\
$h_r$                      &    0.3     &   0.2      &             &   0.3       &    0.2      &             \\
$\langle T \rangle$        & 11.21504   & 11.21577   &  11.21570   & 6.25431     & 6.25464     & 6.25458     \\
$\langle V^{\mathrm{ne}}_{\mathrm{t}} \rangle$ & $-$83.72582& $-$83.72695& $-$83.72685 & $-$37.29933 & $-$37.29987 & $-$37.29979 \\
$\langle V^{\mathrm{ee}}_{\mathrm{t}} \rangle$ & 31.07572   & 31.07594   &  31.07594   & 12.63884    & 12.63903    & 12.63901    \\
$\langle E_{\mathrm{nu}} \rangle$   &  11.66667  & 11.66667   &  11.66667   & 2.91667     & 2.91667     & 2.91667     \\
$\langle V \rangle$        & $-$40.98344& $-$40.98434& $-$40.98424 & $-$21.74382 & $-$21.74417 & $-$21.74411 \\
$\langle E_{\mathrm{el}} \rangle$   & $-$41.43506& $-$41.43524& $-$41.43522 & $-$18.40618 & $-$18.40620 & $-$18.40620 \\
$\langle E \rangle$        & $-$29.76840& $-$29.76857& $-$29.76855\footnotemark[1] & 
                             $-$15.48951 & $-$15.48953 & $-$15.48953\footnotemark[2] \\
 $N$                       & 14.00006   & 14.00000   & 13.99998    & 8.00002     & 8.00000     &  8.00000    \\
\end{tabular}
\end{ruledtabular}
\footnotetext[1]{Corresponding \emph{grid-free} DFT value is $-$29.74755 a.u. \cite{schmidt93}.} 
\footnotetext[2]{Corresponding \emph{grid-free} DFT value is $-$15.48083 a.u. \cite{schmidt93}.} 
\end{table}
\endgroup

Up to this point, we focused on results with LDA functionals only. As mentioned in the Introduction, such calculations
are fraught with some severe limitations. Hence in order to make practical applications, more refined and accurate functionals
are essential. In this context, let us recall that $E_{\mathrm{xc}}$ consists of two components, each arising as a difference: 
(a) between classical and quantum mechanical inter-electronic repulsion (b) between kinetic energy of the hypothetical, 
non-interacting and the real, interacting system concerned. Usually, the second portion is not explicitly incorporated in the
current family of functionals. In 
this sub-section, we present a cross-section of CCG results within the familiar non-local Becke exchange \cite{becke88a} and 
LYP \cite{lee88} correlation, who converted the Colle-Salvetti formula for the correlation energy from HF second
order density matrix into a density functional form. This will extend the range and scope of applicability of the present scheme
in the context of more realistic non-local functionals which will be crucial for future chemical applications. The latter
is implemented using an alternate equivalent form containing only the first derivative through a procedure due to \cite{miehlich89}.  
Table~II presents a comparison of our BLYP energy components and $N$ at some selective grid sets at same $R$ as in Table~I. 
The same quantities as in the LDA case are reported. Several test calculations were performed in several grids to ensure 
convergence of results; which offered conclusions quite similar to those found in LDA. From this, two grid sets are given 
for each of them. Once again, our CCG results exhibit excellent agreement with literature values, as observed in LDA scenario. In 
Set B, results are slightly better than Set A for obvious reasons of finer grid and an extended length of box. In Cl$_2$, 
this effect is more pronounced than in case of HCl. Absolute deviations in Set B energies are 0.00002 and 0.00000 a.u. respectively 
for Cl$_2$, HCl. Grid-free and grid DFT results are substantially different for both molecules. More details could be 
found in \cite{roy08a}. 

\begingroup
\squeezetable
\begin{table}
\caption{\label{tab:table3} Comparison of calculated negative eigenvalues of Cl$_2$ and HCl with reference values.
BLYP results are given in a.u. See text for more details.} 
\begin{ruledtabular}
\begin{tabular} {lccclccc}
    MO   & \multicolumn{3}{c}{Cl$_2$ ($R=4.2$ a.u.)} & MO  & \multicolumn{3}{c}{HCl ($R=2.4$ a.u.)} \\
\cline{2-4}  \cline{6-8}
Set          & A      &   B     & Ref. \cite{schmidt93} &  &  A       & B       & Ref. \cite{schmidt93} \\ \hline 
$N_r$        & 64     &  128    &                       &  &  64      &  128    &          \\
$h_r$        & 0.3    &  0.2    &                       &  &  0.3     &  0.2    &          \\
 $2\sigma_g$ & 0.8143 & 0.8143  & 0.8143   & $2\sigma$     &  0.7707  & 0.7707  & 0.7707   \\
 $2\sigma_u$ & 0.7094 & 0.7094  & 0.7094   & $3\sigma$     &  0.4168  & 0.4167  & 0.4167   \\
 $3\sigma_g$ & 0.4170 & 0.4171  & 0.4171   & $1\pi_x$      &  0.2786  & 0.2786  & 0.2786   \\
 $1\pi_{xu}$ & 0.3405 & 0.3405  & 0.3405   & $1\pi_y$      &  0.2786  & 0.2786  & 0.2786   \\
 $1\pi_{yu}$ & 0.3405 & 0.3405  & 0.3405   &  &            &          &                    \\ 
 $1\pi_{xg}$ & 0.2778 & 0.2778  & 0.2778   &  &            &          &                    \\
 $1\pi_{yg}$ & 0.2778 & 0.2778  & 0.2778   &  &            &          &                     \\
\end{tabular}
\end{ruledtabular}
\end{table}
\endgroup

In order to justify the dependability and usefulness of CCG, Table~III now samples calculated negative eigenvalues for 
Cl$_2$ and HCl using BLYP XC functional at same geometries of previous tables. These are provided in two selected grid 
sets. Comparison with reference values again indicates excellent agreement for both molecules for \emph{all} orbital energies 
except the lone case of $3\sigma_g$ for Cl$_2$ (Set A) and $3\sigma$ for HCl (Set B); even then the absolute deviation remains 
well within 0.0001 a.u. More details could be found in \cite{roy08}.  
 
\begingroup
\squeezetable
\begin{table}      
\caption{\label{tab:table4} Comparison of various energies and $N$ for several atoms and molecules for different XC functionals. 
For more details, see text.}
\begin{ruledtabular}
\begin{tabular} {lcccccccccc}
    & \multicolumn{4}{c}{$\left\langle T \right\rangle $}  &  \multicolumn{4}{c}{$-\left\langle V \right\rangle$} \\
\cline{2-5} \cline{7-10}
System  & LDA & Ref.~\citep{schmidt93} & BLYP & Ref.~\citep{schmidt93} &  & LDA &  Ref.~\citep{schmidt93} & BLYP & 
Ref.~ \citep{schmidt93} &  \\
\cline{1-10}
P & 2.35430 & 2.35334 & 2.38891 & 2.38890 & & 8.73501 & 8.73404 & 8.78249 & 8.78248 \\
H$_2$S & 4.90204 & 4.90197 & 4.98071 & 4.98066 & & 16.10707 & 16.10698 & 16.21919 & 16.21913 \\
MgCl$_2$ & 11.62114 & 11.62208 & 11.75947 & 11.75999 & & 42.34513 & 42.34621 & 42.54049 & 42.54103 \\
SiH$_2$Cl$_2$ & 13.95036 & 13.94989 & 14.14948 & 14.14945 & & 48.78729 & 48.78685 & 49.04463 & 49.04461 \\
\cline{1-10}
    & \multicolumn{4}{c}{$-\left\langle E \right\rangle $}  &  \multicolumn{4}{c}{N} \\
\cline{2-5} \cline{7-10}
System  & LDA & Ref.~\citep{schmidt93} & BLYP & Ref.~\citep{schmidt93} &  & LDA &  Ref.~\citep{schmidt93} & BLYP & 
Ref.~\citep{schmidt93} &  \\
\cline{1-10} 
P & 6.38070 & 6.38071 & 6.39358 & 6.39358 & & 5.00000 & 4.99999 & 4.99999 & 4.99999 \\
H$_2$S & 11.20503 & 11.20501 & 11.23848 & 11.23846 & & 8.00000 & 7.99999 & 8.00000 & 7.99999 \\
MgCl$_2$ & 30.72399 & 30.72413 & 30.78102 & 30.78104 & & 16.00004 & 15.99957 & 16.00004 & 15.99999 \\
SiH$_2$Cl$_2$ & 34.83693 & 34.83696 & 34.89515 & 34.89516 & & 19.99999 & 20.00015 & 19.99999 & 20.00000 \\    
\end{tabular}
\end{ruledtabular}
\end{table}
\endgroup 

Now, Table~IV offers a comparative chart of $\left\langle T \right\rangle $, $\left\langle V \right\rangle$, 
total (E) energies and $N$ for few chosen atoms and molecules using both LDA and XC functionals. LDA results employ Set E grid 
($N_r=64, h_r =0.4$), while for BLYP calculations, except P and H$_2$S (using Set E), grid Set I ($N_r=128, h_r=0.3$) was used. 
Overall, the present calculated results agree with literature values more or less up to two decimal places and $N$ 
completely matches with the reference values with few exceptions. Results for more atoms and molecules can be found in 
\citep{roy08, roy08a}.

For further examination, Fig.~(1) displays the changes in above LDA energies in Cl$_2$ (relative to $-$29 a.u.) for
four selected Sets D, E, G, I and HCl (relative to $-$15 a.u.) for Sets B, C, D respectively. They cover a span of 3.5-5.0 and 
1.6-3.1 a.u., for Cl$_2$ and HCl. Respective reference plots are also given for easy comparison. It is quite gratifying that all 
four sets replicate the shape of the potential energy curve quite well for the entire range of $R$, in Cl$_2$. A thorough analysis 
of results (not given here and the reader may be referred to \cite{roy08}) reveals that, until $R=4$ a.u., Set D
energies are quite close to the reference (higher by only 0.0001 a.u.), while implying a gradual tendency to deviate 
thereafter. However, the maximum deviation still remains quite small overall (only 0.0007 a.u.) occurring for $R=5$ a.u. Sets G, I
computed energies either completely match with reference or provide a maximum difference of 0.0001 a.u. Further in all sets, our 
CCG energies have always been above reference except in two occasions ($R=$4 and 4.3 for Set G). Now we turn to the 
bottom panel in the left side for HCl. 
Once again, we observe excellent qualitative agreement for whole range of $R$, as evident from their nearly identical shapes. 
Careful examination admits that maximum absolute deviation for three sets B, C, D remains rather small, \emph{viz.}, 0.0014, 0.0007,
0.0002 for $R=$3, 2.3, 2.3 respectively. It is possible to fine-tune these results further by enforcing more rigorous 
convergence criteria. For more details, see \cite{roy08}. 
Similar plots for BLYP functionals are given in right side for Cl$_2$, HCl in top, bottom panels, for two grids A, B.  
In both cases, again respective grid-DFT plots are provided for comparison, which practically coincide.
A careful analysis \citep{roy08a} shows that maximum disagreements are 0.0002 a.u. (only for 2 instances) and 0.0001 a.u. 
for Sets A, B respectively in case of Cl$_2$. However, for HCl, the same for both sets are 0.0001 a.u. (only for 2 instances). It 
is a matter of fact that the two sets confirm to each other more for HCl than Cl$_2$.

\begin{figure}            
\centering
\includegraphics[scale=0.9]{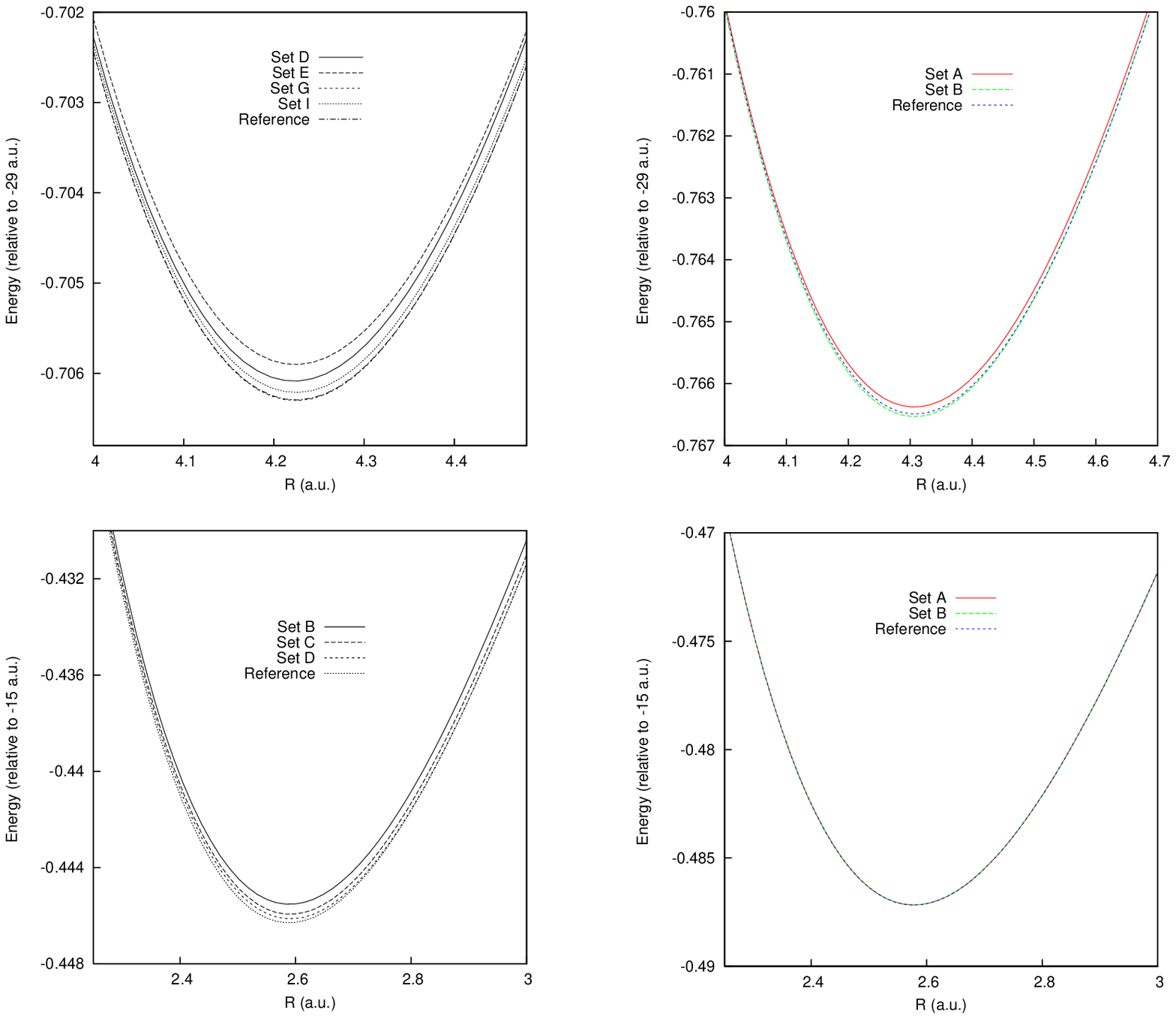}
\caption[optional]{Potential energy curves for Cl$_2$ (top), HCl (bottom) for different grid sets. Left, right panels 
correspond to LDA, LYP functionals. Reference grid-DFT results are also given for comparison. Grid sets in left, right panels 
refer to corresponding $(N_r, h_r)$ values of Tables~I, II respectively.}
\end{figure}

Now results are given for four selected species in Table~V, for two other functionals which are used in applications of 
many-electron systems with considerable success, namely, PBE \cite{perdew96} and Filatov-Thiel (FT97) \cite{filatov97a, filatov97b}.
These non-local functionals are implemented with the help of Density Functional Repository program \cite{repository}.  
For sake of completeness, respective LDA and BLYP values are also provided. Experimental geometries are taken from NIST 
computational chemistry database \cite{johnson06}. Set E grid was used for this table. Since correctness of our results
has already been established in previous tables, here we omit the reference values; all entries are very close to them.
Total energies decrease as one passes from LDA to non-local functionals. Within the latter category, there is some 
agreement amongst the three, with FT97 offering the lowest. $N$ maintains same accuracy for all functionals.  

\begingroup
\squeezetable
\begin{table}[t]      
\caption{\label{tab:table5} Comparison of kinetic ($\left\langle T \right\rangle $), potential ($\left\langle V \right\rangle$), 
total (E) energies and $N$ for several atoms and molecules for different XC functionals. See text for details.}
\begin{ruledtabular}
\begin{tabular} {lcccccccccc}
    & \multicolumn{4}{c}{$\left\langle T \right\rangle $}  &  \multicolumn{4}{c}{$-\left\langle V \right\rangle$} \\
\cline{2-5} \cline{7-10}
System  & LDA & BLYP & PBE & FT97 &  & LDA &  BLYP & PBE & FT97 &\\
\cline{1-10}
Br & 4.22038 & 4.27022 & 4.17153 & 4.22828 & & 17.28157 & 17.36122 & 17.29168 & 17.37615 \\
Al$_2$ & 1.35409 & 1.37642 & 1.32292 & 1.34896 & & 5.21624 & 5.24844 & 5.20904 & 5.25881 \\
SiH$_4$ & 3.65832 & 3.76989 & 3.53302 & 3.60707 & & 9.82333 & 9.97905 & 9.75741 & 9.86882 \\
S$_2$ & 7.58306 & 7.68138 & 7.51586 & 7.59269 & & 27.62272 & 27.77409 & 27.64063 & 27.76097 \\
\cline{1-10}
    & \multicolumn{4}{c}{$-\left\langle E \right\rangle $}  &  \multicolumn{4}{c}{$N$} \\
\cline{2-5} \cline{7-10}
System  & LDA & BLYP & PBE & FT97 &  & LDA &  BLYP & PBE & FT97 &\\
\cline{1-10} 
Br & 13.06119 & 13.09100 & 13.12015 & 13.14787 & & 7.00000 & 7.00000 & 6.99999 & 6.99999 \\
Al$_2$ & 3.86215 & 3.87203 & 3.88612 & 3.90985 & & 5.99999 & 5.99999 & 5.99999 & 5.99999 \\
SiH$_4$ & 6.16500 & 6.20916 & 6.22438 & 6.26175 & & 7.99999 & 8.00000 & 7.99999 & 7.99999 \\
S$_2$ & 20.03966 & 20.09271 & 20.12477 & 20.16828 & & 12.00000 & 12.00000 & 12.00000 & 12.00000 \\    
\end{tabular}
\end{ruledtabular}
\end{table}
\endgroup 

Finally Table~VI, provides a comparison of calculated $-\epsilon_{\mathrm{HOMO}}$ and atomization energies with 
experimental results \citep{afeefy05} for 4 selected molecules at their experimental geometries \citep{johnson06}, with LDA and 
BLYP XC functionals. For both quantities, reference theoretical results \citep{schmidt93} are practically identical to those
found from CCG, and so omitted here to avoid crowding. It is observed that calculated ionization energies generally signal large 
deviations (typically off by 30-50\%) from experimental results for both functionals. Estimated ionization 
energies from PBE and FT97 are also reported, which also deliver similar performances. This leads to a general trend, which some
of the frequently used functionals suffer from, i.e., even though energies are reproduced properly, ionization energies are 
described rather poorly. Now, it is a crucial requirement for real-time electron dynamics to have a proper description of orbital 
energies, especially the higher-lying ones. For this, we present the computed HOMO energies with modified Leeuwen-Baerends (LB) 
exchange potential \citep{leeuwen94,schipper00} along with the other four functionals. It is known to give a more satisfactory 
representation of the long-range nature of the XC potential, as reflected from a better matching with experimental results. 
This is formulated as:
\begin{equation}
v_{\mathrm{xc\sigma}}^{\mathrm{LB\alpha}} (\alpha, \beta: \rvec) = \alpha v_{\mathrm{x\sigma}}^{\mathrm{LDA}} (\rvec) + 
v_{\mathrm{c\sigma}}^{\mathrm{LDA}} (\rvec) +
\frac{\beta x_{\sigma}^2 (\rvec) \rho_{\sigma}^{1/3} (\rvec)} 
{1+3\beta x_{\sigma}(\rvec) \mathrm{ln} \{x_{\sigma}(\rvec)+[x_{\sigma}^2 (\rvec)+1]^{1/2}\}}, 
\end{equation}
where $\alpha=1.19$, $ \beta=0.01$ are two empirical parameters, $\sigma$ signifies up, down spins and last term containing 
gradient correction is reminiscent of the exchange functional of \citep{becke88a}, while 
$x_{\sigma}(\rvec) = |\nabla \rho_{\sigma}(\rvec)|[\rho_{\sigma}(\rvec)]^{-4/3}$ is a dimensionless quantity. This ensures the
required long-range nature, i.e., $v_{\mathrm{xc\sigma}}^{\mathrm{LB\alpha}} (\rvec) \rightarrow -1/r, r \rightarrow \infty.$ 
A combination of LB exchange with the VWN correlation functional produces the HOMO energies of  
column 6, which abundantly establishes its superiority over all the other four functionals considered here. These are  
substantially lower and more accurate than four functionals. In case of atomization energy, the present results reveal a
considerable deviation from experimental values for both LDA and BLYP functionals. Note that latter contains zero-point 
vibrational corrections and relativistic effects. Interestingly, LDA atomization energies are apparently better than BLYP 
counterparts in some cases, but that should not be misconstrued to identify the former as a better choice. There could be some 
cancellation of errors. Moreover there are other important factors, such as using more complete accurate basis 
set or more appropriate pseudopotential (the current work uses one designed for HF calculations; those developed
for DFT would be more suitable), which could circumvent some of these discrepancies. However, such deviations
are not uncommon and found, even in all-electron calculations with more sophisticated, extended basis sets. More discussion 
on these issues could be found in \cite{roy08a}. Nevertheless, this is aside the primary objective of this chapter and 
does not directly impact the main conclusions drawn herein.

\begingroup
\squeezetable
\begin{table}      
\caption{\label{tab:table6} Comparison of $-\epsilon_{\mathrm{HOMO}}$ (in a.u.) and atomization energies (in kcal/mole) for 
several molecules for different XC functionals with literature experimental data. }
\begin{ruledtabular}
\begin{tabular} {lcccccccccc}
    & \multicolumn{6}{c}{$-\epsilon_{\mathrm{HOMO}}$ (a.u.)} &\multicolumn{3}{c}{Atomization energies (kcal/mol)} \\
\cline{2-7} \cline{8-10}
System  & LDA & BLYP & PBE & FT97 & LBVWN  & Expt. \citep{afeefy05} & LDA & BLYP & Expt. \citep{afeefy05} &\\
\cline{1-10} 
SiH$_4$ & 0.3188 & 0.3156 & 0.2919  & 0.2702 & 0.4624 & 0.4042 & 339.43 & 312.02 & 302.6 \\
PH$_2$ & 0.2170 & 0.2111 & 0.1970 & 0.1741 & 0.3504 & 0.3610 & 152.77 & 139.92 & 149.2 \\
P$_4$ & 0.2712 & 0.2575 & 0.2525 & 0.2369 & 0.3964 & 0.3432 & 200.77 & 142.99 & 285.9 \\
Al$_2$ & 0.1407 & 0.1400 & 0.1318 & 0.1161 & 0.2371 & 0.1984 & 22.92 & 21.42 & 37.0 \\    
\end{tabular}
\end{ruledtabular}
\end{table}
\endgroup 

So far, we were mainly concerned with valence-shell electrons using some sort of effective core potentials to incorporate the effects 
of frozen core electrons. Undoubtedly, while the pseudopotential approximation is indispensable for larger systems (especially those
containing heavy atoms), ``full" calculations provide a more detailed and accurate result at the expense of a heavy computational 
cost. Nowadays, it is all about a balance between cost and accuracy. If the cost accuracy ratio permits, unquestionably, the latter 
remains the preferred choice for most chemical and physical problems. Thus, as a further extension of our method, we present some 
sample results on ``full" calculations in Table~VII. The electronic energy including the relevant components and $N$, 
are produced for two representative atoms, namely, N and O in their open-shell ground-state configuration. We employed an 
STO-3G basis set, LDA XC potential and grid Set I ($N_r = 128, h_r = 0.3$). In all cases computed quantities, once again, 
establish excellent agreement with reference values. This validates the effectiveness and level of performance of our method in 
the context
of all-electron calculations. It also gives a comparison of orbital energies for both $\alpha$ and $\beta$ spins for the same 
two atoms, which again proves the usefulness. It is a matter of fact that we use very crude basis functions and LDA functional for 
the purpose of illustration; involvement of better basis sets and non-local XC functionals should enhance the applicability and 
accuracy of full calculations.

Before passing, a few comments are in order. The computational scaling goes as follows: (a) construction of the localized basis 
set scales as $N_g$, where $N_g$ is total number of grid points (b) formation of the basis set in grid scales as $N_b \times 
N_g$, with $N_b$ giving the number of basis functions (c) generation of the electron density in grid scales as $N_b^2 \times N_g$ (d)
one- and two-body matrix elements of the KS matrix scale as $N_b^2$ and $N_b^2 \times N_g$ respectively. One obvious way
to achieve better scaling is by reducing $N_b$ and $N_g$. Now, up to this point, all our calculations were performed in a uniform 
Cartesian grid; further there was a restriction of $2^{2N}$ grid points due to the FFT algorithm employed. To lift these 
restrictions, we pursue some calculations in a non-uniform, unequal box of arbitrary size other than a cubic shape. For treatment 
of larger systems, it is highly desirable, as it offers much flexibility. At this stage, we provide some preliminary results 
on a general non-uniform grid with arbitrary number of points. For this, Cl$_2$ is chosen as a representative in Table~VIII. 
These are performed with LDA XC functional, using 
a grid spacing of $h_r=0.3$, keeping the same pseudopotential. A snapshot of computed energies, from two sets (labelled I, II) is 
offered for illustration. In Set~I in the left-hand side, we first vary $N_z$, the number of grid points along internuclear axis 
($z$) 
keeping same along $xy$ plane static at certain reasonable value ($N_x=N_y=30$). This can be appreciated from a glance at Table~I; 
the calculated total electronic energy of Cl$_2$ is $-$28.41626 a.u. in case of Set A using the LDA XC functional. As $N_z$ is 
increased from 30 to 40, there is a huge jump in energy value ($\approx 4.16$ a.u); with any further increase, it slowly converges to 
$-$29.66505 a.u., at around $N_z=70$. Applying the same procedure for fixed $N_x, N_y$ at 36, offers an energy value of $-$29.70388, 
converging again nearly at $N_z =70$. There is no need to show all these details. Rather it suffices to mention that for each such 
$(N_x, N_y)$ pair, 
energies attain convergence for a certain $N_z$, which in this case happens to be around 70. Then Set~II, in the right side, varies   
the number of points, $N_x, N_y$ along $xy$ plane keeping $N_z$ constant at 70. It is apparent that formal convergence of energy 
takes place for $N_x= N_y=48$. A re-look at Table~I reveals that the comparable convergence in a uniform grid occurs for Set~D 
corresponding to $N_x=N_y=N_z=64$ and $h_r=0.3$. This leads to a drastic reduction in the total number of points, which eventually
helps to achieve an enhanced     
computational scaling compared to the uniform grid. This will have a tremendous bearing on routine calculations of larger systems.  
A more thorough analysis of these aspects will be undertaken later.

\begingroup
\squeezetable
\begin{table}      
\caption{\label{tab:table7} Energy components and orbital energies for two atoms (N, O) using Cartesian grid. All-electron calculations with 
LDA XC functionals using STO-3G basis set are given along with reference data. PR implies Present Result.}
\begin{ruledtabular}
\begin{tabular} {cccccccccc}
 Energy & \multicolumn{2}{c}{N}  &  \multicolumn{2}{c}{O}  &  Orbital & \multicolumn{2}{c}{N}  & \multicolumn{2}{c}{O} \\
\cline{2-3} \cline{4-5} \cline{7-8} \cline{9-10}

& PR & Ref.~\citep{schmidt93} & PR & Ref.~\citep{schmidt93} & & PR & Ref.~\citep{schmidt93} & PR & Ref.~\citep{schmidt93} \\
\hline
$\langle T \rangle$ & 53.66407 & 53.66407 & 73.44497 & 73.44497 & $\epsilon^{\alpha}_{1s}$ & $-$13.6312 & $-$13.6311 & $-$18.3331 & $-$18.1330 \\
$\langle V^{\mathrm{ne}} \rangle$ & $-$127.32649 & $-$127.32649 & $-$176.32432 & $-$176.32432 & $\epsilon^{\alpha}_{2s}$ & $-$0.6152 & $-$0.6153 & $-$0.7538 & $-$0.7537 \\
$\langle V^{\mathrm{ee}} \rangle$ & 20.25538 & 20.25536 & 29.42801 & 29.42799 & $\epsilon^{\alpha}_{2p_{x}}$ & $-$0.1671 & $-$0.1672 & $-$0.1941 & $-$0.1942 \\
$\langle V \rangle$ & $-$107.07111 & $-$107.07113 & $-$146.89631 & $-$146.89634 & $\epsilon^{\alpha}_{2p_{y}}$ & $-$0.1671 & $-$0.1672 & $-$0.1941 & $-$0.1942  \\
$\langle E \rangle$ & $-$53.40704 & $-$53.40701 & $-$73.45134 & $-$73.45137 & $\epsilon^{\alpha}_{2p_{z}}$ & $-$0.1671 & $-$0.1672 & $-$0.1085 & $-$0.1085 \\
$N$ & 6.99999 & 6.99999 & 7.99999 & 7.99999 & $\epsilon^{\beta}_{1s}$ & $-$13.5837 & $-$13.5836 & $-$18.2972 & $-$18.2971 \\
 &  &  &  & & $\epsilon^{\beta}_{2s}$ & $-$0.4449 & $-$0.4450 & $-$0.6301 & $-$0.6302 \\
 &  &  &  & & $\epsilon^{\beta}_{2p_{x}}$ & & & $-$0.0378 & $-$0.0379 \\
\end{tabular}
\end{ruledtabular}
\end{table}
\endgroup 

\begingroup
\squeezetable
\begin{table}      
\caption{\label{tab:table8} Electronic energy$^a$ of Cl$_2$ in non-uniform grid ($h_r=0.3$). LDA results in a.u.}
\begin{ruledtabular}
\begin{tabular} {rrrrrrrrr}
 \multicolumn{4}{c}{Set~I }  & & \multicolumn{4}{c}{Set~II}      \\
\cline{1-4} \cline{6-9} 
$N_x$   &   $N_y$   &    $N_z$  &  $ \langle E \rangle $ &    & $N_x$   &   $N_y$   &    $N_z$  &  $ \langle E \rangle$ \\
\cline{1-9}
30 & 30 & 30 & $-$25.32542 & &30 & 30 & 70 & $-$29.66504 \\
 - & - & 40 & $-$29.48297 & &32 & 32 & - & $-$29.68722 \\
- & - & 50 & $-$29.65381 & &34 & 34 & - & $-$29.69838 \\
- & - & 60 & $-$29.66461 & &36 & 36 & - & $-$29.70387 \\
- & - & 70 & $-$29.66504  & &38 & 38 & - & $-$29.70651 \\
- & - & 80 & $-$29.66505  & &40 & 40 & - & $-$29.70777 \\
- & - & 90 & $-$29.66505  & &42 & 42 & - & $-$29.70837 \\
36 & 36 & 40 & $-$29.52104 & &44 & 44 & - & $-$29.70865 \\
- & - & 60 & $-$29.70343  & &46 & 46 & - & $-$29.70877 \\
- & - & 70 & $-$29.70387   & &48 & 48 & - & $-$29.70884 \\
- & - & 80 & $-$29.70388  & &50 & 50 & - & $-$29.70886 \\
- & - & 90 & $-$29.70388  & &60 & 60 & - & $-$29.70888 \\
\end{tabular}
\end{ruledtabular}
\footnotetext[1]{Corresponding \emph{grid}-DFT value is $-$29.70899 a.u. \cite{schmidt93}.} 
\end{table}
\endgroup

\section{Future and outlook}
We have demonstrated the validity and feasibility of a Gaussian-based LCAO-MO approach to DFT using CCG in the context of atomic and 
molecular properties. This produces practically identical results with those obtained from other grid-based/grid-free 
quantum chemistry programs available in literature. The classical Coulomb potential is obtained by means of a Fourier convolution 
technique, accurately and efficiently in the real-space grid. The matrix elements of all two-body potentials are computed through 
numerical integration. Quantities such as HOMO energies, potential energy curves, orbital energies, atomization energies are 
reproduced very well. The method performs quite decently for both local and non-local XC density functionals. While we focused 
mainly on pseudopotential situations, it is equally applicable for all-electron calculations as well. So far, these works 
were restricted to a uniform grid; here we report some exploratory calculations for Cl$_2$ in a more general \emph{non-uniform} 
grid, for the first time. A more thorough analysis is in progress. Energies are variationally well-founded.

The success of the CCG method, as delineated above, encourages us to employ (i) more appropriate effective core potentials suitable 
for DFT calculations (ii) elaborate, extended and sophisticated basis sets (iii) superior quality density functionals in our 
future works.   
It would be worthwhile to assess its merit and suitability for geometry optimization of molecules using CCG. In this regard, a 
non-uniform grid, could be quite useful, for which some early results have been presented here and needs further inspection. 
An important concern would be to lessen the computational cost 
by incorporating a linear scaling approach; apparently the most  promising will be real-space multi-grid techniques. 
A useful application of this methodology could be in the real-time dynamical studies, particularly laser-atom/molecule interactions 
in intense/super-intense regimes within a TDDFT framework. Some of these works are currently being investigated by us. 

\section{Acknowledgement}
We express our gratitude to the Editors Prof.~Michael Springborg and Prof.~Jan-Ole Joswig for their kind invitation to contribute 
in this issue. We also thank them for extending the deadline of submission. Mr.~Siladitya Jana provided help in getting some of the 
references. Computer support from Mr.~Suman Chakraborty is duly thanked. AKR acknowledges financial support from DST 
(Grant No.~EMR/2014/000838) and AG from UGC, New Delhi, for a Junior Research Fellowship.


\providecommand*{\mcitethebibliography}{\thebibliography}
\csname @ifundefined\endcsname{endmcitethebibliography}
{\let\endmcitethebibliography\endthebibliography}{}
\begin{mcitethebibliography}{189}
\providecommand*{\natexlab}[1]{#1}
\providecommand*{\mciteSetBstSublistMode}[1]{}
\providecommand*{\mciteSetBstMaxWidthForm}[2]{}
\providecommand*{\mciteBstWouldAddEndPuncttrue}
  {\def\EndOfBibitem{\unskip.}}
\providecommand*{\mciteBstWouldAddEndPunctfalse}
  {\let\EndOfBibitem\relax}
\providecommand*{\mciteSetBstMidEndSepPunct}[3]{}
\providecommand*{\mciteSetBstSublistLabelBeginEnd}[3]{}
\providecommand*{\EndOfBibitem}{}
\mciteSetBstSublistMode{f}
\mciteSetBstMaxWidthForm{subitem}
{(\emph{\alph{mcitesubitemcount}})}
\mciteSetBstSublistLabelBeginEnd{\mcitemaxwidthsubitemform\space}
{\relax}{\relax}

\bibitem[(Ed.)(1995)]{yarkony95}
D.~R.~Y. (Ed.), \emph{Modern Electronic Structure Theory}, {World Scientific},
  {Singapore}, 1995\relax
\mciteBstWouldAddEndPuncttrue
\mciteSetBstMidEndSepPunct{\mcitedefaultmidpunct}
{\mcitedefaultendpunct}{\mcitedefaultseppunct}\relax
\EndOfBibitem
\bibitem[Szabo and Ostlund(1996)]{szabo96}
A.~Szabo and N.~S. Ostlund, \emph{Modern Quantum Chemistry}, {Dover}, {New
  York}, 1996\relax
\mciteBstWouldAddEndPuncttrue
\mciteSetBstMidEndSepPunct{\mcitedefaultmidpunct}
{\mcitedefaultendpunct}{\mcitedefaultseppunct}\relax
\EndOfBibitem
\bibitem[Simons and Nichols(1997)]{simon97}
J.~Simons and J.~Nichols, \emph{Quantum Mechanics in Chemistry}, {Oxford
  University Press}, {New York}, 1997\relax
\mciteBstWouldAddEndPuncttrue
\mciteSetBstMidEndSepPunct{\mcitedefaultmidpunct}
{\mcitedefaultendpunct}{\mcitedefaultseppunct}\relax
\EndOfBibitem
\bibitem[Kohn(1999)]{kohn99}
W.~Kohn, \emph{Rev.~Mod.~Phys.}, 1999, \textbf{71}, 1253\relax
\mciteBstWouldAddEndPuncttrue
\mciteSetBstMidEndSepPunct{\mcitedefaultmidpunct}
{\mcitedefaultendpunct}{\mcitedefaultseppunct}\relax
\EndOfBibitem
\bibitem[Helgaker \emph{et~al.}(2000)Helgaker, J{\o}rgensen, and
  Olsen]{helgaker00}
T.~Helgaker, P.~J{\o}rgensen and J.~Olsen, \emph{Modern Electronic Structure
  Theory}, {John Wiley}, {New York}, 2000\relax
\mciteBstWouldAddEndPuncttrue
\mciteSetBstMidEndSepPunct{\mcitedefaultmidpunct}
{\mcitedefaultendpunct}{\mcitedefaultseppunct}\relax
\EndOfBibitem
\bibitem[Springborg(2000)]{springborg00}
M.~Springborg, \emph{Methods of Electronic-Structure Calculations}, {Wiley
  Chichester}, {New York}, 2000\relax
\mciteBstWouldAddEndPuncttrue
\mciteSetBstMidEndSepPunct{\mcitedefaultmidpunct}
{\mcitedefaultendpunct}{\mcitedefaultseppunct}\relax
\EndOfBibitem
\bibitem[Young(2001)]{young01}
D.~C. Young, \emph{Computational Chemistry: A Practical Guide for Applying
  Techniques to Real-World Problems}, {John Wiley}, {New York}, 2001\relax
\mciteBstWouldAddEndPuncttrue
\mciteSetBstMidEndSepPunct{\mcitedefaultmidpunct}
{\mcitedefaultendpunct}{\mcitedefaultseppunct}\relax
\EndOfBibitem
\bibitem[Foulkes \emph{et~al.}(2001)Foulkes, Mitas, Needs, and
  Rajagopal]{foulkes01}
W.~M.~C. Foulkes, L.~Mitas, R.~J. Needs and G.~Rajagopal,
  \emph{Rev.~Mod.~Phys.}, 2001, \textbf{73}, 33\relax
\mciteBstWouldAddEndPuncttrue
\mciteSetBstMidEndSepPunct{\mcitedefaultmidpunct}
{\mcitedefaultendpunct}{\mcitedefaultseppunct}\relax
\EndOfBibitem
\bibitem[Lewars(2003)]{lewars03}
E.~Lewars, \emph{Computational Chemistry: Introduction to the Theory and
  Applications of Molecular and Quantum Mechanics}, {Kluwer Academic},
  {Netherlands}, 2003\relax
\mciteBstWouldAddEndPuncttrue
\mciteSetBstMidEndSepPunct{\mcitedefaultmidpunct}
{\mcitedefaultendpunct}{\mcitedefaultseppunct}\relax
\EndOfBibitem
\bibitem[Cramer(2004)]{cramer04}
C.~J. Cramer, \emph{Essentials of Computational Chemistry: Theories and
  Models}, {John Wiley}, {New York}, 2004\relax
\mciteBstWouldAddEndPuncttrue
\mciteSetBstMidEndSepPunct{\mcitedefaultmidpunct}
{\mcitedefaultendpunct}{\mcitedefaultseppunct}\relax
\EndOfBibitem
\bibitem[Martin(2004)]{martin04}
R.~M. Martin, \emph{Electronic Structure: Basic Theory and Practical Methods},
  {Cambridge University Press}, {Cambridge, UK}, 2004\relax
\mciteBstWouldAddEndPuncttrue
\mciteSetBstMidEndSepPunct{\mcitedefaultmidpunct}
{\mcitedefaultendpunct}{\mcitedefaultseppunct}\relax
\EndOfBibitem
\bibitem[Hoffman(2007)]{hoffman07}
E.~O. Hoffman, \emph{Progress in Quantum Chemistry Research}, {Nova Science
  Publishers}, {New York}, 2007\relax
\mciteBstWouldAddEndPuncttrue
\mciteSetBstMidEndSepPunct{\mcitedefaultmidpunct}
{\mcitedefaultendpunct}{\mcitedefaultseppunct}\relax
\EndOfBibitem
\bibitem[Jensen(2007)]{jensen07}
F.~Jensen, \emph{Introduction to Computational Chemistry}, {John Wiley}, {New
  York}, 2007\relax
\mciteBstWouldAddEndPuncttrue
\mciteSetBstMidEndSepPunct{\mcitedefaultmidpunct}
{\mcitedefaultendpunct}{\mcitedefaultseppunct}\relax
\EndOfBibitem
\bibitem[Tolosa \emph{et~al.}(2007)Tolosa, Sans{\'o}n, and Hidalgo]{kaisas07}
S.~Tolosa, J.~A. Sans{\'o}n and A.~Hidalgo, Quantum Chemistry Research Trends,
  {New York}, 2007\relax
\mciteBstWouldAddEndPuncttrue
\mciteSetBstMidEndSepPunct{\mcitedefaultmidpunct}
{\mcitedefaultendpunct}{\mcitedefaultseppunct}\relax
\EndOfBibitem
\bibitem[Borden(2011)]{borden11}
W.~T. Borden, \emph{J.~Am.~Chem.~Soc.}, 2011, \textbf{133}, 14841\relax
\mciteBstWouldAddEndPuncttrue
\mciteSetBstMidEndSepPunct{\mcitedefaultmidpunct}
{\mcitedefaultendpunct}{\mcitedefaultseppunct}\relax
\EndOfBibitem
\bibitem[(Ed.)(2012)]{springborg12a}
M.~S. (Ed.), \emph{Specialist Periodical Reports: Chemical Modelling,
  Applications and Theory, Vol.~9}, {Royal Society of Chemistry}, {London},
  2012\relax
\mciteBstWouldAddEndPuncttrue
\mciteSetBstMidEndSepPunct{\mcitedefaultmidpunct}
{\mcitedefaultendpunct}{\mcitedefaultseppunct}\relax
\EndOfBibitem
\bibitem[Springborg and (Eds.)(2014)]{springborg14}
M.~Springborg and J.-O.~J. (Eds.), \emph{Specialist Periodical Reports:
  Chemical Modelling, Applications and Theory, Vol.~10}, {Royal Society of
  Chemistry}, {London}, 2014\relax
\mciteBstWouldAddEndPuncttrue
\mciteSetBstMidEndSepPunct{\mcitedefaultmidpunct}
{\mcitedefaultendpunct}{\mcitedefaultseppunct}\relax
\EndOfBibitem
\bibitem[Saebo and Pulay(1993)]{saebo93}
S.~Saebo and P.~Pulay, \emph{Annu.~Rev.~Phys.~Chem.}, 1993, \textbf{44},
  213\relax
\mciteBstWouldAddEndPuncttrue
\mciteSetBstMidEndSepPunct{\mcitedefaultmidpunct}
{\mcitedefaultendpunct}{\mcitedefaultseppunct}\relax
\EndOfBibitem
\bibitem[Bartlett(1989)]{bartlett89}
R.~J. Bartlett, \emph{J.~Phys.~Chem.}, 1989, \textbf{93}, 1697\relax
\mciteBstWouldAddEndPuncttrue
\mciteSetBstMidEndSepPunct{\mcitedefaultmidpunct}
{\mcitedefaultendpunct}{\mcitedefaultseppunct}\relax
\EndOfBibitem
\bibitem[Roos \emph{et~al.}(1996)Roos, Andersson, F{\"u}lscher, Malmqvist,
  Serrano-Andr{\'e}s, Pierloot, and Merchan]{roos96}
B.~O. Roos, K.~Andersson, M.~P. F{\"u}lscher, P.-{\AA}. Malmqvist,
  L.~Serrano-Andr{\'e}s, K.~Pierloot and M.~Merchan, \emph{Adv.~Chem.~Phys},
  1996, \textbf{93}, 219\relax
\mciteBstWouldAddEndPuncttrue
\mciteSetBstMidEndSepPunct{\mcitedefaultmidpunct}
{\mcitedefaultendpunct}{\mcitedefaultseppunct}\relax
\EndOfBibitem
\bibitem[Parr and Yang(1989)]{parr89}
R.~G. Parr and W.~Yang, \emph{Density Functional Theory of Atoms and
  Molecules}, {Oxford University Press}, {New York}, 1989\relax
\mciteBstWouldAddEndPuncttrue
\mciteSetBstMidEndSepPunct{\mcitedefaultmidpunct}
{\mcitedefaultendpunct}{\mcitedefaultseppunct}\relax
\EndOfBibitem
\bibitem[(Ed.)(1995)]{chong95}
D.~P.~C. (Ed.), \emph{Recent Advances in Density Functional Methods, Vol.~I},
  {World Scientific}, {Singapore}, 1995\relax
\mciteBstWouldAddEndPuncttrue
\mciteSetBstMidEndSepPunct{\mcitedefaultmidpunct}
{\mcitedefaultendpunct}{\mcitedefaultseppunct}\relax
\EndOfBibitem
\bibitem[(Ed.)(1996)]{seminario96}
J.~M.~S. (Ed.), \emph{Recent Developments and Applications of Modern DFT},
  {Elsevier}, {Amsterdam}, 1996\relax
\mciteBstWouldAddEndPuncttrue
\mciteSetBstMidEndSepPunct{\mcitedefaultmidpunct}
{\mcitedefaultendpunct}{\mcitedefaultseppunct}\relax
\EndOfBibitem
\bibitem[(Ed.)(1998)]{joubert98}
D.~J. (Ed.), \emph{Density Functionals: Theory and Applications}, {Springer},
  {Berlin}, 1998\relax
\mciteBstWouldAddEndPuncttrue
\mciteSetBstMidEndSepPunct{\mcitedefaultmidpunct}
{\mcitedefaultendpunct}{\mcitedefaultseppunct}\relax
\EndOfBibitem
\bibitem[Dobson \emph{et~al.}(1998)Dobson, Vignale, and (Eds.)]{dobson98}
J.~F. Dobson, G.~Vignale and M.~P.~D. (Eds.), \emph{Density Functional Theory:
  Recent Progress and New Directions}, {Plenum}, {New York}, 1998\relax
\mciteBstWouldAddEndPuncttrue
\mciteSetBstMidEndSepPunct{\mcitedefaultmidpunct}
{\mcitedefaultendpunct}{\mcitedefaultseppunct}\relax
\EndOfBibitem
\bibitem[Koch and Holthausen(2001)]{koch01}
W.~Koch and M.~C. Holthausen, \emph{A Chemist's Guide to Density Functional
  Theory}, {John Wiley}, {New York}, 2001\relax
\mciteBstWouldAddEndPuncttrue
\mciteSetBstMidEndSepPunct{\mcitedefaultmidpunct}
{\mcitedefaultendpunct}{\mcitedefaultseppunct}\relax
\EndOfBibitem
\bibitem[Fiolhais \emph{et~al.}(2003)Fiolhais, Nogueira, and
  Marques]{fiolhais03}
C.~Fiolhais, F.~Nogueira and M.~Marques, \emph{A Primer in Density Functional
  Theory}, {Springer}, {Berlin}, 2003\relax
\mciteBstWouldAddEndPuncttrue
\mciteSetBstMidEndSepPunct{\mcitedefaultmidpunct}
{\mcitedefaultendpunct}{\mcitedefaultseppunct}\relax
\EndOfBibitem
\bibitem[Gidopoulos and Wilson(2003)]{gidopoulos03}
N.~I. Gidopoulos and S.~Wilson, \emph{The Fundamentals of Electron Density,
  Density Matrix and Density Functional Theory in Atoms, Molecules and the
  Solid State}, {Springer}, {Berlin}, 2003\relax
\mciteBstWouldAddEndPuncttrue
\mciteSetBstMidEndSepPunct{\mcitedefaultmidpunct}
{\mcitedefaultendpunct}{\mcitedefaultseppunct}\relax
\EndOfBibitem
\bibitem[Hu and Yang(2008)]{hu08}
H.~Hu and W.~Yang, \emph{Annu.~Rev.~Phys.~Chem.}, 2008, \textbf{59}, 573\relax
\mciteBstWouldAddEndPuncttrue
\mciteSetBstMidEndSepPunct{\mcitedefaultmidpunct}
{\mcitedefaultendpunct}{\mcitedefaultseppunct}\relax
\EndOfBibitem
\bibitem[Cramer and Truhlar(2009)]{cramer09}
C.~J. Cramer and D.~J. Truhlar, \emph{Phys.~Chem.~Chem.~Phys.}, 2009,
  \textbf{11}, 10757\relax
\mciteBstWouldAddEndPuncttrue
\mciteSetBstMidEndSepPunct{\mcitedefaultmidpunct}
{\mcitedefaultendpunct}{\mcitedefaultseppunct}\relax
\EndOfBibitem
\bibitem[Cohen \emph{et~al.}(2011)Cohen, Mori-S{\'a}nchez, and Yang]{cohen11}
A.~J. Cohen, P.~Mori-S{\'a}nchez and W.~Yang, \emph{Chem.~Rev.}, 2011,
  \textbf{112}, 289\relax
\mciteBstWouldAddEndPuncttrue
\mciteSetBstMidEndSepPunct{\mcitedefaultmidpunct}
{\mcitedefaultendpunct}{\mcitedefaultseppunct}\relax
\EndOfBibitem
\bibitem[Burke(2012)]{burke12}
K.~Burke, \emph{J.~Chem.~Phys.}, 2012, \textbf{136}, 150901\relax
\mciteBstWouldAddEndPuncttrue
\mciteSetBstMidEndSepPunct{\mcitedefaultmidpunct}
{\mcitedefaultendpunct}{\mcitedefaultseppunct}\relax
\EndOfBibitem
\bibitem[(Ed.)(2012)]{roy12}
A.~K.~R. (Ed.), \emph{Theoretical and Computational Developments in Modern
  Density Functional Theory}, {Nova Science Publishers}, {New York}, 2012\relax
\mciteBstWouldAddEndPuncttrue
\mciteSetBstMidEndSepPunct{\mcitedefaultmidpunct}
{\mcitedefaultendpunct}{\mcitedefaultseppunct}\relax
\EndOfBibitem
\bibitem[Becke(2014)]{becke14}
A.~D. Becke, \emph{J.~Chem.~Phys.}, 2014, \textbf{140}, 18A301\relax
\mciteBstWouldAddEndPuncttrue
\mciteSetBstMidEndSepPunct{\mcitedefaultmidpunct}
{\mcitedefaultendpunct}{\mcitedefaultseppunct}\relax
\EndOfBibitem
\bibitem[Jones(2015)]{jones15}
R.~O. Jones, \emph{Rev.~Mod.~Phys.}, 2015, \textbf{87}, 897\relax
\mciteBstWouldAddEndPuncttrue
\mciteSetBstMidEndSepPunct{\mcitedefaultmidpunct}
{\mcitedefaultendpunct}{\mcitedefaultseppunct}\relax
\EndOfBibitem
\bibitem[Dirac(1930)]{dirac30}
P.~A.~M. Dirac, \emph{Mathematical Proceedings of the Cambridge Philosophical
  Society}, 1930, \textbf{26}, 376\relax
\mciteBstWouldAddEndPuncttrue
\mciteSetBstMidEndSepPunct{\mcitedefaultmidpunct}
{\mcitedefaultendpunct}{\mcitedefaultseppunct}\relax
\EndOfBibitem
\bibitem[Vosko \emph{et~al.}(1980)Vosko, Wilk, and Nusair]{vosko80}
S.~H. Vosko, L.~Wilk and M.~Nusair, \emph{Can.~J.~Phys.}, 1980, \textbf{58},
  1200\relax
\mciteBstWouldAddEndPuncttrue
\mciteSetBstMidEndSepPunct{\mcitedefaultmidpunct}
{\mcitedefaultendpunct}{\mcitedefaultseppunct}\relax
\EndOfBibitem
\bibitem[Perdew and Wang(1992)]{perdew92}
J.~P. Perdew and Y.~Wang, \emph{Phys.~Rev.~B}, 1992, \textbf{45}, 13244\relax
\mciteBstWouldAddEndPuncttrue
\mciteSetBstMidEndSepPunct{\mcitedefaultmidpunct}
{\mcitedefaultendpunct}{\mcitedefaultseppunct}\relax
\EndOfBibitem
\bibitem[Becke(1988)]{becke88a}
A.~D. Becke, \emph{Phys.~Rev.~A}, 1988, \textbf{38}, 3098\relax
\mciteBstWouldAddEndPuncttrue
\mciteSetBstMidEndSepPunct{\mcitedefaultmidpunct}
{\mcitedefaultendpunct}{\mcitedefaultseppunct}\relax
\EndOfBibitem
\bibitem[Lee \emph{et~al.}(1988)Lee, Yang, and Parr]{lee88}
C.~Lee, W.~Yang and R.~G. Parr, \emph{Phys.~Rev.~B}, 1988, \textbf{37},
  785\relax
\mciteBstWouldAddEndPuncttrue
\mciteSetBstMidEndSepPunct{\mcitedefaultmidpunct}
{\mcitedefaultendpunct}{\mcitedefaultseppunct}\relax
\EndOfBibitem
\bibitem[Perdew \emph{et~al.}(1996)Perdew, Burke, and Ernzerhof]{perdew96}
J.~P. Perdew, K.~Burke and M.~Ernzerhof, \emph{Phys.~Rev.~Lett.}, 1996,
  \textbf{77}, 3865\relax
\mciteBstWouldAddEndPuncttrue
\mciteSetBstMidEndSepPunct{\mcitedefaultmidpunct}
{\mcitedefaultendpunct}{\mcitedefaultseppunct}\relax
\EndOfBibitem
\bibitem[Becke(1988)]{becke88}
A.~D. Becke, \emph{J.~Chem.~Phys.}, 1988, \textbf{88}, 1053\relax
\mciteBstWouldAddEndPuncttrue
\mciteSetBstMidEndSepPunct{\mcitedefaultmidpunct}
{\mcitedefaultendpunct}{\mcitedefaultseppunct}\relax
\EndOfBibitem
\bibitem[Becke and Roussel(1989)]{becke89}
A.~D. Becke and M.~R. Roussel, \emph{Phys. Rev. A}, 1989, \textbf{39},
  3761\relax
\mciteBstWouldAddEndPuncttrue
\mciteSetBstMidEndSepPunct{\mcitedefaultmidpunct}
{\mcitedefaultendpunct}{\mcitedefaultseppunct}\relax
\EndOfBibitem
\bibitem[Tao \emph{et~al.}(2003)Tao, Perdew, Staroverov, and Scuseria]{tao03}
J.~Tao, J.~P. Perdew, V.~N. Staroverov and G.~E. Scuseria,
  \emph{Phys.~Rev.~Lett.}, 2003, \textbf{91}, 146401\relax
\mciteBstWouldAddEndPuncttrue
\mciteSetBstMidEndSepPunct{\mcitedefaultmidpunct}
{\mcitedefaultendpunct}{\mcitedefaultseppunct}\relax
\EndOfBibitem
\bibitem[Roy(2008)]{roy08}
A.~K. Roy, \emph{Int.~J.~Quant.~Chem.}, 2008, \textbf{108}, 837\relax
\mciteBstWouldAddEndPuncttrue
\mciteSetBstMidEndSepPunct{\mcitedefaultmidpunct}
{\mcitedefaultendpunct}{\mcitedefaultseppunct}\relax
\EndOfBibitem
\bibitem[Roy(2008)]{roy08a}
A.~K. Roy, \emph{Chem.~Phys.~Lett.}, 2008, \textbf{461}, 142\relax
\mciteBstWouldAddEndPuncttrue
\mciteSetBstMidEndSepPunct{\mcitedefaultmidpunct}
{\mcitedefaultendpunct}{\mcitedefaultseppunct}\relax
\EndOfBibitem
\bibitem[Roy(2009)]{roy09}
A.~K. Roy, Handbook of Computational Chemistry Research, New York, 2009\relax
\mciteBstWouldAddEndPuncttrue
\mciteSetBstMidEndSepPunct{\mcitedefaultmidpunct}
{\mcitedefaultendpunct}{\mcitedefaultseppunct}\relax
\EndOfBibitem
\bibitem[Roy(2010)]{roy10}
A.~K. Roy, \emph{Trends in Phys.~Chem.}, 2010, \textbf{14}, 27\relax
\mciteBstWouldAddEndPuncttrue
\mciteSetBstMidEndSepPunct{\mcitedefaultmidpunct}
{\mcitedefaultendpunct}{\mcitedefaultseppunct}\relax
\EndOfBibitem
\bibitem[Roy(2011)]{roy11}
A.~K. Roy, \emph{J.~Math.~Chem.}, 2011, \textbf{49}, 1687\relax
\mciteBstWouldAddEndPuncttrue
\mciteSetBstMidEndSepPunct{\mcitedefaultmidpunct}
{\mcitedefaultendpunct}{\mcitedefaultseppunct}\relax
\EndOfBibitem
\bibitem[Martyna and Tuckerman(1999)]{martyna99}
G.~J. Martyna and M.~E. Tuckerman, \emph{J.~Chem.~Phys.}, 1999, \textbf{110},
  2810\relax
\mciteBstWouldAddEndPuncttrue
\mciteSetBstMidEndSepPunct{\mcitedefaultmidpunct}
{\mcitedefaultendpunct}{\mcitedefaultseppunct}\relax
\EndOfBibitem
\bibitem[Minary \emph{et~al.}(2002)Minary, Tuckerman, Pihakari, and
  Martyna]{minary02}
P.~Minary, M.~E. Tuckerman, K.~A. Pihakari and G.~J. Martyna,
  \emph{J.~Chem.~Phys.}, 2002, \textbf{116}, 5351\relax
\mciteBstWouldAddEndPuncttrue
\mciteSetBstMidEndSepPunct{\mcitedefaultmidpunct}
{\mcitedefaultendpunct}{\mcitedefaultseppunct}\relax
\EndOfBibitem
\bibitem[Wadt and Hay(1985)]{wadt85}
W.~R. Wadt and P.~J. Hay, \emph{J.~Chem.~Phys.}, 1985, \textbf{82}, 284\relax
\mciteBstWouldAddEndPuncttrue
\mciteSetBstMidEndSepPunct{\mcitedefaultmidpunct}
{\mcitedefaultendpunct}{\mcitedefaultseppunct}\relax
\EndOfBibitem
\bibitem[Hay and Wadt(1985)]{hay85}
P.~J. Hay and W.~R. Wadt, \emph{J.~Chem.~Phys.}, 1985, \textbf{82}, 299\relax
\mciteBstWouldAddEndPuncttrue
\mciteSetBstMidEndSepPunct{\mcitedefaultmidpunct}
{\mcitedefaultendpunct}{\mcitedefaultseppunct}\relax
\EndOfBibitem
\bibitem[L{\"o}wdin(1955)]{lowdin95}
P.~O. L{\"o}wdin, \emph{Phys.~Rev.}, 1955, \textbf{97}, 1474\relax
\mciteBstWouldAddEndPuncttrue
\mciteSetBstMidEndSepPunct{\mcitedefaultmidpunct}
{\mcitedefaultendpunct}{\mcitedefaultseppunct}\relax
\EndOfBibitem
\bibitem[Thomas(1927)]{thomas27}
L.~H. Thomas, \emph{Mathematical Proceedings of the Cambridge Philosophical
  Society}, 1927, \textbf{23}, 542\relax
\mciteBstWouldAddEndPuncttrue
\mciteSetBstMidEndSepPunct{\mcitedefaultmidpunct}
{\mcitedefaultendpunct}{\mcitedefaultseppunct}\relax
\EndOfBibitem
\bibitem[Fermi(1927)]{fermi27}
E.~Fermi, \emph{Rend. Accad. Naz. Lincei}, 1927, \textbf{6}, 32\relax
\mciteBstWouldAddEndPuncttrue
\mciteSetBstMidEndSepPunct{\mcitedefaultmidpunct}
{\mcitedefaultendpunct}{\mcitedefaultseppunct}\relax
\EndOfBibitem
\bibitem[Fermi(1928)]{fermi28}
E.~Fermi, \emph{Z.~Phys.}, 1928, \textbf{48}, 73\relax
\mciteBstWouldAddEndPuncttrue
\mciteSetBstMidEndSepPunct{\mcitedefaultmidpunct}
{\mcitedefaultendpunct}{\mcitedefaultseppunct}\relax
\EndOfBibitem
\bibitem[Hohenberg and Kohn(1964)]{hohenberg64}
P.~Hohenberg and W.~Kohn, \emph{Phys.~Rev.}, 1964, \textbf{136}, B864\relax
\mciteBstWouldAddEndPuncttrue
\mciteSetBstMidEndSepPunct{\mcitedefaultmidpunct}
{\mcitedefaultendpunct}{\mcitedefaultseppunct}\relax
\EndOfBibitem
\bibitem[Roy and Chu(2002)]{roy02b}
A.~K. Roy and S.-I. Chu, \emph{J.~Phys.~B.}, 2002, \textbf{35}, 2075\relax
\mciteBstWouldAddEndPuncttrue
\mciteSetBstMidEndSepPunct{\mcitedefaultmidpunct}
{\mcitedefaultendpunct}{\mcitedefaultseppunct}\relax
\EndOfBibitem
\bibitem[Deb and Chattaraj(1989)]{deb89}
B.~M. Deb and P.~K. Chattaraj, \emph{Phys.~Rev.~A}, 1989, \textbf{39},
  1696\relax
\mciteBstWouldAddEndPuncttrue
\mciteSetBstMidEndSepPunct{\mcitedefaultmidpunct}
{\mcitedefaultendpunct}{\mcitedefaultseppunct}\relax
\EndOfBibitem
\bibitem[Dey and Deb(1998)]{dey98a}
B.~K. Dey and B.~M. Deb, \emph{Int.~J.~Quant.~Chem}, 1998, \textbf{67},
  251\relax
\mciteBstWouldAddEndPuncttrue
\mciteSetBstMidEndSepPunct{\mcitedefaultmidpunct}
{\mcitedefaultendpunct}{\mcitedefaultseppunct}\relax
\EndOfBibitem
\bibitem[Dey and Deb(1998)]{dey98}
B.~K. Dey and B.~M. Deb, \emph{Int.~J.~Quant.~Chem}, 1998, \textbf{70},
  441\relax
\mciteBstWouldAddEndPuncttrue
\mciteSetBstMidEndSepPunct{\mcitedefaultmidpunct}
{\mcitedefaultendpunct}{\mcitedefaultseppunct}\relax
\EndOfBibitem
\bibitem[Roy and Chu(2002)]{roy02a}
A.~K. Roy and S.-I. Chu, \emph{Phys.~Rev.~A}, 2002, \textbf{65}, 043402\relax
\mciteBstWouldAddEndPuncttrue
\mciteSetBstMidEndSepPunct{\mcitedefaultmidpunct}
{\mcitedefaultendpunct}{\mcitedefaultseppunct}\relax
\EndOfBibitem
\bibitem[Sadhukhan and Deb(2011)]{sadhukhan11}
M.~Sadhukhan and B.~M. Deb, \emph{Eur.~Phys.~Lett.}, 2011, \textbf{94},
  50008\relax
\mciteBstWouldAddEndPuncttrue
\mciteSetBstMidEndSepPunct{\mcitedefaultmidpunct}
{\mcitedefaultendpunct}{\mcitedefaultseppunct}\relax
\EndOfBibitem
\bibitem[Vikas(2011)]{vikas11}
Vikas, \emph{J.~Comput.~Chem.}, 2011, \textbf{32}, 2404\relax
\mciteBstWouldAddEndPuncttrue
\mciteSetBstMidEndSepPunct{\mcitedefaultmidpunct}
{\mcitedefaultendpunct}{\mcitedefaultseppunct}\relax
\EndOfBibitem
\bibitem[Vikas(2013)]{vikas13}
Vikas, \emph{Int.~J.~Quant.~Chem.}, 2013, \textbf{113}, 139\relax
\mciteBstWouldAddEndPuncttrue
\mciteSetBstMidEndSepPunct{\mcitedefaultmidpunct}
{\mcitedefaultendpunct}{\mcitedefaultseppunct}\relax
\EndOfBibitem
\bibitem[Sadhukhan \emph{et~al.}(2016)Sadhukhan, Roy, Panigrahi, and
  Deb]{sadhukhan16}
M.~Sadhukhan, A.~K. Roy, P.~K. Panigrahi and B.~M. Deb,
  \emph{Int.~J.~Quant.~Chem.}, 2016, \textbf{116}, 377\relax
\mciteBstWouldAddEndPuncttrue
\mciteSetBstMidEndSepPunct{\mcitedefaultmidpunct}
{\mcitedefaultendpunct}{\mcitedefaultseppunct}\relax
\EndOfBibitem
\bibitem[Wang and Carter(2000)]{wang00}
Y.~A. Wang and E.~A. Carter, \emph{Theoretical Methods in Condensed Phase
  Chemistry,~S.~D.~Schwartz (Ed.)}, Kluwer, Dordrecht, 2000, p. 117\relax
\mciteBstWouldAddEndPuncttrue
\mciteSetBstMidEndSepPunct{\mcitedefaultmidpunct}
{\mcitedefaultendpunct}{\mcitedefaultseppunct}\relax
\EndOfBibitem
\bibitem[Wang \emph{et~al.}(1998)Wang, Govind, and Carter]{wang98}
Y.~A. Wang, N.~Govind and E.~A. Carter, \emph{Phys.~Rev.~B}, 1998, \textbf{58},
  13465\relax
\mciteBstWouldAddEndPuncttrue
\mciteSetBstMidEndSepPunct{\mcitedefaultmidpunct}
{\mcitedefaultendpunct}{\mcitedefaultseppunct}\relax
\EndOfBibitem
\bibitem[Wang \emph{et~al.}(1999)Wang, Govind, and Carter]{wang99}
Y.~A. Wang, N.~Govind and E.~A. Carter, \emph{Phys.~Rev.~B}, 1999, \textbf{60},
  16350\relax
\mciteBstWouldAddEndPuncttrue
\mciteSetBstMidEndSepPunct{\mcitedefaultmidpunct}
{\mcitedefaultendpunct}{\mcitedefaultseppunct}\relax
\EndOfBibitem
\bibitem[Ho \emph{et~al.}(2008)Ho, Ligneres, and Carter]{ho08}
G.~S. Ho, V.~L. Ligneres and E.~A. Carter, \emph{Phys.~Rev.~B}, 2008,
  \textbf{78}, 045105\relax
\mciteBstWouldAddEndPuncttrue
\mciteSetBstMidEndSepPunct{\mcitedefaultmidpunct}
{\mcitedefaultendpunct}{\mcitedefaultseppunct}\relax
\EndOfBibitem
\bibitem[Gavini \emph{et~al.}(2007)Gavini, Bhattacharya, and Oritz]{gavini07}
V.~Gavini, K.~Bhattacharya and M.~Oritz, \emph{J.~Mech.~Phys.~Solids}, 2007,
  \textbf{55}, 697\relax
\mciteBstWouldAddEndPuncttrue
\mciteSetBstMidEndSepPunct{\mcitedefaultmidpunct}
{\mcitedefaultendpunct}{\mcitedefaultseppunct}\relax
\EndOfBibitem
\bibitem[Huang and Carter(2010)]{huang10}
C.~Huang and E.~A. Carter, \emph{Phys.~Rev.~B}, 2010, \textbf{81}, 045206\relax
\mciteBstWouldAddEndPuncttrue
\mciteSetBstMidEndSepPunct{\mcitedefaultmidpunct}
{\mcitedefaultendpunct}{\mcitedefaultseppunct}\relax
\EndOfBibitem
\bibitem[Kohn and Sham(1965)]{kohn65}
W.~Kohn and L.~J. Sham, \emph{Phys.~Rev.}, 1965, \textbf{140}, A1133\relax
\mciteBstWouldAddEndPuncttrue
\mciteSetBstMidEndSepPunct{\mcitedefaultmidpunct}
{\mcitedefaultendpunct}{\mcitedefaultseppunct}\relax
\EndOfBibitem
\bibitem[Velde and Baerends(1992)]{te92}
G.~T. Velde and E.~J. Baerends, \emph{J.~Comput.~Phys.}, 1992, \textbf{99},
  84\relax
\mciteBstWouldAddEndPuncttrue
\mciteSetBstMidEndSepPunct{\mcitedefaultmidpunct}
{\mcitedefaultendpunct}{\mcitedefaultseppunct}\relax
\EndOfBibitem
\bibitem[Gill \emph{et~al.}(1993)Gill, Johnson, and Pople]{gill93}
P.~M.~W. Gill, B.~G. Johnson and J.~A. Pople, \emph{Chem.~Phys.~Lett.}, 1993,
  \textbf{209}, 506\relax
\mciteBstWouldAddEndPuncttrue
\mciteSetBstMidEndSepPunct{\mcitedefaultmidpunct}
{\mcitedefaultendpunct}{\mcitedefaultseppunct}\relax
\EndOfBibitem
\bibitem[Treutler and Ahlrichs(1995)]{treutler95}
O.~Treutler and R.~Ahlrichs, \emph{J.~Chem.~Phys.}, 1995, \textbf{102},
  346\relax
\mciteBstWouldAddEndPuncttrue
\mciteSetBstMidEndSepPunct{\mcitedefaultmidpunct}
{\mcitedefaultendpunct}{\mcitedefaultseppunct}\relax
\EndOfBibitem
\bibitem[Delley(1990)]{delley90}
B.~Delley, \emph{J.~Chem.~Phys.}, 1990, \textbf{92}, 508\relax
\mciteBstWouldAddEndPuncttrue
\mciteSetBstMidEndSepPunct{\mcitedefaultmidpunct}
{\mcitedefaultendpunct}{\mcitedefaultseppunct}\relax
\EndOfBibitem
\bibitem[Lindh \emph{et~al.}(2001)Lindh, Malmqvist, and Gagliardi]{lindh01}
R.~Lindh, P.~Malmqvist and L.~Gagliardi, \emph{Theor.~Chem.~Acc.}, 2001,
  \textbf{106}, 178\relax
\mciteBstWouldAddEndPuncttrue
\mciteSetBstMidEndSepPunct{\mcitedefaultmidpunct}
{\mcitedefaultendpunct}{\mcitedefaultseppunct}\relax
\EndOfBibitem
\bibitem[Franchini \emph{et~al.}(2013)Franchini, Philipsen, and
  Visscher]{franchini13}
M.~Franchini, P.~H.~T. Philipsen and L.~Visscher, \emph{J.~Comput.~Chem.},
  2013, \textbf{34}, 1819\relax
\mciteBstWouldAddEndPuncttrue
\mciteSetBstMidEndSepPunct{\mcitedefaultmidpunct}
{\mcitedefaultendpunct}{\mcitedefaultseppunct}\relax
\EndOfBibitem
\bibitem[Lebedev and Skorokhodov(1992)]{lebedev92}
V.~I. Lebedev and A.~L. Skorokhodov, \emph{Russ.~Acad.~Sci.~Dokl.~Math.}, 1992,
  \textbf{45}, 587\relax
\mciteBstWouldAddEndPuncttrue
\mciteSetBstMidEndSepPunct{\mcitedefaultmidpunct}
{\mcitedefaultendpunct}{\mcitedefaultseppunct}\relax
\EndOfBibitem
\bibitem[Lebedev and Laikov(1999)]{lebedev99}
V.~I. Lebedev and D.~N. Laikov, \emph{Dokl.~Akad.~Nauk.~}, 1999, \textbf{366},
  741\relax
\mciteBstWouldAddEndPuncttrue
\mciteSetBstMidEndSepPunct{\mcitedefaultmidpunct}
{\mcitedefaultendpunct}{\mcitedefaultseppunct}\relax
\EndOfBibitem
\bibitem[McLaren(1963)]{mclaren63}
A.~D. McLaren, \emph{Math.~Comput.}, 1963, \textbf{17}, 361\relax
\mciteBstWouldAddEndPuncttrue
\mciteSetBstMidEndSepPunct{\mcitedefaultmidpunct}
{\mcitedefaultendpunct}{\mcitedefaultseppunct}\relax
\EndOfBibitem
\bibitem[Murray \emph{et~al.}(1993)Murray, Handy, and Laming]{murray93}
C.~W. Murray, N.~C. Handy and G.~J. Laming, \emph{Mol.~Phys.}, 1993,
  \textbf{78}, 997\relax
\mciteBstWouldAddEndPuncttrue
\mciteSetBstMidEndSepPunct{\mcitedefaultmidpunct}
{\mcitedefaultendpunct}{\mcitedefaultseppunct}\relax
\EndOfBibitem
\bibitem[El-Sherbiny and Poirier(2004)]{el04}
A.~El-Sherbiny and R.~A. Poirier, \emph{J.~Comput.~Chem.}, 2004, \textbf{25},
  1378\relax
\mciteBstWouldAddEndPuncttrue
\mciteSetBstMidEndSepPunct{\mcitedefaultmidpunct}
{\mcitedefaultendpunct}{\mcitedefaultseppunct}\relax
\EndOfBibitem
\bibitem[Chien and Gill(2006)]{chien06}
S.-H. Chien and P.~M.~W. Gill, \emph{J.~Comput.~Chem.}, 2006, \textbf{27},
  730\relax
\mciteBstWouldAddEndPuncttrue
\mciteSetBstMidEndSepPunct{\mcitedefaultmidpunct}
{\mcitedefaultendpunct}{\mcitedefaultseppunct}\relax
\EndOfBibitem
\bibitem[Mura and Knowles(1996)]{mura96}
M.~E. Mura and P.~J. Knowles, \emph{J.~Chem.~Phys.}, 1996, \textbf{104},
  9848\relax
\mciteBstWouldAddEndPuncttrue
\mciteSetBstMidEndSepPunct{\mcitedefaultmidpunct}
{\mcitedefaultendpunct}{\mcitedefaultseppunct}\relax
\EndOfBibitem
\bibitem[Gill and Chien(2003)]{gill03}
P.~M.~W. Gill and S.-H. Chien, \emph{J.~Comput.~Chem.}, 2003, \textbf{24},
  732\relax
\mciteBstWouldAddEndPuncttrue
\mciteSetBstMidEndSepPunct{\mcitedefaultmidpunct}
{\mcitedefaultendpunct}{\mcitedefaultseppunct}\relax
\EndOfBibitem
\bibitem[Kakhiani \emph{et~al.}(2009)Kakhiani, Tsereteli, and
  Tsereteli]{kakhiani09}
K.~Kakhiani, K.~Tsereteli and P.~Tsereteli, \emph{Comput.~Phys.~Commun.}, 2009,
  \textbf{180}, 256\relax
\mciteBstWouldAddEndPuncttrue
\mciteSetBstMidEndSepPunct{\mcitedefaultmidpunct}
{\mcitedefaultendpunct}{\mcitedefaultseppunct}\relax
\EndOfBibitem
\bibitem[Mori(1985)]{mori85}
M.~Mori, \emph{J.~Comput.~Appl.~Math.}, 1985, \textbf{12}, 119\relax
\mciteBstWouldAddEndPuncttrue
\mciteSetBstMidEndSepPunct{\mcitedefaultmidpunct}
{\mcitedefaultendpunct}{\mcitedefaultseppunct}\relax
\EndOfBibitem
\bibitem[Mori and Sugihara(2001)]{mori01}
M.~Mori and M.~Sugihara, \emph{J.~Comput.~Appl.~Math.}, 2001, \textbf{127},
  287\relax
\mciteBstWouldAddEndPuncttrue
\mciteSetBstMidEndSepPunct{\mcitedefaultmidpunct}
{\mcitedefaultendpunct}{\mcitedefaultseppunct}\relax
\EndOfBibitem
\bibitem[Muhammad and Mori(2003)]{muhammad03}
M.~Muhammad and M.~Mori, \emph{J.~Comput.~Appl.~Math.}, 2003, \textbf{161},
  431\relax
\mciteBstWouldAddEndPuncttrue
\mciteSetBstMidEndSepPunct{\mcitedefaultmidpunct}
{\mcitedefaultendpunct}{\mcitedefaultseppunct}\relax
\EndOfBibitem
\bibitem[Krack and Koster(1998)]{krack98}
M.~Krack and A.~M. Koster, \emph{J.~Chem.~Phys.}, 1998, \textbf{108},
  3226\relax
\mciteBstWouldAddEndPuncttrue
\mciteSetBstMidEndSepPunct{\mcitedefaultmidpunct}
{\mcitedefaultendpunct}{\mcitedefaultseppunct}\relax
\EndOfBibitem
\bibitem[P{\'e}rez-Jord{\'a} \emph{et~al.}(1994)P{\'e}rez-Jord{\'a}, Becke, and
  San-Fabi{\'a}n]{perez94}
J.~M. P{\'e}rez-Jord{\'a}, A.~D. Becke and E.~San-Fabi{\'a}n,
  \emph{J.~Chem.~Phys.}, 1994, \textbf{100}, 6520\relax
\mciteBstWouldAddEndPuncttrue
\mciteSetBstMidEndSepPunct{\mcitedefaultmidpunct}
{\mcitedefaultendpunct}{\mcitedefaultseppunct}\relax
\EndOfBibitem
\bibitem[Termath and Sauer(1996)]{termath96}
V.~Termath and J.~Sauer, \emph{Chem.~Phys.~Lett.}, 1996, \textbf{255},
  187\relax
\mciteBstWouldAddEndPuncttrue
\mciteSetBstMidEndSepPunct{\mcitedefaultmidpunct}
{\mcitedefaultendpunct}{\mcitedefaultseppunct}\relax
\EndOfBibitem
\bibitem[Challacombe(2000)]{challacombe00}
M.~Challacombe, \emph{J.~Chem.~Phys.}, 2000, \textbf{113}, 10037\relax
\mciteBstWouldAddEndPuncttrue
\mciteSetBstMidEndSepPunct{\mcitedefaultmidpunct}
{\mcitedefaultendpunct}{\mcitedefaultseppunct}\relax
\EndOfBibitem
\bibitem[Rodriguez \emph{et~al.}(2008)Rodriguez, Thompson, Ayers, and
  K{\"o}ster]{rodriguez08}
J.~I. Rodriguez, D.~C. Thompson, P.~W. Ayers and A.~M. K{\"o}ster,
  \emph{J.~Chem.~Phys.}, 2008, \textbf{128}, 224103\relax
\mciteBstWouldAddEndPuncttrue
\mciteSetBstMidEndSepPunct{\mcitedefaultmidpunct}
{\mcitedefaultendpunct}{\mcitedefaultseppunct}\relax
\EndOfBibitem
\bibitem[Fusti-Molnar and Pulay(2002)]{fustimolnar02}
L.~Fusti-Molnar and P.~Pulay, \emph{J.~Chem.~Phys.}, 2002, \textbf{117},
  7827\relax
\mciteBstWouldAddEndPuncttrue
\mciteSetBstMidEndSepPunct{\mcitedefaultmidpunct}
{\mcitedefaultendpunct}{\mcitedefaultseppunct}\relax
\EndOfBibitem
\bibitem[Fusti-Molnar(2003)]{fustimolnar03}
L.~Fusti-Molnar, \emph{J.~Chem.~Phys.}, 2003, \textbf{119}, 11080\relax
\mciteBstWouldAddEndPuncttrue
\mciteSetBstMidEndSepPunct{\mcitedefaultmidpunct}
{\mcitedefaultendpunct}{\mcitedefaultseppunct}\relax
\EndOfBibitem
\bibitem[Brown \emph{et~al.}(2006)Brown, F{\"u}sti-Moln{\'a}r, and
  Kong]{brown06}
S.~T. Brown, L.~F{\"u}sti-Moln{\'a}r and J.~Kong, \emph{Chem.~Phys.~Lett.},
  2006, \textbf{418}, 490\relax
\mciteBstWouldAddEndPuncttrue
\mciteSetBstMidEndSepPunct{\mcitedefaultmidpunct}
{\mcitedefaultendpunct}{\mcitedefaultseppunct}\relax
\EndOfBibitem
\bibitem[Saad(2003)]{saad03}
Y.~Saad, \emph{Iterative Methods for Sparse Linear Systems}, {Siam}, 2003\relax
\mciteBstWouldAddEndPuncttrue
\mciteSetBstMidEndSepPunct{\mcitedefaultmidpunct}
{\mcitedefaultendpunct}{\mcitedefaultseppunct}\relax
\EndOfBibitem
\bibitem[Brandt(1977)]{brandt77}
A.~Brandt, \emph{Math.~Comput.}, 1977, \textbf{31}, 333\relax
\mciteBstWouldAddEndPuncttrue
\mciteSetBstMidEndSepPunct{\mcitedefaultmidpunct}
{\mcitedefaultendpunct}{\mcitedefaultseppunct}\relax
\EndOfBibitem
\bibitem[Skylaris \emph{et~al.}(2002)Skylaris, Mostofi, Haynes, Di{\'e}guez,
  and Payne]{skylaris02}
C.-K. Skylaris, A.~A. Mostofi, P.~D. Haynes, O.~Di{\'e}guez and M.~C. Payne,
  \emph{Phys.~Rev.~B}, 2002, \textbf{66}, 035119\relax
\mciteBstWouldAddEndPuncttrue
\mciteSetBstMidEndSepPunct{\mcitedefaultmidpunct}
{\mcitedefaultendpunct}{\mcitedefaultseppunct}\relax
\EndOfBibitem
\bibitem[Hine \emph{et~al.}(2011)Hine, Dziedzic, Haynes, and Skylaris]{hine11}
N.~D.~M. Hine, J.~Dziedzic, P.~D. Haynes and C.-K. Skylaris,
  \emph{J.~Chem.~Phys.}, 2011, \textbf{135}, 204103\relax
\mciteBstWouldAddEndPuncttrue
\mciteSetBstMidEndSepPunct{\mcitedefaultmidpunct}
{\mcitedefaultendpunct}{\mcitedefaultseppunct}\relax
\EndOfBibitem
\bibitem[Chang \emph{et~al.}(2012)Chang, Shao, and Kong]{chang12}
C.-M. Chang, Y.~Shao and J.~Kong, \emph{J.~Chem.~Phys.}, 2012, \textbf{136},
  114112\relax
\mciteBstWouldAddEndPuncttrue
\mciteSetBstMidEndSepPunct{\mcitedefaultmidpunct}
{\mcitedefaultendpunct}{\mcitedefaultseppunct}\relax
\EndOfBibitem
\bibitem[Beatson and Greengard(1997)]{beatson97}
R.~Beatson and L.~Greengard, \emph{Wavelets, Multilevel Methods and Elliptic
  PDEs}, 1997, \textbf{1}, 1\relax
\mciteBstWouldAddEndPuncttrue
\mciteSetBstMidEndSepPunct{\mcitedefaultmidpunct}
{\mcitedefaultendpunct}{\mcitedefaultseppunct}\relax
\EndOfBibitem
\bibitem[Skeel \emph{et~al.}(2002)Skeel, Tezcan, and Hardy]{skeel02}
R.~D. Skeel, I.~Tezcan and D.~J. Hardy, \emph{J.~Comput.~Chem.}, 2002,
  \textbf{23}, 673\relax
\mciteBstWouldAddEndPuncttrue
\mciteSetBstMidEndSepPunct{\mcitedefaultmidpunct}
{\mcitedefaultendpunct}{\mcitedefaultseppunct}\relax
\EndOfBibitem
\bibitem[York and Yang(1994)]{york94}
D.~York and W.~Yang, \emph{J.~Chem.~Phys.}, 1994, \textbf{101}, 3298\relax
\mciteBstWouldAddEndPuncttrue
\mciteSetBstMidEndSepPunct{\mcitedefaultmidpunct}
{\mcitedefaultendpunct}{\mcitedefaultseppunct}\relax
\EndOfBibitem
\bibitem[Chelikowsky \emph{et~al.}(1994)Chelikowsky, Troullier, and
  Saad]{chelikowsky94a}
J.~R. Chelikowsky, N.~Troullier and Y.~Saad, \emph{Phys.~Rev.~Lett.}, 1994,
  \textbf{72}, 1240\relax
\mciteBstWouldAddEndPuncttrue
\mciteSetBstMidEndSepPunct{\mcitedefaultmidpunct}
{\mcitedefaultendpunct}{\mcitedefaultseppunct}\relax
\EndOfBibitem
\bibitem[Chelikowsky \emph{et~al.}(1994)Chelikowsky, Troullier, Wu, and
  Saad]{chelikowsky94b}
J.~R. Chelikowsky, N.~Troullier, K.~Wu and Y.~Saad, \emph{Phys.~Rev.~B}, 1994,
  \textbf{50}, 11355\relax
\mciteBstWouldAddEndPuncttrue
\mciteSetBstMidEndSepPunct{\mcitedefaultmidpunct}
{\mcitedefaultendpunct}{\mcitedefaultseppunct}\relax
\EndOfBibitem
\bibitem[Mundt and K{\"u}mmel(2007)]{mundt07}
M.~Mundt and S.~K{\"u}mmel, \emph{Phys.~Rev.~B}, 2007, \textbf{76},
  035413\relax
\mciteBstWouldAddEndPuncttrue
\mciteSetBstMidEndSepPunct{\mcitedefaultmidpunct}
{\mcitedefaultendpunct}{\mcitedefaultseppunct}\relax
\EndOfBibitem
\bibitem[Marques \emph{et~al.}(2012)Marques, .Maitra, Nogueira, Gross, and
  Rubio]{marques12}
M.~A.~L. Marques, N.~T. .Maitra, F.~M.~S. Nogueira, N.~K.~U. Gross and
  A.~Rubio, \emph{Fundamentals of Time-Dependent Density Functional Theory},
  {Springer Science \& Business Media}, 2012\relax
\mciteBstWouldAddEndPuncttrue
\mciteSetBstMidEndSepPunct{\mcitedefaultmidpunct}
{\mcitedefaultendpunct}{\mcitedefaultseppunct}\relax
\EndOfBibitem
\bibitem[Andrade and Aspuru-Guzik(2013)]{andrade13}
X.~Andrade and A.~Aspuru-Guzik, \emph{J.~Chem.~Theory~Comput.}, 2013,
  \textbf{9}, 4360\relax
\mciteBstWouldAddEndPuncttrue
\mciteSetBstMidEndSepPunct{\mcitedefaultmidpunct}
{\mcitedefaultendpunct}{\mcitedefaultseppunct}\relax
\EndOfBibitem
\bibitem[Chelikowsky \emph{et~al.}(2009)Chelikowsky, Zayak, Chan, Tiago, Zhou,
  and Saad]{chelikowsky09}
J.~R. Chelikowsky, A.~T. Zayak, T.~Chan, M.~L. Tiago, Y.~Zhou and Y.~Saad,
  \emph{J.~Phys.:~Condens.~Matter}, 2009, \textbf{21}, 064207\relax
\mciteBstWouldAddEndPuncttrue
\mciteSetBstMidEndSepPunct{\mcitedefaultmidpunct}
{\mcitedefaultendpunct}{\mcitedefaultseppunct}\relax
\EndOfBibitem
\bibitem[Beck(2000)]{beck00}
T.~L. Beck, \emph{Rev.~Mod.~Phys.}, 2000, \textbf{72}, 1041\relax
\mciteBstWouldAddEndPuncttrue
\mciteSetBstMidEndSepPunct{\mcitedefaultmidpunct}
{\mcitedefaultendpunct}{\mcitedefaultseppunct}\relax
\EndOfBibitem
\bibitem[Natan \emph{et~al.}(2008)Natan, Benjamini, Naveh, Kronik, Tiago,
  Beckman, and Chelikowsky]{natan08}
A.~Natan, A.~Benjamini, D.~Naveh, L.~Kronik, M.~L. Tiago, S.~P. Beckman and
  J.~R. Chelikowsky, \emph{Phys.~Rev.~B}, 2008, \textbf{78}, 075109\relax
\mciteBstWouldAddEndPuncttrue
\mciteSetBstMidEndSepPunct{\mcitedefaultmidpunct}
{\mcitedefaultendpunct}{\mcitedefaultseppunct}\relax
\EndOfBibitem
\bibitem[Bowler and Miyazaki(2012)]{bowler12}
D.~R. Bowler and T.~Miyazaki, \emph{Rep.~Prog.~Phys.}, 2012, \textbf{75},
  036503\relax
\mciteBstWouldAddEndPuncttrue
\mciteSetBstMidEndSepPunct{\mcitedefaultmidpunct}
{\mcitedefaultendpunct}{\mcitedefaultseppunct}\relax
\EndOfBibitem
\bibitem[Bernholc \emph{et~al.}(2008)Bernholc, Hodak, and Lu]{bernholc08}
J.~Bernholc, M.~Hodak and W.~Lu, \emph{J.~Phys.:~Condens.~Matter}, 2008,
  \textbf{20}, 294205\relax
\mciteBstWouldAddEndPuncttrue
\mciteSetBstMidEndSepPunct{\mcitedefaultmidpunct}
{\mcitedefaultendpunct}{\mcitedefaultseppunct}\relax
\EndOfBibitem
\bibitem[Mohr \emph{et~al.}(2014)Mohr, Ratcliff, Boulanger, Genovese, Caliste,
  Deutsch, and Goedecker]{mohr14}
S.~Mohr, L.~E. Ratcliff, P.~Boulanger, L.~Genovese, D.~Caliste, T.~Deutsch and
  S.~Goedecker, \emph{J.~Chem.~Phys.}, 2014, \textbf{140}, 204110\relax
\mciteBstWouldAddEndPuncttrue
\mciteSetBstMidEndSepPunct{\mcitedefaultmidpunct}
{\mcitedefaultendpunct}{\mcitedefaultseppunct}\relax
\EndOfBibitem
\bibitem[Genovese \emph{et~al.}(2008)Genovese, Neelov, Goedecker, Deutsch,
  Ghasemi, Willand, Caliste, Zilberberg, Rayson,
  Bergman,\emph{et~al.}]{genovese08}
L.~Genovese, A.~Neelov, S.~Goedecker, T.~Deutsch, S.~A. Ghasemi, A.~Willand,
  D.~Caliste, O.~Zilberberg, M.~Rayson, A.~Bergman \emph{et~al.},
  \emph{J.~Chem.~Phys.}, 2008, \textbf{129}, 014109\relax
\mciteBstWouldAddEndPuncttrue
\mciteSetBstMidEndSepPunct{\mcitedefaultmidpunct}
{\mcitedefaultendpunct}{\mcitedefaultseppunct}\relax
\EndOfBibitem
\bibitem[Briggs \emph{et~al.}(1995)Briggs, Sullivan, and Bernholc]{briggs96}
E.~L. Briggs, D.~J. Sullivan and J.~Bernholc, \emph{Phys.~Rev.~B}, 1995,
  \textbf{52}, R5471\relax
\mciteBstWouldAddEndPuncttrue
\mciteSetBstMidEndSepPunct{\mcitedefaultmidpunct}
{\mcitedefaultendpunct}{\mcitedefaultseppunct}\relax
\EndOfBibitem
\bibitem[Fattebert and Bernholc(2000)]{fattebert00}
J.-L. Fattebert and J.~Bernholc, \emph{Phys.~Rev.~B}, 2000, \textbf{62},
  1713\relax
\mciteBstWouldAddEndPuncttrue
\mciteSetBstMidEndSepPunct{\mcitedefaultmidpunct}
{\mcitedefaultendpunct}{\mcitedefaultseppunct}\relax
\EndOfBibitem
\bibitem[Alemany \emph{et~al.}(2007)Alemany, Jain, Tiago, Zhou, Saad, and
  Chelikowsky]{alemany07}
M.~M.~G. Alemany, M.~Jain, M.~L. Tiago, Y.~Zhou, Y.~Saad and J.~R. Chelikowsky,
  \emph{Comput.~Phys.~Commun.}, 2007, \textbf{177}, 339\relax
\mciteBstWouldAddEndPuncttrue
\mciteSetBstMidEndSepPunct{\mcitedefaultmidpunct}
{\mcitedefaultendpunct}{\mcitedefaultseppunct}\relax
\EndOfBibitem
\bibitem[Iwata \emph{et~al.}(2010)Iwata, Takahashi, Oshiyama, Boku, Shiraishi,
  Okada, and Yabana]{Iwata10}
J.-I. Iwata, D.~Takahashi, A.~Oshiyama, T.~Boku, K.~Shiraishi, S.~Okada and
  K.~Yabana, \emph{J.~Comput.~Phys.}, 2010, \textbf{229}, 2339\relax
\mciteBstWouldAddEndPuncttrue
\mciteSetBstMidEndSepPunct{\mcitedefaultmidpunct}
{\mcitedefaultendpunct}{\mcitedefaultseppunct}\relax
\EndOfBibitem
\bibitem[Fujimoto and Oshiyama(2010)]{fujimoto10}
Y.~Fujimoto and A.~Oshiyama, \emph{Phys.~Rev.~B}, 2010, \textbf{81},
  205309\relax
\mciteBstWouldAddEndPuncttrue
\mciteSetBstMidEndSepPunct{\mcitedefaultmidpunct}
{\mcitedefaultendpunct}{\mcitedefaultseppunct}\relax
\EndOfBibitem
\bibitem[Heiskanen \emph{et~al.}(2001)Heiskanen, Torsti, Puska, and
  Nieminen]{heiskanen01}
M.~Heiskanen, T.~Torsti, M.~J. Puska and R.~M. Nieminen, \emph{Phys.~Rev.~B},
  2001, \textbf{63}, 245106\relax
\mciteBstWouldAddEndPuncttrue
\mciteSetBstMidEndSepPunct{\mcitedefaultmidpunct}
{\mcitedefaultendpunct}{\mcitedefaultseppunct}\relax
\EndOfBibitem
\bibitem[Mortensen \emph{et~al.}(2005)Mortensen, Hansen, and
  Jacobsen]{mortensen05}
J.~J. Mortensen, L.~B. Hansen and K.~W. Jacobsen, \emph{Phys.~Rev.~B}, 2005,
  \textbf{71}, 035109\relax
\mciteBstWouldAddEndPuncttrue
\mciteSetBstMidEndSepPunct{\mcitedefaultmidpunct}
{\mcitedefaultendpunct}{\mcitedefaultseppunct}\relax
\EndOfBibitem
\bibitem[Modine \emph{et~al.}(1997)Modine, Zumbach, and Kaxiras]{modine95}
N.~A. Modine, G.~Zumbach and E.~Kaxiras, \emph{Phys.~Rev.~B}, 1997,
  \textbf{55}, 10289\relax
\mciteBstWouldAddEndPuncttrue
\mciteSetBstMidEndSepPunct{\mcitedefaultmidpunct}
{\mcitedefaultendpunct}{\mcitedefaultseppunct}\relax
\EndOfBibitem
\bibitem[Lee \emph{et~al.}(2000)Lee, Kim, and Martin]{lee00}
I.-H. Lee, Y.-H. Kim and R.~M. Martin, \emph{Phys.~Rev.~B}, 2000, \textbf{61},
  4397\relax
\mciteBstWouldAddEndPuncttrue
\mciteSetBstMidEndSepPunct{\mcitedefaultmidpunct}
{\mcitedefaultendpunct}{\mcitedefaultseppunct}\relax
\EndOfBibitem
\bibitem[Pask and Sterne(2005)]{pask05}
J.~E. Pask and P.~A. Sterne, \emph{Model.~Simul.~Mater.~Sci.~Eng.}, 2005,
  \textbf{13}, R71\relax
\mciteBstWouldAddEndPuncttrue
\mciteSetBstMidEndSepPunct{\mcitedefaultmidpunct}
{\mcitedefaultendpunct}{\mcitedefaultseppunct}\relax
\EndOfBibitem
\bibitem[Tsuchida and Tsukada(1995)]{tsuchida95}
E.~Tsuchida and M.~Tsukada, \emph{Phys.~Rev.~B}, 1995, \textbf{52}, 5573\relax
\mciteBstWouldAddEndPuncttrue
\mciteSetBstMidEndSepPunct{\mcitedefaultmidpunct}
{\mcitedefaultendpunct}{\mcitedefaultseppunct}\relax
\EndOfBibitem
\bibitem[White \emph{et~al.}(1989)White, Wilkins, and Teter]{white89}
S.~R. White, J.~W. Wilkins and M.~P. Teter, \emph{Phys.~Rev.~B}, 1989,
  \textbf{39}, 11355\relax
\mciteBstWouldAddEndPuncttrue
\mciteSetBstMidEndSepPunct{\mcitedefaultmidpunct}
{\mcitedefaultendpunct}{\mcitedefaultseppunct}\relax
\EndOfBibitem
\bibitem[Ackermann \emph{et~al.}(1994)Ackermann, Erdmann, and
  Roitzsch]{ackermann94}
J.~Ackermann, B.~Erdmann and R.~Roitzsch, \emph{J.~Chem.~Phys.}, 1994,
  \textbf{101}, 7643\relax
\mciteBstWouldAddEndPuncttrue
\mciteSetBstMidEndSepPunct{\mcitedefaultmidpunct}
{\mcitedefaultendpunct}{\mcitedefaultseppunct}\relax
\EndOfBibitem
\bibitem[Pask \emph{et~al.}(1999)Pask, Klein, Fong, and Sterne]{pask99}
J.~E. Pask, B.~M. Klein, C.~Y. Fong and P.~A. Sterne, \emph{Phys.~Rev.~B},
  1999, \textbf{59}, 12352\relax
\mciteBstWouldAddEndPuncttrue
\mciteSetBstMidEndSepPunct{\mcitedefaultmidpunct}
{\mcitedefaultendpunct}{\mcitedefaultseppunct}\relax
\EndOfBibitem
\bibitem[Yu \emph{et~al.}(1994)Yu, Bandrauk, and Sonnad]{yu94}
H.~Yu, A.~D. Bandrauk and V.~Sonnad, \emph{Chem.~Phys.~lett.}, 1994,
  \textbf{222}, 387\relax
\mciteBstWouldAddEndPuncttrue
\mciteSetBstMidEndSepPunct{\mcitedefaultmidpunct}
{\mcitedefaultendpunct}{\mcitedefaultseppunct}\relax
\EndOfBibitem
\bibitem[Hern{\'a}ndez \emph{et~al.}(1997)Hern{\'a}ndez, Gillan, and
  Goringe]{goringe97}
E.~Hern{\'a}ndez, M.~J. Gillan and C.~M. Goringe, \emph{Phys.~Rev.~B}, 1997,
  \textbf{55}, 13485\relax
\mciteBstWouldAddEndPuncttrue
\mciteSetBstMidEndSepPunct{\mcitedefaultmidpunct}
{\mcitedefaultendpunct}{\mcitedefaultseppunct}\relax
\EndOfBibitem
\bibitem[Briggs \emph{et~al.}(1996)Briggs, Sullivan, and Bernholc]{briggs95}
E.~L. Briggs, D.~J. Sullivan and J.~Bernholc, \emph{Phys.~Rev.~B}, 1996,
  \textbf{54}, 14362\relax
\mciteBstWouldAddEndPuncttrue
\mciteSetBstMidEndSepPunct{\mcitedefaultmidpunct}
{\mcitedefaultendpunct}{\mcitedefaultseppunct}\relax
\EndOfBibitem
\bibitem[Gygi and Galli(1995)]{gygi95}
F.~Gygi and G.~Galli, \emph{Phys.~Rev.~B}, 1995, \textbf{52}, R2229\relax
\mciteBstWouldAddEndPuncttrue
\mciteSetBstMidEndSepPunct{\mcitedefaultmidpunct}
{\mcitedefaultendpunct}{\mcitedefaultseppunct}\relax
\EndOfBibitem
\bibitem[Braess and Verf{\"u}rth(1990)]{braess90}
D.~Braess and R.~Verf{\"u}rth, \emph{J.~Numer.~Anal.}, 1990, \textbf{27},
  979\relax
\mciteBstWouldAddEndPuncttrue
\mciteSetBstMidEndSepPunct{\mcitedefaultmidpunct}
{\mcitedefaultendpunct}{\mcitedefaultseppunct}\relax
\EndOfBibitem
\bibitem[Brenner and Scott(2007)]{brenner94}
S.~Brenner and R.~Scott, \emph{The Mathematical Theory of Finite Element
  Methods}, {Springer Science \& Business Media}, 2007\relax
\mciteBstWouldAddEndPuncttrue
\mciteSetBstMidEndSepPunct{\mcitedefaultmidpunct}
{\mcitedefaultendpunct}{\mcitedefaultseppunct}\relax
\EndOfBibitem
\bibitem[Arias(1999)]{arias99}
T.~A. Arias, \emph{Rev.~Mod.~Phys.}, 1999, \textbf{71}, 267\relax
\mciteBstWouldAddEndPuncttrue
\mciteSetBstMidEndSepPunct{\mcitedefaultmidpunct}
{\mcitedefaultendpunct}{\mcitedefaultseppunct}\relax
\EndOfBibitem
\bibitem[Slater(1930)]{slater30}
J.~C. Slater, \emph{Phys.~Rev.}, 1930, \textbf{36}, 57\relax
\mciteBstWouldAddEndPuncttrue
\mciteSetBstMidEndSepPunct{\mcitedefaultmidpunct}
{\mcitedefaultendpunct}{\mcitedefaultseppunct}\relax
\EndOfBibitem
\bibitem[Boys(1950)]{boys50}
S.~F. Boys, \emph{Proceedings of the Royal Society of London A: Mathematical,
  Physical and Engineering Sciences}, 1950, \textbf{200}, 542\relax
\mciteBstWouldAddEndPuncttrue
\mciteSetBstMidEndSepPunct{\mcitedefaultmidpunct}
{\mcitedefaultendpunct}{\mcitedefaultseppunct}\relax
\EndOfBibitem
\bibitem[Cho \emph{et~al.}(1993)Cho, Arias, Joannopoulos, and Lam]{cho93}
K.~Cho, T.~A. Arias, J.~D. Joannopoulos and P.~K. Lam, \emph{Phys.~Rev.~Lett.},
  1993, \textbf{71}, 1808\relax
\mciteBstWouldAddEndPuncttrue
\mciteSetBstMidEndSepPunct{\mcitedefaultmidpunct}
{\mcitedefaultendpunct}{\mcitedefaultseppunct}\relax
\EndOfBibitem
\bibitem[Ihm \emph{et~al.}(1979)Ihm, Zunger, and Cohen]{ihm79}
J.~Ihm, A.~Zunger and M.~L. Cohen, \emph{J.~Phys.~C}, 1979, \textbf{12},
  4409\relax
\mciteBstWouldAddEndPuncttrue
\mciteSetBstMidEndSepPunct{\mcitedefaultmidpunct}
{\mcitedefaultendpunct}{\mcitedefaultseppunct}\relax
\EndOfBibitem
\bibitem[Obara and Saika(1986)]{obara86}
S.~Obara and A.~Saika, \emph{J.~Chem.~Phys.}, 1986, \textbf{84}, 3963\relax
\mciteBstWouldAddEndPuncttrue
\mciteSetBstMidEndSepPunct{\mcitedefaultmidpunct}
{\mcitedefaultendpunct}{\mcitedefaultseppunct}\relax
\EndOfBibitem
\bibitem[Sambe and Felton(1975)]{sambe75}
H.~Sambe and R.~H. Felton, \emph{J.~Chem.~Phys.}, 1975, \textbf{62}, 1122\relax
\mciteBstWouldAddEndPuncttrue
\mciteSetBstMidEndSepPunct{\mcitedefaultmidpunct}
{\mcitedefaultendpunct}{\mcitedefaultseppunct}\relax
\EndOfBibitem
\bibitem[Dunlap \emph{et~al.}(1979)Dunlap, Connolly, and Sabin]{dunlap79a}
B.~I. Dunlap, J.~W.~D. Connolly and J.~R. Sabin, \emph{J.~Chem.~Phys.}, 1979,
  \textbf{71}, 3396\relax
\mciteBstWouldAddEndPuncttrue
\mciteSetBstMidEndSepPunct{\mcitedefaultmidpunct}
{\mcitedefaultendpunct}{\mcitedefaultseppunct}\relax
\EndOfBibitem
\bibitem[Dunlap \emph{et~al.}(1979)Dunlap, Connolly, and Sabin]{dunlap79b}
B.~I. Dunlap, J.~W.~D. Connolly and J.~R. Sabin, \emph{J.~Chem.~Phys.}, 1979,
  \textbf{71}, 4993\relax
\mciteBstWouldAddEndPuncttrue
\mciteSetBstMidEndSepPunct{\mcitedefaultmidpunct}
{\mcitedefaultendpunct}{\mcitedefaultseppunct}\relax
\EndOfBibitem
\bibitem[Pople \emph{et~al.}(1992)Pople, Gill, and Johnson]{pople92}
J.~A. Pople, P.~M.~W. Gill and B.~G. Johnson, \emph{Chem.~Phys.~Lett.}, 1992,
  \textbf{199}, 557\relax
\mciteBstWouldAddEndPuncttrue
\mciteSetBstMidEndSepPunct{\mcitedefaultmidpunct}
{\mcitedefaultendpunct}{\mcitedefaultseppunct}\relax
\EndOfBibitem
\bibitem[Huang and Yang(1957)]{huang57}
K.~Huang and C.~N. Yang, \emph{Phys.~Rev.}, 1957, \textbf{105}, 767\relax
\mciteBstWouldAddEndPuncttrue
\mciteSetBstMidEndSepPunct{\mcitedefaultmidpunct}
{\mcitedefaultendpunct}{\mcitedefaultseppunct}\relax
\EndOfBibitem
\bibitem[Cohen \emph{et~al.}(1975)Cohen, Schl{\"u}ter, Chelikowsky, and
  Louie]{cohen75}
M.~L. Cohen, M.~Schl{\"u}ter, J.~R. Chelikowsky and S.~G. Louie,
  \emph{Phys.~Rev.~B}, 1975, \textbf{12}, 5575\relax
\mciteBstWouldAddEndPuncttrue
\mciteSetBstMidEndSepPunct{\mcitedefaultmidpunct}
{\mcitedefaultendpunct}{\mcitedefaultseppunct}\relax
\EndOfBibitem
\bibitem[Leggett(2001)]{leggett01}
A.~J. Leggett, \emph{Rev.~Mod.~Phys.}, 2001, \textbf{73}, 307\relax
\mciteBstWouldAddEndPuncttrue
\mciteSetBstMidEndSepPunct{\mcitedefaultmidpunct}
{\mcitedefaultendpunct}{\mcitedefaultseppunct}\relax
\EndOfBibitem
\bibitem[Cohen(1984)]{cohen84}
M.~L. Cohen, \emph{Phys.~Rep.}, 1984, \textbf{110}, 293\relax
\mciteBstWouldAddEndPuncttrue
\mciteSetBstMidEndSepPunct{\mcitedefaultmidpunct}
{\mcitedefaultendpunct}{\mcitedefaultseppunct}\relax
\EndOfBibitem
\bibitem[Schwerdtfeger \emph{et~al.}(2011)Schwerdtfeger, Assadollahzadeh,
  Rohrmann, Sch{\"a}fer, and Cheeseman]{schwerdtfeger11}
P.~Schwerdtfeger, B.~Assadollahzadeh, U.~Rohrmann, R.~Sch{\"a}fer and J.~R.
  Cheeseman, \emph{J.~Chem.~Phys.}, 2011, \textbf{134}, 204102\relax
\mciteBstWouldAddEndPuncttrue
\mciteSetBstMidEndSepPunct{\mcitedefaultmidpunct}
{\mcitedefaultendpunct}{\mcitedefaultseppunct}\relax
\EndOfBibitem
\bibitem[Maron and Teichteil(1998)]{maron98}
L.~Maron and C.~Teichteil, \emph{Chem.~Phys.}, 1998, \textbf{237}, 105\relax
\mciteBstWouldAddEndPuncttrue
\mciteSetBstMidEndSepPunct{\mcitedefaultmidpunct}
{\mcitedefaultendpunct}{\mcitedefaultseppunct}\relax
\EndOfBibitem
\bibitem[Schwerdtfeger \emph{et~al.}(1995)Schwerdtfeger, Fischer, Dolg,
  Igel-Mann, Nicklass, Stoll, and Haaland]{schwerdtfeger95}
P.~Schwerdtfeger, T.~Fischer, M.~Dolg, G.~Igel-Mann, A.~Nicklass, H.~Stoll and
  A.~Haaland, \emph{J.~Chem.~Phys.}, 1995, \textbf{102}, 2050\relax
\mciteBstWouldAddEndPuncttrue
\mciteSetBstMidEndSepPunct{\mcitedefaultmidpunct}
{\mcitedefaultendpunct}{\mcitedefaultseppunct}\relax
\EndOfBibitem
\bibitem[Schwerdtfeger \emph{et~al.}(2000)Schwerdtfeger, Brown, Laerdahl, and
  Stoll]{schwerdtfeger00}
P.~Schwerdtfeger, J.~R. Brown, J.~K. Laerdahl and H.~Stoll,
  \emph{J.~Chem.~Phys.}, 2000, \textbf{113}, 7110\relax
\mciteBstWouldAddEndPuncttrue
\mciteSetBstMidEndSepPunct{\mcitedefaultmidpunct}
{\mcitedefaultendpunct}{\mcitedefaultseppunct}\relax
\EndOfBibitem
\bibitem[Batyrev \emph{et~al.}(2001)Batyrev, Cho, and Kleinman]{batyrev01}
I.~G. Batyrev, J.-H. Cho and L.~Kleinman, \emph{Phys.~Rev.~B}, 2001,
  \textbf{63}, 172420\relax
\mciteBstWouldAddEndPuncttrue
\mciteSetBstMidEndSepPunct{\mcitedefaultmidpunct}
{\mcitedefaultendpunct}{\mcitedefaultseppunct}\relax
\EndOfBibitem
\bibitem[Huzinaga and Cantu(1971)]{huzinaga71}
S.~Huzinaga and A.~A. Cantu, \emph{J.~Chem.~Phys.}, 1971, \textbf{55},
  5543\relax
\mciteBstWouldAddEndPuncttrue
\mciteSetBstMidEndSepPunct{\mcitedefaultmidpunct}
{\mcitedefaultendpunct}{\mcitedefaultseppunct}\relax
\EndOfBibitem
\bibitem[Zeng and Klobukowski(2009)]{zeng09}
T.~Zeng and M.~Klobukowski, \emph{J.~Chem.~Phys.}, 2009, \textbf{130},
  204107\relax
\mciteBstWouldAddEndPuncttrue
\mciteSetBstMidEndSepPunct{\mcitedefaultmidpunct}
{\mcitedefaultendpunct}{\mcitedefaultseppunct}\relax
\EndOfBibitem
\bibitem[Fedorov and Klobukowski(2002)]{fedorov02}
D.~G. Fedorov and M.~Klobukowski, \emph{Chem.~Phys.~Lett.}, 2002, \textbf{360},
  223\relax
\mciteBstWouldAddEndPuncttrue
\mciteSetBstMidEndSepPunct{\mcitedefaultmidpunct}
{\mcitedefaultendpunct}{\mcitedefaultseppunct}\relax
\EndOfBibitem
\bibitem[Andzelm \emph{et~al.}(1985)Andzelm, Radzio, Barandiar{\'a}n, and
  Seijo]{andzelm85}
J.~Andzelm, E.~Radzio, Z.~Barandiar{\'a}n and L.~Seijo, \emph{J.~Chem.~Phys.},
  1985, \textbf{83}, 4565\relax
\mciteBstWouldAddEndPuncttrue
\mciteSetBstMidEndSepPunct{\mcitedefaultmidpunct}
{\mcitedefaultendpunct}{\mcitedefaultseppunct}\relax
\EndOfBibitem
\bibitem[Schwarz(1971)]{schwarz71}
W.~H.~E. Schwarz, \emph{Theor.~Chim.~Acta}, 1971, \textbf{23}, 147\relax
\mciteBstWouldAddEndPuncttrue
\mciteSetBstMidEndSepPunct{\mcitedefaultmidpunct}
{\mcitedefaultendpunct}{\mcitedefaultseppunct}\relax
\EndOfBibitem
\bibitem[Kahn and III(1972)]{kahn72}
L.~R. Kahn and W.~A.~G. III, \emph{J.~Chem.~Phys.}, 1972, \textbf{56},
  2685\relax
\mciteBstWouldAddEndPuncttrue
\mciteSetBstMidEndSepPunct{\mcitedefaultmidpunct}
{\mcitedefaultendpunct}{\mcitedefaultseppunct}\relax
\EndOfBibitem
\bibitem[Redondo \emph{et~al.}(1977)Redondo, III, and McGill]{redondo77}
A.~Redondo, W.~A.~G. III and T.~C. McGill, \emph{Phys.~Rev.~B}, 1977,
  \textbf{15}, 5038\relax
\mciteBstWouldAddEndPuncttrue
\mciteSetBstMidEndSepPunct{\mcitedefaultmidpunct}
{\mcitedefaultendpunct}{\mcitedefaultseppunct}\relax
\EndOfBibitem
\bibitem[Preuss \emph{et~al.}(1981)Preuss, Stoll, Wedig, and
  Kr{\"u}ger]{preuss81}
H.~Preuss, H.~Stoll, U.~Wedig and T.~Kr{\"u}ger, \emph{Int.~J.~Quant.~Chem.},
  1981, \textbf{19}, 113\relax
\mciteBstWouldAddEndPuncttrue
\mciteSetBstMidEndSepPunct{\mcitedefaultmidpunct}
{\mcitedefaultendpunct}{\mcitedefaultseppunct}\relax
\EndOfBibitem
\bibitem[Schwerdtfeger \emph{et~al.}(1982)Schwerdtfeger, Stoll, and
  Preuss]{schwerdtfeger82}
P.~Schwerdtfeger, H.~Stoll and H.~Preuss, \emph{J.~Phys.~B}, 1982, \textbf{15},
  1061\relax
\mciteBstWouldAddEndPuncttrue
\mciteSetBstMidEndSepPunct{\mcitedefaultmidpunct}
{\mcitedefaultendpunct}{\mcitedefaultseppunct}\relax
\EndOfBibitem
\bibitem[Kahn(1984)]{kahn84}
L.~R. Kahn, \emph{Int.~J.~Quant.~Chem.}, 1984, \textbf{25}, 149\relax
\mciteBstWouldAddEndPuncttrue
\mciteSetBstMidEndSepPunct{\mcitedefaultmidpunct}
{\mcitedefaultendpunct}{\mcitedefaultseppunct}\relax
\EndOfBibitem
\bibitem[McMurchie and Davidson(1981)]{mcmurchie81}
L.~E. McMurchie and E.~R. Davidson, \emph{J.~Comput.~Phys.}, 1981, \textbf{44},
  289\relax
\mciteBstWouldAddEndPuncttrue
\mciteSetBstMidEndSepPunct{\mcitedefaultmidpunct}
{\mcitedefaultendpunct}{\mcitedefaultseppunct}\relax
\EndOfBibitem
\bibitem[Fuentealba \emph{et~al.}(1983)Fuentealba, Stoll, Szentpaly,
  Schwerdtfeger, and Preuss]{fuentealba83}
P.~Fuentealba, H.~Stoll, L.~V. Szentpaly, P.~Schwerdtfeger and H.~Preuss,
  \emph{J.~Phys.~B}, 1983, \textbf{16}, L323\relax
\mciteBstWouldAddEndPuncttrue
\mciteSetBstMidEndSepPunct{\mcitedefaultmidpunct}
{\mcitedefaultendpunct}{\mcitedefaultseppunct}\relax
\EndOfBibitem
\bibitem[Lee \emph{et~al.}(2009)Lee, Jeung, and Lee]{lee09}
D.-K. Lee, G.-H. Jeung and Y.~S. Lee, \emph{Int.~J.~Quant.~Chem.}, 2009,
  \textbf{109}, 1975\relax
\mciteBstWouldAddEndPuncttrue
\mciteSetBstMidEndSepPunct{\mcitedefaultmidpunct}
{\mcitedefaultendpunct}{\mcitedefaultseppunct}\relax
\EndOfBibitem
\bibitem[Thierfelder and Schwerdtfeger(2010)]{thierfelder10}
C.~Thierfelder and P.~Schwerdtfeger, \emph{Phys.~Rev.~A}, 2010, \textbf{82},
  062503\relax
\mciteBstWouldAddEndPuncttrue
\mciteSetBstMidEndSepPunct{\mcitedefaultmidpunct}
{\mcitedefaultendpunct}{\mcitedefaultseppunct}\relax
\EndOfBibitem
\bibitem[Hay and Wadt(1985)]{hay85a}
P.~J. Hay and W.~R. Wadt, \emph{J.~Chem.~Phys.}, 1985, \textbf{82}, 270\relax
\mciteBstWouldAddEndPuncttrue
\mciteSetBstMidEndSepPunct{\mcitedefaultmidpunct}
{\mcitedefaultendpunct}{\mcitedefaultseppunct}\relax
\EndOfBibitem
\bibitem[Figgen \emph{et~al.}(2009)Figgen, Peterson, Dolg, and Stoll]{figgen09}
D.~Figgen, K.~A. Peterson, M.~Dolg and H.~Stoll, \emph{J.~Chem.~Phys.}, 2009,
  \textbf{130}, 164108\relax
\mciteBstWouldAddEndPuncttrue
\mciteSetBstMidEndSepPunct{\mcitedefaultmidpunct}
{\mcitedefaultendpunct}{\mcitedefaultseppunct}\relax
\EndOfBibitem
\bibitem[Hay and Wadt(1985)]{hay85b}
P.~J. Hay and W.~R. Wadt, \emph{J.~Chem.~Phys.}, 1985, \textbf{82}, 299\relax
\mciteBstWouldAddEndPuncttrue
\mciteSetBstMidEndSepPunct{\mcitedefaultmidpunct}
{\mcitedefaultendpunct}{\mcitedefaultseppunct}\relax
\EndOfBibitem
\bibitem[Hamann \emph{et~al.}(1979)Hamann, Schl{\"u}ter, and Chiang]{hamann79}
D.~R. Hamann, M.~Schl{\"u}ter and C.~Chiang, \emph{Phys.~Rev.~Lett.}, 1979,
  \textbf{43}, 1494\relax
\mciteBstWouldAddEndPuncttrue
\mciteSetBstMidEndSepPunct{\mcitedefaultmidpunct}
{\mcitedefaultendpunct}{\mcitedefaultseppunct}\relax
\EndOfBibitem
\bibitem[Anderson \emph{et~al.}(1999)Anderson, Bai, Bischof, Blackford,
  Dongarra, Greenbaum, Hammarling, A.~McKenney, and Sorensen]{anderson99}
E.~Anderson, Z.~Bai, C.~Bischof, S.~Blackford, J.~Dongarra, J.~D.~C.~A.
  Greenbaum, S.~Hammarling, A.~A.~McKenney and D.~Sorensen, \emph{LAPACK Users'
  Guide}, SIAM, 1999, vol.~9\relax
\mciteBstWouldAddEndPuncttrue
\mciteSetBstMidEndSepPunct{\mcitedefaultmidpunct}
{\mcitedefaultendpunct}{\mcitedefaultseppunct}\relax
\EndOfBibitem
\bibitem[Frigo and Johnson(2005)]{fftw05}
M.~Frigo and S.~G. Johnson, \emph{Proceedings of the IEEE}, 2005, \textbf{93},
  216\relax
\mciteBstWouldAddEndPuncttrue
\mciteSetBstMidEndSepPunct{\mcitedefaultmidpunct}
{\mcitedefaultendpunct}{\mcitedefaultseppunct}\relax
\EndOfBibitem
\bibitem[Schmidt \emph{et~al.}(1993)Schmidt, Baldridge, Boatz, Elbert, Gordon,
  Hensen, Koseki, Matsunaga, Nguyen, Su, Windus, Dupuis, and
  Montgomery]{schmidt93}
M.~W. Schmidt, K.~K. Baldridge, J.~A. Boatz, S.~T. Elbert, M.~S. Gordon, J.~H.
  Hensen, S.~Koseki, N.~Matsunaga, K.~A. Nguyen, S.~J. Su, T.~L. Windus,
  M.~Dupuis and J.~A. Montgomery, \emph{J.~Comput.~Chem.}, 1993, \textbf{14},
  1347\relax
\mciteBstWouldAddEndPuncttrue
\mciteSetBstMidEndSepPunct{\mcitedefaultmidpunct}
{\mcitedefaultendpunct}{\mcitedefaultseppunct}\relax
\EndOfBibitem
\bibitem[Glaesemann and Gordon(1998)]{glaesemann98}
K.~R. Glaesemann and M.~S. Gordon, \emph{J.~Chem.~Phys.}, 1998, \textbf{108},
  9959\relax
\mciteBstWouldAddEndPuncttrue
\mciteSetBstMidEndSepPunct{\mcitedefaultmidpunct}
{\mcitedefaultendpunct}{\mcitedefaultseppunct}\relax
\EndOfBibitem
\bibitem[Miehlich \emph{et~al.}(1989)Miehlich, Savin, Stoll, and
  Preuss]{miehlich89}
B.~Miehlich, A.~Savin, H.~Stoll and H.~Preuss, \emph{Chem.~Phys.~Lett.}, 1989,
  \textbf{157}, 200\relax
\mciteBstWouldAddEndPuncttrue
\mciteSetBstMidEndSepPunct{\mcitedefaultmidpunct}
{\mcitedefaultendpunct}{\mcitedefaultseppunct}\relax
\EndOfBibitem
\bibitem[Filatov and Thiel(1997)]{filatov97a}
M.~Filatov and W.~Thiel, \emph{Int.~J.~Quant.~Chem.}, 1997, \textbf{62},
  603\relax
\mciteBstWouldAddEndPuncttrue
\mciteSetBstMidEndSepPunct{\mcitedefaultmidpunct}
{\mcitedefaultendpunct}{\mcitedefaultseppunct}\relax
\EndOfBibitem
\bibitem[Filatov and Thiel(1997)]{filatov97b}
M.~Filatov and W.~Thiel, \emph{Mol.~Phys.}, 1997, \textbf{91}, 847\relax
\mciteBstWouldAddEndPuncttrue
\mciteSetBstMidEndSepPunct{\mcitedefaultmidpunct}
{\mcitedefaultendpunct}{\mcitedefaultseppunct}\relax
\EndOfBibitem
\bibitem[Repository(2001)]{repository}
D.~F. Repository, \emph{Quantum Chemistry Group}, CCLRC Daresbury Laboratory,
  {Daresbury, Cheshire, UK}, 2001\relax
\mciteBstWouldAddEndPuncttrue
\mciteSetBstMidEndSepPunct{\mcitedefaultmidpunct}
{\mcitedefaultendpunct}{\mcitedefaultseppunct}\relax
\EndOfBibitem
\bibitem[(Eds.)(2006)]{johnson06}
R.~D.~J.~I. (Eds.), \emph{NIST Computational Chemistry Comparisons and
  Benchmark Database, NIST Standard Reference Database, Number, Release 14},
  NIST, {Gaithersburg, MD}, 2006\relax
\mciteBstWouldAddEndPuncttrue
\mciteSetBstMidEndSepPunct{\mcitedefaultmidpunct}
{\mcitedefaultendpunct}{\mcitedefaultseppunct}\relax
\EndOfBibitem
\bibitem[Afeefy \emph{et~al.}(2005)Afeefy, Liebman, S.~E.~Stein, and
  (Eds.)]{afeefy05}
H.~Y. Afeefy, J.~E. Liebman, i.~P.~J.~L. S.~E.~Stein and W.~G.~M. (Eds.),
  \emph{NIST Chemistry Webbook, NIST Standard Reference Database, Number 69},
  NIST, {Gaithersburg, MD}, 2005\relax
\mciteBstWouldAddEndPuncttrue
\mciteSetBstMidEndSepPunct{\mcitedefaultmidpunct}
{\mcitedefaultendpunct}{\mcitedefaultseppunct}\relax
\EndOfBibitem
\bibitem[Leeuwen and Baerends(1994)]{leeuwen94}
R.~V. Leeuwen and E.~J. Baerends, \emph{Phys.~Rev.~A}, 1994, \textbf{49},
  2421\relax
\mciteBstWouldAddEndPuncttrue
\mciteSetBstMidEndSepPunct{\mcitedefaultmidpunct}
{\mcitedefaultendpunct}{\mcitedefaultseppunct}\relax
\EndOfBibitem
\bibitem[Schipper \emph{et~al.}(2000)Schipper, Gritsenko, Gisbergen, and
  Baerends]{schipper00}
P.~R.~T. Schipper, O.~V. Gritsenko, S.~J.~A.~V. Gisbergen and E.~J. Baerends,
  \emph{J.~Chem.~Phys}, 2000, \textbf{112}, 1344\relax
\mciteBstWouldAddEndPuncttrue
\mciteSetBstMidEndSepPunct{\mcitedefaultmidpunct}
{\mcitedefaultendpunct}{\mcitedefaultseppunct}\relax
\EndOfBibitem
\end{mcitethebibliography}

\end{document}